\documentclass[aps,prd,preprint,floatfix,superscriptaddress,nofootinbib,longbibliography,a4paper]{revtex4-1}
\usepackage{xcolor}
\usepackage{orcidlink}
\usepackage{graphicx}
\usepackage{textcomp}
\usepackage{float}
\usepackage{multirow}
\usepackage{subfigure}
\usepackage{feynmp}
\usepackage{amsmath,amssymb,array}
\usepackage{enumerate}
\usepackage{hhline}
\usepackage{geometry}
\usepackage{hyperref}
\hypersetup{
    citecolor= red,
    colorlinks=true,
    linkcolor=blue,
    filecolor=magenta,      
    urlcolor=blue}

\usepackage{xfrac}
\usepackage{slashed}
\usepackage{rotating}
\newcommand{\nn}{\nonumber}
\newcommand{\beqa}{\begin{eqnarray}}
\newcommand{\eeqa}{\end{eqnarray}}
\newcommand{\be}{\begin{equation}}
\newcommand{\ee}{\end{equation}}
\newcommand{\ba}{\begin{array}} 
\newcommand{\ea}{\end{array}}

\let \bar \overline

\newcommand{\hc}{\ensuremath{\text{H.c.}}}

\def \YL {{Y_1^L}}
\def \YR {{Y_1^R}}
\def \Y2 {{Y_2}}

\newcommand{\yl}[1]{{y^{L}_{#1}}} 
\newcommand{\yr}[1]{{y^{R}_{#1}}}

\newcommand{\Yl}[1]{{\left( Y^{L}_1\right)_{#1}}} 
\newcommand{\Yr}[1]{{\left(Y^{R}_1\right)_{#1}}} 
\newcommand{\Yt}[1]{{\left(Y_{2}\right)_{#1}}}

\newcommand{\Ylt}[1]{{\left(\widetilde{Y_1^{L}}\right)_{#1}}}

\begin{document} 
\vspace*{0.5cm}
\preprint{CETUP-2024-004}
\title{Neutrinoless Double Beta Decay from Scalar Leptoquarks: Interplay with Neutrino Mass and Flavor Physics}
\bigskip
\author{P.~S. Bhupal Dev \orcidlink{0000-0003-4655-2866}}
\email{bdev@wustl.edu}
\affiliation{Department of Physics and McDonnell Center for the Space Sciences, \\
Washington University, St.~Louis, Missouri 63130, USA}
\author{Srubabati Goswami \orcidlink{0000-0002-5614-4092}}
\email{sruba@prl.res.in}
\affiliation{Theoretical Physics Division, Physical Research Laboratory\\ Navrangpura, Ahmedabad 380009, India}
\author{Chayan Majumdar \orcidlink{0000-0003-3505-5485}}
\email{c.majumdar@ucl.ac.uk}
\affiliation{Department of Physics \& Astronomy, University College London, WC1E 6BT London, UK}
\author{Debashis Pachhar \orcidlink{0000-0001-8931-5321}}
\email{debashispachhar@prl.res.in}
\affiliation{Theoretical Physics Division, Physical Research Laboratory\\ Navrangpura, Ahmedabad 380009, India}
\affiliation{Indian Institute of Technology Gandhinagar, Palaj 382055, India}

\begin{abstract}
We perform a comprehensive analysis of  neutrinoless double beta ($0\nu\beta\beta$) decay and its interplay with low-energy flavor observables in a radiative neutrino mass model with scalar leptoquarks $S_1(\bar{3},1,1/3)$ and $\widetilde{R}_2(3,2,1/6)$. 
We carve out the parameter region consistent with constraints from neutrino mass and mixing, collider searches, as well as measurements of several flavor observables, such as  muon and electron  anomalous magnetic moments, charged lepton flavor violation and rare (semi)leptonic kaon and $B$-meson decays, including the recent anomalies in $R_{D^{(*)}}$ and $B\to K\nu\bar{\nu}$ observables.  We perform a global analysis  to all the existing constraints and show the (anti)correlations between all relevant Yukawa couplings satisfying these restrictions.
We find that the most stringent constraint on the parameter space comes from 
$\mu \to e$ conversion in nuclei and $K^{+} \rightarrow\, \pi^{+}\nu \bar{\nu}$ decay. We also point out a tension between the muon and electron $(g-2)$ anomalies in this context. 
Taking benchmark values from the combined allowed regions, we study the implications for $0\nu\beta\beta$ decay including both the canonical light neutrino and the leptoquark contributions. We find that for normal ordering of neutrino masses, the 
leptoquark contribution removes the cancellation region that occurs for the canonical case. The  effective mass in presence of leptoquarks can lie in the desert region between the standard normal and inverted ordering cases, and this can be probed in future ton-scale experiments like LEGEND-1000 and nEXO. 

\end{abstract} 
\maketitle
\tableofcontents
\section{Introduction}
\label{sec:intro}
The Standard Model (SM) of elementary particle physics successfully describes the short-distance physics probed so far in collider experiments up to the TeV scale. Nevertheless, there exist several empirical observations, as well as theoretical arguments, for the existence of some beyond-the-SM (BSM) physics. Perhaps the strongest piece of evidence comes from the discovery of neutrino oscillations~\cite{Super-Kamiokande:1998kpq, SNO:2002tuh}, which implies nonzero neutrino mass. This definitely requires BSM physics because the neutrino mass is exactly zero in the SM due to the absence of right-handed neutrinos and presence of an accidental global $B-L$ symmetry. 

A simple way to generate nonzero neutrino mass is by breaking the $B-L$ symmetry of the SM via higher-dimensional non-renormalizable operators, suppressed by a high energy scale that characterizes lepton number violation (LNV). At dimension five, there is only one such operator, namely, the Weinberg operator~\cite{Weinberg:1979sa} $\mathcal{O}_5=L^iL^jH^kH^l\epsilon_{ik}\epsilon_{jl}$,  
where $\ell_L = \left( \nu_L, \ e_L \right)^T $ is the SM lepton doublet, $H = \left( H^{+}, \ H^0 \right)^T $ is the Higgs doublet, 
and $\epsilon_{ik}$ is the $SU(2)_L$ antisymmetric tensor. 
After electroweak symmetry breaking, once the vacuum expectation value $v$ of the Higgs field is inserted, this operator gives rise to Majorana neutrino masses of order $m_\nu\simeq y_D^2 v^2/\Lambda$, where $y_D$ is the Dirac Yukawa coupling and $\Lambda$ is the scale of new physics. To obtain sub-eV neutrino masses, as required to explain the oscillation data~\cite{nufit, ParticleDataGroup:2022pth}, as well as to be consistent with the cosmological constraints on the sum of neutrino masses \cite{Planck:2018vyg}, the scale $\Lambda$ should be around $10^{14}$ GeV with $y_D \sim \mathcal{O}(1)$. This is the conventional seesaw mechanism~\cite{Minkowski:1977sc, Mohapatra:1979ia, Yanagida:1979as, Gell-Mann:1979vob} for tree-level neutrino mass generation. However, the scale of new physics can in principle be much lower, depending on the mass of the heavy integrated-out particle. For instance, in the type-I seesaw, for fine-tuned Yukawa $y_D\sim 10^{-6}$, the scale $\Lambda$ can be as low as the electroweak scale. There also exists alternative ways to generate low-scale seesaw with $\mathcal{O}(1)$ Yukawa couplings, e.g., in inverse~\cite{Mohapatra:1986aw, Mohapatra:1986bd}, linear~\cite{Akhmedov:1995ip, Malinsky:2005bi} or extended~\cite{Gavela:2009cd, Barry:2011wb, Kang:2006sn, Dev:2012sg} seesaw models. 

An alternative way to naturally lower the new physics scale is by invoking higher-dimensional operators beyond dimension five. This is the case with radiative neutrino mass models~\cite{Zee:1980ai,Babu:1988ki,Cai:2017jrq}, where the tree-level Lagrangian does not generate the operator ${\cal O}_5$ due to the particle spectrum of the considered models, and nonzero neutrino mass is only induced at loop level, which automatically comes with a loop suppression factor, and typically with an additional chiral suppression. This can naturally bring down the new physics scale to the experimentally accessible range, with far-reaching phenomenological implications in both low- and high-energy experiments~\cite{Babu:2019mfe}. 

A prime example of such radiative neutrino mass models with TeV-scale new physics involves  leptoquarks~\cite{Cai:2017jrq}, which are theoretically well-motivated BSM particles that emerge naturally in extensions of the SM such as technicolor and composite models~\cite{Dimopoulos:1979es} or models that unify quarks and leptons~\cite{Pati:1973uk, Fritzsch:1974nn}; see Ref.~\cite{Dorsner:2016wpm} for a review on leptoquarks. Here, we consider one such model with two scalar leptoquarks  $S_1(\bar{3},1,1/3)$ and  $\widetilde{R}_2(3,2,1/6)$~\cite{Zhang:2021dgl, Parashar:2022wrd}, where the charges are given under the SM gauge group $SU(3)_c\times SU(2)_L\times U(1)_Y$. This model is known to generate neutrino mass at one-loop level via a dimension-7 operator of the type ${\cal O}_7=H(LQ)(Ld^c)$~\cite{AristizabalSierra:2007nf, Cai:2014kra, Dorsner:2017wwn}, where $Q=(u_L,d_L)^T$ is the SM quark doublet.\footnote{The same one-loop mechanism is also realized in $R$-parity violating supersymmetric models~\cite{Hall:1983id}, where the $S_1^{1/3}$ and $\widetilde{R}_2^{-1/3}$ components are identified as the down-type squarks $\widetilde{d}^c$ and $\widetilde{d}$, respectively.} At the same time, these leptoquarks
with lepton number violating (LNV) and lepton flavor violating (LFV) interactions also contribute to a variety of low-energy rare processes, namely the LNV process like $0\nu\beta\beta$ decay~\cite{Hirsch:1996ye}, LFV processes~\cite{Lavoura:2003xp, Benbrik:2008si,De:2024foq} like $\mu\to e\gamma$, $\mu \rightarrow e$ conversion in nuclei as well as $\ell \rightarrow \ell^\prime gg$ decay, and semileptonic $K$ and $B$-meson decays~\cite{Fajfer:2018bfj, Descotes-Genon:2020buf, Deppisch:2020oyx,Crivellin:2019dwb,Crivellin:2017zlb,Crivellin:2022mff,Crivellin:2020mjs,ColuccioLeskow:2016dox}. In fact, the recent hints of lepton flavor universality violation (LFUV) in charged-current semileptonic $B$-meson decays, defined by the ratio of branching ratios (BRs), $R_{D^{(*)}}=\frac{{\rm BR}(\bar{B}\to D^{(*)}\tau \bar{\nu})}{{\rm BR}(\bar{B}\to D^{(*)}\ell \bar{\nu})}$ with $\ell=e,\mu$, by BaBar~\cite{BaBar:2013mob}, LHCb~\cite{LHCb:2023zxo},  Belle~\cite{Belle:2019rba} and Belle-II~\cite{Belle-II:2024ami}, and the muon $(g-2)$ anomaly~\cite{Muong-2:2006rrc, Muong-2:2023cdq,Crivellin:2020tsz}, have led to a deluge of papers invoking the $S_1$ leptoquark as a possible explanation of both anomalies~\cite{Sakaki:2013bfa, Bauer:2015knc, Popov:2016fzr, Cai:2017wry, Altmannshofer:2017poe, Angelescu:2021lln} (see Ref.~\cite{Capdevila:2023yhq} for a recent review). Note that the combination of $S_3(\bar{3},3,1/3)$ and $\widetilde{R}_2(3,2,1/6)$ leptoquarks also works for neutrino mass generation~\cite{Cai:2014kra} and gives additional contribution to $0\nu\beta\beta$~\cite{Hirsch:1996ye};  however $S_3$ cannot explain the $R_{D^{(*)}}$ anomaly~\cite{Angelescu:2018tyl}, and therefore, we choose to work with the $(S_1,\widetilde{R}_2)$ combination. 

In this paper, we perform a detailed study of the $0\nu\beta\beta$ predictions in presence of the scalar leptoquarks $S_1$ and  $\widetilde{R}_2$. The observation of $0\nu\beta\beta$ would be a clean signature of the Majorana nature of neutrinos~\cite{Schechter:1981bd}, and hence, of BSM physics. So far, only lower limits on the $0\nu\beta\beta$ half-life ($T^{0\nu}_{1/2}$) have been set using different isotopes~\cite{EXO-200:2019rkq, GERDA:2020xhi, CUORE:2022piu, CUPID:2022puj, Majorana:2022udl,   KamLAND-Zen:2024eml},  with the most stringent one being  $T_{1/2}^{0\nu}> 3.8\times 10^{26}$ yrs from KamLAND-Zen using $^{136}$Xe~\cite{KamLAND-Zen:2024eml}. The next-generation ton-scale $0\nu\beta\beta$ experiments such as nEXO~\cite{nEXO:2021ujk} and LEGEND-1000~\cite{LEGEND:2021bnm} are projected to reach half-life sensitivities up to $10^{28}$ yrs, thus providing an unprecedented opportunity to probe the Majorana nature of the neutrino mass and the associated BSM physics signatures~\cite{Agostini:2022zub}. However, in the presence of BSM contributions to $0\nu\beta\beta$, such as in the leptoquark model under consideration, the interpretation of the experimental result on $T_{1/2}^{0\nu}$ in terms of the effective Majorana mass parameter requires a careful study.

Going beyond earlier works on the study of $0\nu\beta\beta$ in leptoquark models~\cite{Hirsch:1996ye, Helo:2013ika, Pas:2015hca, Graesser:2022nkv, Graf:2022lhj, Scholer:2023bnn,Li:2023wfi}, we analyze the interplay of $0\nu\beta\beta$ with neutrino mass and flavor observables, including the $R_{D^{(*)}}$ and $(g-2)$ anomalies. For $0\nu\beta\beta$, we include two types of long-range contributions via light-neutrino exchange mechanism. One is mediated via two standard model $V-A$ vertices, the other via one standard model and one leptoquark vertex. There can be either constructive or destructive interference between these contributions, depending on the signs of the leptoquark Yukawa couplings involved. This leads to interesting correlations between the $0\nu\beta\beta$ and other low-energy flavor observables, which we study in detail for the first time. In particular, we impose the latest LHC constraints on the leptoquark masses, as well as the low-energy LFV and LFUV constraints on their Yukawa couplings, including the recently updated $R_{K^{(*)}}$ results from LHCb~\cite{LHCb:2022qnv, LHCb:2022vje} and $B^+\to K^+\nu\bar{\nu}$ results from Belle-II~\cite{Belle-II:2023esi}. We then perform a multi-dimensional scan over all the relevant leptoquark couplings and find some nontrivial (anti)correlations between them. We identify the parameter space that could simultaneously address the $R_{D^{(*)}}$ anomaly~\cite{BaBar:2013mob, Belle:2019rba, LHCb:2023zxo, Belle-II:2024ami}, as well as the electron~\cite{Parker:2018vye, Morel:2020dww} and muon~\cite{Muong-2:2023cdq} $(g-2)$ anomalies, while satisfying all other flavor constraints. It turns out that the $\mu\rightarrow e$ conversion in nuclei is the most constraining flavor observable in the parameter space of our interest. However, it still allows the leptoquark contribution to the $0\nu\beta\beta$ process to dominate over the canonical light neutrino contribution, for both normal and inverted orderings for some values of the parameter space.  We find that for our chosen benchmark scenario, the current KamLAND-Zen limit already excludes leptoquark masses below 3.3 TeV and 6.6 TeV for normal ordering (NO) and inverted ordering (IO), respectively when the lightest neutrino have zero mass. The future $0\nu\beta\beta$ experiments like nEXO can probe leptoquark masses up to ${\cal O}(10)$ TeV, beyond the reach of HL-LHC. 

It should be mentioned here that earlier attempts were made to simultaneously explain the neutral-current $B$-decay anomalies, most notably the $R_{K^{(*)}}$ anomaly, along with the $R_{D^{(*)}}$ and muon $(g-2)$ anomalies within the leptoquark model under consideration (or its extensions)~\cite{Bauer:2015knc,Popov:2016fzr, Cai:2017wry, Altmannshofer:2020axr, Dev:2021ipu}; however, since the $R_{K^{(*)}}$ anomaly has now disappeared~\cite{LHCb:2022qnv, LHCb:2022vje}, it gives an additional constraint on the leptoquark couplings, as we will see later. In the same vein, we will also consider a hypothetical situation where the current $R_{D^{(*)}}$ anomaly disappears, and will analyze its implications for the allowed model parameter space. In particular, we find that one cannot simultaneously satisfy both muon and electron $g-2$ anomalies in this $\tilde{R}_2-S_1 $ scenario due to the constraints coming from charged LFV (cLFV) processes, especially $\mu \rightarrow \, e \, \gamma$ and $\mu\to e$ conversion in nuclei, as well as from $K^{+} \rightarrow \, \pi^{+} \, \nu\,\bar{\nu}.$

The rest of the paper is organized as follows: In Section~\ref{sec:model} we briefly describe the leptoquark model under consideration. Section~\ref{sec:massgeneration} discusses how neutrino mass is generated at one-loop level in this model. In Section~\ref{sec:0vbb}, we present the $0\nu\beta\beta$ analysis in this model. In Section~\ref{sec:flavor}, we discuss the leptoquark contributions to various flavor observables, including cLFV, lepton $(g-2)$, neutrino electric and magnetic dipole moment and semileptonic $B$ and $K$-meson decays. Section~\ref{sec:results} presents our numerical analysis and results. In Section~\ref{sec:con}, we conclude with a summary and outlook.   

\section{The Model}
\label{sec:model}
The model we consider consists of the SM particle content augmented with two scalar leptoquarks 
\begin{align}
  S_1(\bar{3},1,1/3) \equiv S_1^{1/3}, \qquad  \widetilde{R}_2(3,2,1/6) \equiv 
  \begin{pmatrix}
      \widetilde{R}_2^{+2/3}\\
      \widetilde{R}_2^{-1/3}
  \end{pmatrix} ,
\end{align}
where we have expanded the $SU(2)_L$ doublet into its components, with the superscripts denoting the electric charges. The corresponding Yukawa interaction terms are given by the  Lagrangian 
\beqa
-\mathcal{L}_{\rm LQ} &\supset& Y_1^L \bar{Q}_L^c \epsilon \ell_L S_1 + Y_1^R \bar{u}_R^c e_R S_1 + Y_2 \bar{d}_R \widetilde{R}_2^T \epsilon \ell_L + \hc  \nn \\
&=& Y_1^L \,\bar{u}_L^c \,e_L \,S_1 - Y_1^L \,\bar{d}_L^c \,\nu_L \,S_1 +  Y_1^R \,\bar{u}_R^c \,e_R \,S_1 + Y_2\, \bar{d}_R\, \widetilde{R}_2^{2/3} \, e_L - Y_2\, \bar{d}_R \,\widetilde{R}_2^{-1/3} \,\nu_L + \hc 
\label{eq:leptoquarklag}
\eeqa
Here the flavor indices are suppressed and ${Y_1^{L,R}}$ and $Y_2 $ are $3\times3$ complex matrices describing the Yukawa interactions of the leptoquarks $S_1$ and $\widetilde{R}_2$, respectively. 

The scalar potential involving the leptoquarks is given by 
\beqa
\mathcal{V}_{\rm LQ} &=& m_1^2 S_1^{\dagger}S_1 \, + m_2^2 \widetilde{R}_2^{\dagger} \widetilde{R}_2 + \alpha_1 \left( H^{\dagger} H \right) \left(S_1^{\dagger} S_1 \right) +  \alpha_2 \left(   H^{\dagger} H \right) \left(\widetilde{R}_2^{\dagger} \widetilde{R}_2 \right) + \alpha_2' \left(H^{\dagger}\widetilde{R}_2\right) \left( \widetilde{R}_2^{\dagger} H \right) \nn \\
&& +\left( \kappa H^{\dagger} \widetilde{R}_2 S_1 + \hc \right) .\label{eq:pot}
\eeqa 
Here $\alpha_1$, $\alpha_2$ and $\alpha_2^\prime$ are real dimensionless couplings describing the strength of quartic interactions between the leptoquarks and the SM Higgs doublet, whereas the trilinear coupling $\kappa$ is in general complex with mass dimension one and plays a crucial role in the phenomenology of this model.

 Note that in Eq.~(\ref{eq:leptoquarklag}), the simultaneous presence of the Yukawa couplings $Y_1^L $, $Y_1^R $ and $Y_2 $, in association with the trilinear scalar coupling $\kappa$ in Eq.~\eqref{eq:pot}, violates lepton number in the model.\footnote{Under the assumption that the scalar sector only comprises the SM Higgs and scalar leptoquarks, this is one of the only two possibilities to break lepton number in the scalar sector~\cite{Chua:1999si, Mahanta:1999xd}, the other one being the combination of $\widetilde{R}_2(3,2,1/6)$ and $S_3(\bar{3},3,1/3)$.} Also, we keep only those terms in the Lagrangian that preserve baryon number, which leads to the absence of the diquark coupling terms with $S_1$, as well as the quartic term of the form $S_1^2 \widetilde{R}_2^\dagger H$. The presence of such terms can lead to fast proton decay~\cite{Dorsner:2012nq}, and naturally we wish to avoid that. To conserve baryon number throughout the framework, we can assign $B= -1/3$ to $S_1$ and $B= +1/3$ to $\widetilde{R}_2$ which ensures the absence of the $B$-violating terms in the Lagrangian at the renormalizable level. 
 
 The trilinear coupling in Eq.~\eqref{eq:pot} results in the mixing between the singlet ($S_1^{1/3}$) and the electromagnetic charge $1/3$ component of the doublet (${\widetilde{R}_2}^{1/3}$) leptoquark after the electroweak symmetry breaking. 
The mass matrix involving $S_1$ and ${\widetilde{R}_2}^{*1/3}$ is given by
\beqa
\mathcal{M}^2_{1/3} &=& \begin{pmatrix} m^2_{S_1}  & \frac{1}{\sqrt{2}} \kappa v  \\  \frac{1}{\sqrt{2}} \kappa v & m^2_{\widetilde{R}_2} 
 \end{pmatrix} ,
 \label{mass1}
\eeqa
where $m^2_{S_1} = m_1^2 + \frac 12 \alpha_1 v^2 $ and $m^2_{\widetilde{R}_2} =  m_2^2 + \frac 12 \left( \alpha_2 + \alpha_2' \right) v^2$, with $v=(\sqrt 2 G_F)^{-1/2}\simeq 246$ GeV, $G_F$ being the Fermi constant. The $2\times 2$ matrix~\eqref{mass1} can be diagonalized by a $2\times 2$ rotational matrix parameterized by one mixing angle ($\theta_{\rm LQ}$). It can be shown that
\beqa
\tan {2\theta_{\rm LQ}} &=& \frac{\sqrt{2} \kappa \, v}{m^{2}_{\widetilde{R}_2} - m^2_{S_1}} \, .
\label{eq:mixing}
\eeqa
The physical mass eigenstates are 
\beqa
X_1^{1/3} &=& \cos \theta_{\rm LQ} S_1^{1/3} - \sin \theta_{\rm LQ} \widetilde{R}_2^{* \, 1/3} ,  \nn \\ 
X_2^{1/3} &=&  \sin \theta_{\rm LQ}S_1^{1/3} + \cos \theta_{\rm LQ} \widetilde{R}_2^{* \, 1/3}  , \label{eq:leptoquark1}
\eeqa
and the corresponding mass eigenvalues are 
\beqa
m_{X_1,X_2}^2 &=& \frac 12 \left[ m^2_{S_1} + m^2_{\widetilde{R}_2} \mp \sqrt{\left(  m^2_{S_1} - m^2_{\widetilde{R}_2}\right)^2 + 2 \kappa^2 v^2}\right] . \label{eq:mx1mx2}
\eeqa
As for the charge 2/3 component of the $\widetilde{R}_2$ leptoquark, its mass is simply given by  
\beqa
m^2_{\widetilde{R}_2^{2/3}} &=& m_2^2 + \frac 12 \alpha_2 v^2 \, . \label{mass2}
\eeqa

The current LHC constraints on the leptoquark masses are of order ${\cal O}(1~{\rm TeV})$, depending on the Yukawa couplings~\cite{CMS:2018lab, CMS:2018iye,CMS:2018ncu, ATLAS:2019ebv, ATLAS:2020dsk, CMS:2020wzx, ATLAS:2021oiz, CMS:2022nty, ATLAS:2023vxj, ATLAS:2023prb}. This requires us to go to the decoupling limit with $m_1,m_2\gg v$. In this case, all three physical mass eigenstates $X_{1,2}, \widetilde{R}_2^{2/3}$ are quasi-degenerate. 
\section{Neutrino Mass Generation} \label{sec:massgeneration}
\begin{figure}[t!]
\centering
\includegraphics[scale=0.8]{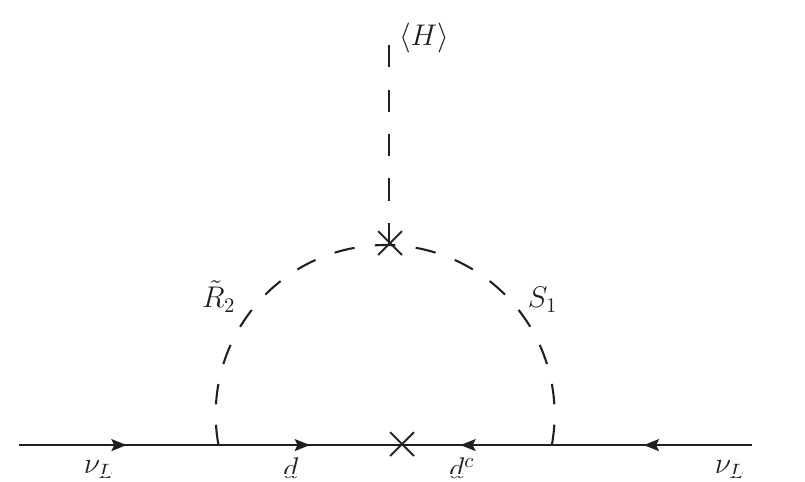}
    \caption{Radiatve Majorana mass generation for active neutrinos via the mixing of singlet-doublet scalar leptoquarks.}
    \label{fig:numass}
\end{figure}  
With a single leptoquark it is not possible to generate Majorana mass for light neutrinos. Therefore, we need two scalar leptoquarks in the model, which will run in the loop for radiative neutrino mass generation. The singlet-doublet leptoquark mixing, parameterized by $\theta_{\rm LQ}$ in Eq.~\eqref{eq:mixing}, which depends on the trilinear coupling $\kappa$, is crucial for this purpose, as noted earlier~\cite{Chua:1999si, Mahanta:1999xd}; see Fig.~\ref{fig:numass}. 

In this singlet-doublet scenario, only $\widetilde{R}_2^{1/3}$ and $S_1^{1/3}$ together can generate one-loop Majorana mass for active neutrinos, given by~\cite{Zhang:2021dgl,Parashar:2022wrd}
\beqa
\mathcal{M}_{\nu} &=& \frac{3\sin 2\theta_{\rm LQ}}{32 \pi^2}\ln\left( \frac{m_{X_1}^2}{m_{X_2}^2}\right) \left(\widetilde{  {Y_1^L}}^{T} \mathcal{M}_d Y_2 +   Y_2^T \mathcal{M}_d \widetilde{ Y_1^L } \right) .
\label{eq:mrel}
\eeqa

For large values of leptoquark masses i.e $m^2_{S_1} \approx m^2_{\tilde{R}_2} \approx m^2_{\rm LQ}$ Eq. (\ref{eq:mrel}) can be approximated as 
\beqa
m_{\nu} &=& 0.01 \,\text{eV} \left(\frac{\kappa}{m_{\rm LQ}} \right) \left( \frac{m_q}{m_b} \right)\, \left( \frac{  1 \, \text{TeV}}{m_{\rm LQ}} \right) \, \left(\frac{y_1^L y_2}{10^{-6}}\right) \, ,  \label{eq:BP_point}
\eeqa
where $y_1^L$ and $y_2$ are typical values of the elements of $\widetilde{Y_1^L}$ (or $Y_1^L)$ and $Y_2$, respectively.  Here $\widetilde{Y_1^L} = V\cdot Y_1^L$ ($V$ being the CKM quark mixing matrix) as we have considered the up-type quarks in their mass basis and $\mathcal{M}_d$ is the diagonal mass matrix for down-type quarks. One can obtain the diagonal mass matrix $\mathcal{M}_\nu^{\rm{diag}}$ (with eigenvalues $m_{\nu_i}$, where $i=1,2,3$) for light neutrinos by diagonalizing $\mathcal{M}_\nu$ with the usual PMNS lepton mixing matrix $U$ which is parameterized by three mixing angles $\theta_{12}, \theta_{13}, \theta_{23}$ and three phases i.e., one Dirac phase ($\delta_{\rm{CP}}$) and two Majorana phases ($\alpha, \beta$).

To understand the origin of the Majorana neutrino mass in this scenario, note that one can define a generalized fermion number for leptoquark states as $F=3B+L$ ~\cite{Dorsner:2016wpm} where $B, L$ correspond to baryon and lepton numbers, respectively. All the SM leptons carry $L=1, B=0$ while the quarks have $L=0, B=1/3$. In our singlet-doublet leptoquark scenario, we can easily deduce from Eq.~\eqref{eq:leptoquarklag} that $S_1$ carries $L=-1, B=-1/3$, while $\widetilde{R}_2$ carries $L=-1, B=1/3$. For $S_1$ the $F$ number is $-2$ while for $\widetilde{R}_2 $ it is 0. For the SM Higgs doublet, $B=L=0$. Therefore, within the loop in Fig.~\ref{fig:numass}, lepton number is violated at the trilinear Higgs-leptoquark-leptoquark vertex, which justifies the Majorana mass generation of neutrinos. The trilinear coupling $\kappa$ governs the size of the lepton number breaking, i.e.~lepton number symmetry is restored in the limit $\kappa\to 0$.  

\section{$0\nu\beta\beta$ Decay}
\label{sec:0vbb}
\begin{figure}[t!]
    \centering
    \includegraphics[scale=0.15]{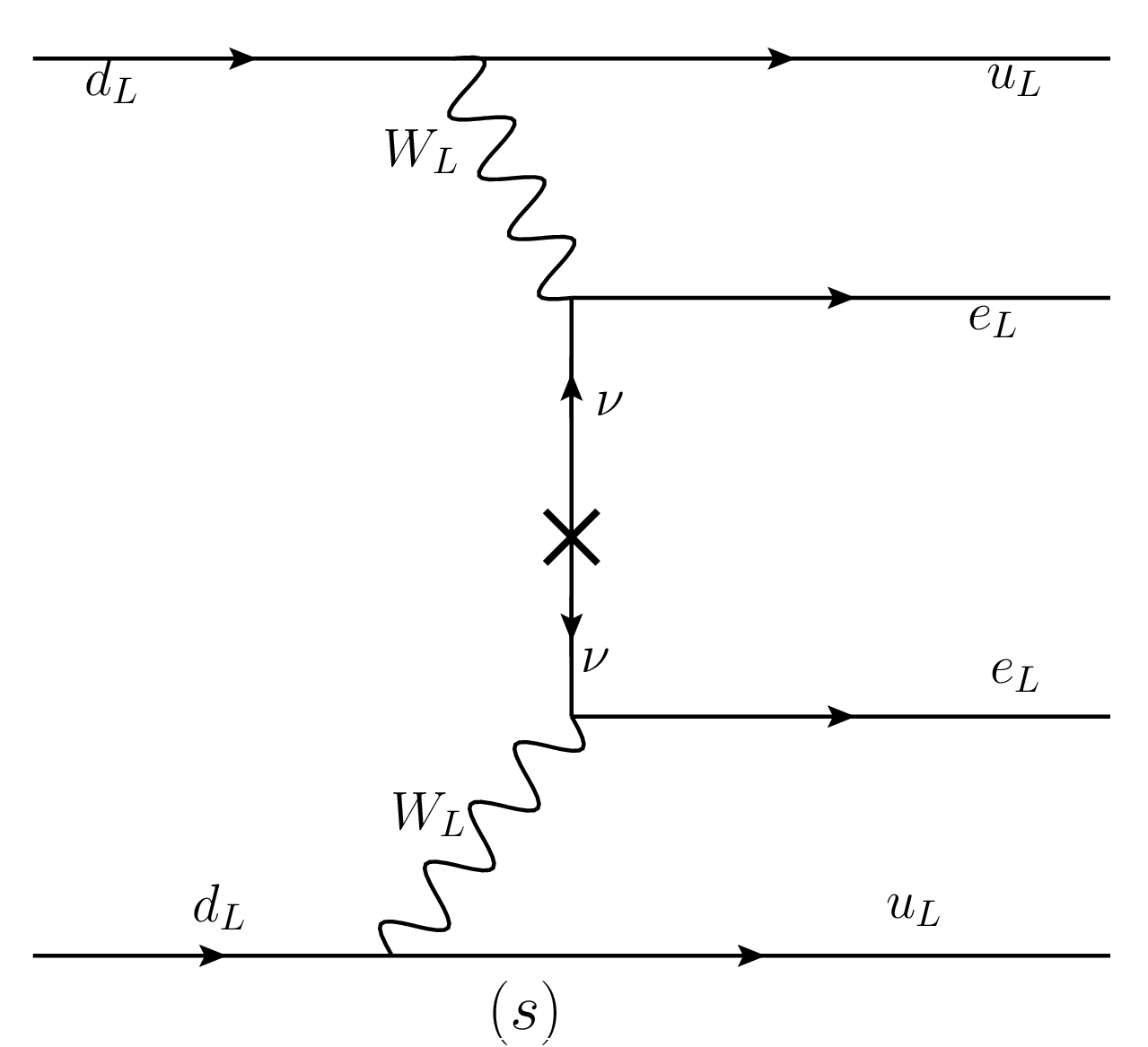}\\
    \includegraphics[scale=0.7]{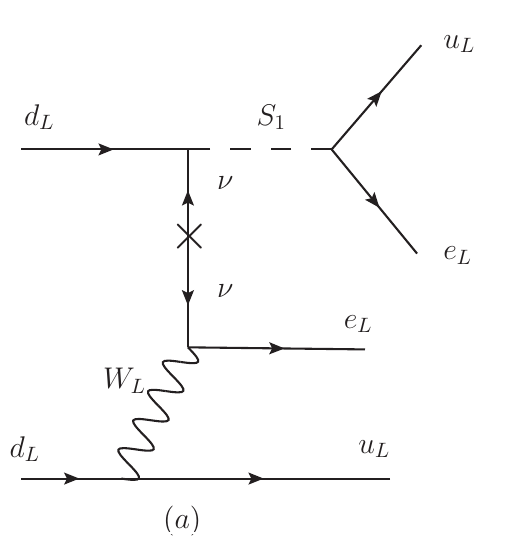}
    \includegraphics[scale=0.7]{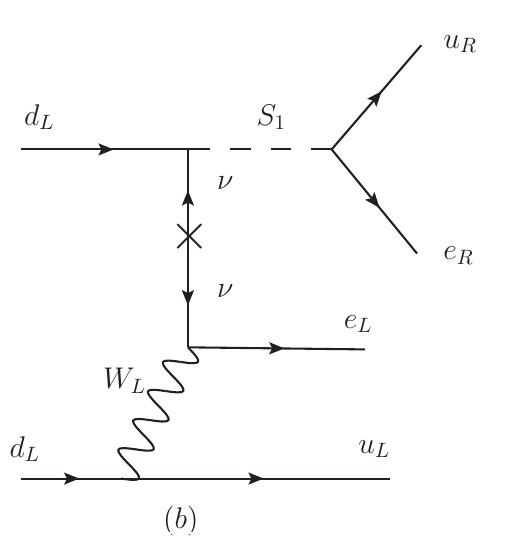}\\
    \includegraphics[scale=0.7]{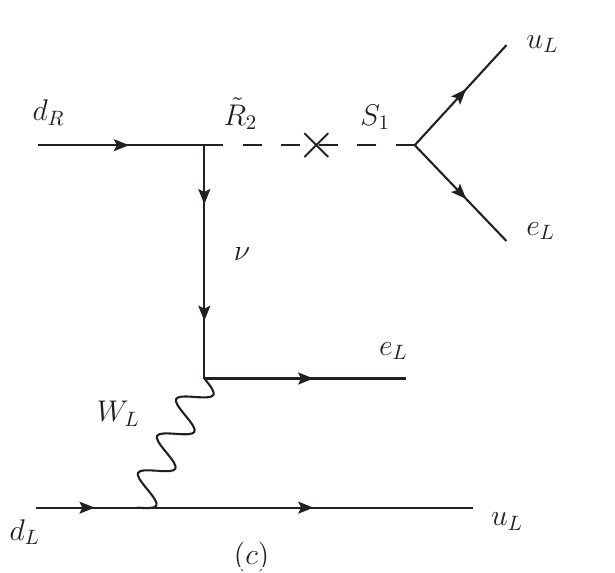}
    \includegraphics[scale=0.7]{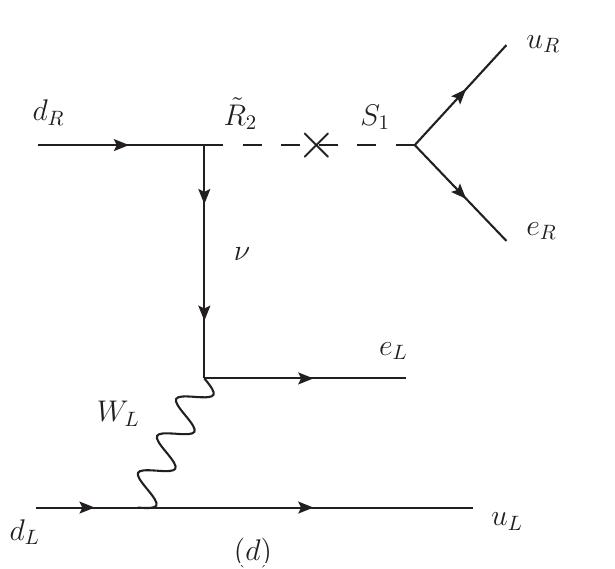}
    \caption{Feynman diagrams for different contributions to the $0\nu\beta\beta$ process in the leptoquark model. The top panel shows the canonical $W_L-W_L$ mediation for light Majorana neutrinos. The middle panel shows the purely $S_1$-mediated diagrams. The bottom panel shows the diagrams mediated by $S_1-\widetilde{R}_2$ mixing. Diagram (a) is purely dependent on $Y_1^L$ while diagram (c) arises from the combination of both $Y_1^L$ and $Y_2$ couplings, whereas (b) and (d) arise from the combination of $Y_1^L, Y_1^R$ and $Y_1^R, Y_2$ couplings, respectively. }
    \label{fig:0nbbfeynman}
    \end{figure}

With Majorana neutrinos, we have the canonical long-range $W_L-W_L$ contribution to $0\nu\beta\beta$, as shown in Fig.~\ref{fig:0nbbfeynman}(s). In the presence of $S_1$ and $\widetilde{R}_2$ leptoquarks, we have new long-range contributions to $0\nu\beta\beta$~\cite{Hirsch:1996ye} mediated by $W_L-S_1$ exchange, as shown in Figs.~\ref{fig:0nbbfeynman}(a) and (b), as well as from the mixing between $S_1^{1/3}$ and $\widetilde{R}_2^{1/3}$, as shown in Figs.~\ref{fig:0nbbfeynman}(c) and (d). Diagram (a) is purely dependent on $Y_1^L$ while diagram (c) arises from the combination of both $Y_1^L$ and $Y_2$ couplings, whereas (b) and (d) arise from the combination of $Y_1^L, Y_1^R$ and $Y_1^R, Y_2$ couplings, respectively [cf.~Eq.~\eqref{eq:leptoquarklag}].

The most general Low Energy Effective Theory (LEFT) Lagrangian for long range mechanisms can be written as \cite{Pas:1997fx,Ali:2007ec,Kotila:2021xgw}
\beqa
-{\cal L}_{\rm eff} \supset \frac{G_F \,V_{ud}}{\sqrt{2}} \, \Big[ j^{\mu}_{V-A}J_{V-A,\mu} + \sum_{\alpha,\beta} \epsilon_{\alpha}^{\beta} \, j_{\beta} \, J_{\alpha} + \hc .\Big], \label{eq:LEFT}
\eeqa
where the leptonic and hadronic currents are defined as $j_{\beta} = \bar e {\cal O}_{\beta} \nu_e$ and $J_{\alpha} = \bar u {\cal O}_{\alpha} d$. The Greek indices $\beta ,\alpha$ can be $V\mp A, \, S\mp P,\, T\mp T_5$, where $V,A,S,P,T,T_5$ correspond to vector, axial-vector, scalar, pseudo-scalar, tensor, and axial-tensor respectively. 
In Eq.~(\ref{eq:LEFT}), we have separated the standard canonical long range mechanism (first term) from the non-standard contributions, with the $\epsilon_\alpha^\beta$ being the corresponding Wilson coefficients of the nonstandard operators.  

For the standard $W_L-W_L$ mediation channel, the leptonic and hadronic currents are both connected via the SM gauge boson $W_L$, and two electrons of the same helicity ($e_Le_L$) are emitted, as shown in Fig.~\ref{fig:0nbbfeynman}(s). In case of standard $0\nu\beta\beta$ mechanism, the half life of the decay is written as
\beqa \left(T^{0\nu}_{1/2}\right)_{\rm std.}^{-1} &=& g_A^4 G_{01} \, \left| {\cal M}_{\nu}^{(3)}\right|^2 \,\frac{\left|m_{ee}^{\rm{std}}\right|^2}{m^2_e},
\label{eq:12}
\eeqa
where $g_A=1.27$ is the axial coupling, $G_{01}$ is the phase space integral and $M_{\nu}^{(3)}$ is the nuclear matrix element (NME) for the standard light neutrino exchange mechanism, $m_e$ is mass of electron and $m_{ee}^{\rm{std}} = |\sum_i U_{ei}^2 m_{\nu_i}|$ represents the effective Majorana mass parameter for the canonical case. 

In the leptoquark-mediated channels, depending on whether the $Y_1^L$ or $Y_1^R$ couplings are involved, we can have either $e_Le_L$ or $e_Le_R$ in the final state, as shown in Fig.~\ref{fig:0nbbfeynman}(a-d). It is also seen from these figures that the amplitude of these diagrams are ${\cal O}(\epsilon)$, as one of the standard operators in Fig.~\ref{fig:0nbbfeynman}(s) is replaced by the new physics (NP) operator mediated by leptoquarks. 
We have calculated these NP operators below and also identified the corresponding Wilson coefficients which are matched directly to the LEFT scale $(\mu = \mu_{\rm ew})$ i.e. after integrating out the $W$ boson and the massive leptoquark ($X_1,X_2$) states.

In Fig.~\ref{fig:0nbbfeynman}, the lepton number violation (LNV) in diagrams (s)-(b) is introduced by the majorana mass of the neutrinos (shown by the mass insertion in the middle) and in diagrams (c)-(d) LNV is introduced through the $S_1-\Tilde{R}_2$ mixing (shown as a cross in the middle of the leptoquark line).
It can be seen from the middle panel of Fig.~\ref{fig:0nbbfeynman} that the amplitudes for diagrams (a) and (b) are proportional to the small neutrino mass and inversely proportional to the leptoquark mass squared, which gives an extra suppression relative to the standard mechanism. Hence, the contribution of these diagrams to $0\nu\beta\beta$ can be simply ignored. 
On the other hand, diagrams (c) and (d), being dependent on the neutrino momentum, give important contribution to $0\nu\beta\beta$ which, depending on the leptoquark masses and couplings, can be  dominant over the standard canonical contribution~\cite{Hirsch:1996ye}. 

The NP operator that contributes to the upper half of diagram (c) in Fig.~\ref{fig:0nbbfeynman} is given by
\beqa
\mathcal{O}^{(c)} =  \left( \frac{G_F}{\sqrt{2}} V_{ud} \right)\, \Big[\epsilon_{S+P}^{S+P}  \left( \bar{u} P_R d \right) \left( \bar{e} P_R \nu_e^c \right) + \epsilon_{T+T_5}^{T+T_5}   \left( \bar{u} \sigma^{\alpha\beta}P_R d \right) \left(\bar{e} \sigma_{\alpha\beta}P_R \nu_e^c \right) \Big],
\eeqa
where the dimensionless parameters $\epsilon_{S+P}^{S+P}$ and $\epsilon_{T+T_5}^{T+T_5}$ at the LEFT scale are given by
\beqa
\epsilon_{S+P}^{S+P} &=&   \frac{\cos\theta_{\rm LQ}\sin\theta_{\rm LQ}}{\sqrt{2} \, G_F} \left( Y_2 ^{*} \right)_{11}\, \left( {Y_1^L}^{*}\right)_{11}  \, \left( \frac{ 1}{m_{X_1}^2} -\frac{ 1}{m_{X_2}^2} \right) ,\label{eq:S+P}\\
\epsilon_{T+T_5}^{T+T_5} &=& - \frac 14 \,\epsilon_{S+P}^{S+P} ,\label{eq:TR}
\eeqa
corresponding to the $S+P$ and $T+T_5$ contributions, respectively. We can also recast Eq.~(\ref{eq:S+P}) using Eq.~(\ref{eq:mx1mx2}) in the $m_{S_1} = m_{\Tilde{R}_2} = m_{\rm LQ}$ limit,
\beqa
\epsilon_{S+P}^{S+P} &\simeq & \frac{\kappa\,v^3}{m_{\rm LQ}^4} \left( Y_2 ^{*} \right)_{11}\, \left( {Y_1^L}^{*}\right)_{11}.  \label{SMEFT_matching}
\eeqa

Similarly, the effective NP operator that generates diagram (d) in Fig.~\ref{fig:0nbbfeynman} is given by
\beqa
\mathcal{O}^{(d)} &=&  \left( \frac{G_F}{\sqrt{2}} V_{ud} \right) \, \epsilon_{V+A}^{V+A} \left(\bar{u} \gamma_{\alpha} P_R d \right)\left( \bar{e} \gamma^{\alpha} P_R \nu_e^c \right),
\eeqa
where the $V+A$ term is given by 
\beqa
\epsilon_{V+A}^{V+A} &=&  - \frac{\cos\theta_{\rm LQ}\sin\theta_{\rm LQ}}{ \sqrt{2} \, G_F} \left( Y_2 ^{*} \right)_{11}\, \left(Y_1^{R}\right)^\ast_{11}  \left( \frac{ 1}{m_{X_1}^2} -\frac{ 1}{m_{X_2}^2} \right) , \label{eq:v+A}
\eeqa
which can also be recast into a form similar to Eq.~\eqref{SMEFT_matching}.

After identifying the relevant LEFT operators and their Wilson coefficients, we can now calculate the half life using the master formula outlined in~\cite{Cirigliano:2018yza}: 
\beqa
\left(T^{0\nu}_{1/2}\right)_{\rm total}^{-1} &=& g_A^4 \, \sum_{k} G_{0k} \, \left| {\cal A}_k\left( \left\{ C_i\right\}\right)\right|^2,
\label{eq:half-life}
\eeqa
where $G_{0k}$ denote the atomic phase space factors and ${\cal A}_k\left(\left\{ C_i\right\} \right)$ are the sub-amplitudes which depend on the NMEs, low-energy constants (LECs) and the Wilson coefficients of the relevant operators; see Appendix~\ref{app:sub-amplitudes} for details. Here, we have used $\tt \nu DoBe $ \cite{Scholer:2023bnn}, a python package to calculate the half life for $^{136}$Xe, with the  NMEs  from the IBM2 model~\cite{Deppisch:2020ztt}. The running of the Wilson coefficients (especially $\epsilon_{S+P}^{S+P}$ and $\epsilon_{T+T_5}^{T+T_5}$) from $\mu=\mu_{\rm ew}$ to $\mu = 2$ GeV is taken care of by the $\tt \nu DoBe$ package. Note that in $\tt \nu DoBe$, the decay-rate formula is expressed in terms of the Wilson coefficients at the chiral symmetry-breaking scale $\mu\sim 2$ GeV, following Ref.~\cite{Cirigliano:2018yza}. 
From Eq.~(\ref{eq:half-life}), the effective Majorana mass can be extracted in a form silimar to Eq.~\eqref{eq:12}, i.e. 
\beqa
\left(T_{1/2}^{0\nu}\right)_{\rm total}^{-1} & = & g_A^4 G_{01}\,\left| {\cal M}_{\nu}^{(3)} \right|^2\, \frac{ \left| m_{ee}^{\rm eff}\right|^2}{m^2_e} \, .
\label{eq:thalf1}
\eeqa
where we have defined $ m_{ee}^{\rm eff} =  m_{ee}^{\rm std} +  m_{ee}^{\rm LQ}$. The advantage of writing it this way is that when the leptoquark contribution becomes sub-dominant i.e., $m_{ee}^{\rm eff}\to m_{ee}^{\rm std}$, we would recover the canonical result. In other words, the deviation of $m_{ee}^{\rm eff}$ from  $m_{ee}^{\rm std}$ gives a measure of the leptoquark contribution, including interference with the SM contribution.   
The lower limit on the half-life of $^{136}$Xe from the KamLAND-Zen experiment translates to an upper bound on $m_{ee}<(28-122)$ meV at 90\% CL~\cite{KamLAND-Zen:2024eml}. The uncertainty band in $m_{ee}$ is taking into account different NME calculations. The next-generation ton-scale experiment nEXO will reach a sensitivity of $T_{1/2}^{0\nu}> 1.35\times 10^{28}$ yrs at 90\% CL with 10 years of data taking~\cite{nEXO:2021ujk}. This translates into an upper limit on $m_{ee}< (4.7-20.7)$ meV. The LEGEND-1000 experiment using $^{76}$Ge isotope will reach a similar design sensitivity of $T_{1/2}^{0\nu}> 1.6\times 10^{28}$ yrs at 90\% CL~\cite{LEGEND:2021bnm}, which translates into an upper limit on $m_{ee}< (9-21)$ meV. We will use these numbers in our numerical analysis.   
\section{Flavor Observables}
\label{sec:flavor}
In this section, we present the relevant low-energy LFV and LFUV observables which constrain the leptoquark parameter space. There are additional constraints, such as those from perturbative unitarity~\cite{Lee:1977eg} and electroweak $T$-parameter~\cite{Peskin:1991sw}, which however are negligible for our choice of small trilinear coupling $\kappa$, and hence, are not shown here. 
\subsection{Charged Lepton Sector}
\label{sec:cLFV}
Here we consider the cLFV processes namely, $\mu\to e$ conversion, $\ell_\alpha\to \ell_\beta\gamma$, $\ell_\alpha\to \ell_\beta\ell_\gamma\ell_\delta$, and lepton anomalous magnetic moments. 
\subsubsection{\texorpdfstring{$\mu \rightarrow e$ Conversion in Nuclei}{mu-e Conversion in Nuclei}}
In this model, leptoquark can mediate $\mu \rightarrow \, e$ conversion inside nuclei, i.e. $\mu\, N \rightarrow \, e\, N$.  As this process occurs at tree level, it provides one of the most stringent constraints among the flavor observables on the Yukawa couplings involving first and second generation leptons. The conversion ratio  is denoted as 
\beqa
\left.\mathcal{R} \right|_{\mu \rightarrow \, e}^{N} &=& \frac{\Gamma^N \left(\mu \rightarrow \, e \right)}{\Gamma^N_{\rm capture}} \, , \label{eq:mutoeconversion}
 \eeqa
  where $N$ denotes a particular nucleus and $\Gamma^N_{\rm capture}$ implies the muon capture rate of that nucleus. 
  The low energy effective Lagrangian describing $\mu \, N \rightarrow \, e\, N$ can be written as~\cite{Kitano:2002mt}
  \beqa
  \mathcal{L}_{\bar{q}q\bar{e}\mu} &\supset & - \sqrt{2} G_F \sum_{q=u,d,s} \sum_{X,Y=L,R} \left[  C_{V_{XY}}^{q} \left(\bar{e} \, \gamma^{\mu} \, P_X \,  \mu \right)\, \left(\bar{q}  \, \gamma_{\mu} \, \, P_Y q\right) \, + C_{S_{XY}}^{q} \left(\bar{e} \, P_X \,  \mu \right)\, \left(\bar{q}  \, P_Y q\right) \, \right. \nn \\
  &  &  +  \left. C_{T_{XY}}^{q} \left(\bar{e} \, \sigma^{\mu\nu} \, P_X \,  \mu \right)\, \left(\bar{q}  \, \sigma_{\mu\nu} \, \, P_Y q\right) \, \right] + \hc
  \eeqa
  Here, the scalar operators and operators with the gluon field strength tensor are neglected because they are suppressed in this model~\cite{Plakias:2023esq}. The relevant Wilson coefficients in our framework are given by  
  \beqa
  C_{V_{LL}}^u &=& \frac{v^2}{2} \left(Y_1^L\right)_{12} \left(Y_1^L\right)_{11}^{\ast} \, \left( \frac{\cos^2\theta_{\rm LQ}}{m^2_{X_1}} + \frac{\sin^2\theta_{\rm LQ}}{m^2_{X_2}} \right) ,\label{eq:cVLL} \\
  C_{V_{LR}}^d &=& -\frac{v^2}{2} \frac{\left(Y_2 \right)_{12} \left(Y_2 \right)_{11}^{\ast}}{m^2_{\widetilde{R}_2^{2/3}} }  , \label{eq:cVLR}\\
    C_{V_{RR}}^u &=& \frac{v^2}{2} \left(Y_1^R\right)_{12} \left(Y_1^R\right)_{11}^{\ast} \left( \frac{\cos^2\theta_{\rm LQ}}{m^2_{X_1}} + \frac{\sin^2\theta_{\rm LQ}}{m^2_{X_2}} \right) \label{eq:cVRR} \, ,\\
       C_{S_{LL}}^u &=& -\frac{v^2}{4} \left(Y_1^L\right)_{12} \left(Y_1^R\right)_{11}^{\ast} \left( \frac{\cos^2\theta_{\rm LQ}}{m^2_{X_1}} + \frac{\sin^2\theta_{\rm LQ}}{m^2_{X_2}} \right) \label{eq:cSLL} \, ,\\
       C_{S_{RR}}^u &=& -\frac{v^2}{4} \left(Y_1^R\right)_{12} \left(Y_1^L \right)_{11}^{\ast} \left( \frac{\cos^2\theta_{\rm LQ}}{m^2_{X_1}} + \frac{\sin^2\theta_{\rm LQ}}{m^2_{X_2}} \right) \label{eq:cSRR} \, , \\
       C_{T_{LL}}^u &=&  - \frac 14 C_{S_{LL}}^u \, \, , \qquad  C_{T_{RR}}^u = -   \frac 14  C_{S_{RR}}^u  \, . \label{eq:cTLL}
  \eeqa
For the coherent conversion process in which the initial and final nucleus are the same, the tensor contribution vanishes and the total conversion ratio can be written as ~\cite{Kitano:2002mt} 
  \beqa
 \left. \mathcal{R}\right|_{\mu \rightarrow \, e}^{N} &=& \frac{2\, G_F^2 \, m_{\mu}^5}{\Gamma_{\rm capture}^N \, } \, \left| C_{SL}^{p} \, S^p + C_{SL}^n \, S^n + C_{VL}^{p} \, V^p + C_{VL}^n \, V^n  \, \right|^2 + L \leftrightarrow R , \label{eq:mueconversion}
\eeqa
where the $S^p$ , $S^n$ , $V^p$ and $V^n$ are called the overlap integrals whose values along with $\Gamma_{\rm{capture}}^N$ are given in table (\ref{tab:conv}) for two nuclei. Here the coefficients are defined as :
  \beqa
  C_{SL}^{p} &=& \sum_{q=u,d} G_{S}^{q,p} \, \frac 12 C_{S_{LL}}^{q}  \\ 
   C_{SL}^{n} &=& \sum_{q=u,d} G_{S}^{q,n} \, \frac 12 C_{S_{LL}}^{q} \\
  C_{VL}^p &=& \left[ C_{V_{LL}}^u + \frac 12 C_{V_{LR}}^d \right], \\
   C_{VL}^n &=& \left[ \frac 12 C_{V_{LL}}^u +  C_{V_{LR}}^d \right], \\
  \eeqa
  and similarly for $L\rightarrow \, R $. The coefficients of the scalar operators are $G_{S}^{u,p}= G_{S}^{d,n}=5.1$ and $G_{S}^{u,n}= G_{S}^{d,p}= 4.3$. Currently, the most stringent limit on $\mu \rightarrow \,e $ conversion comes from the SINDRUM experiment using $^{197}_{79}$Au nucleus: $\left.\mathcal{R} \right|_{\mu \rightarrow \, e}^{\rm Au} \, < 7.0 \times 10^{-13}$~\cite{SINDRUMII:2006dvw}, whereas in the future, the Mu2e experiment at Fermilab~\cite{Mu2e:2014fns} and COMET experiment at J-PARC~\cite{COMET:2018auw} are expected to improve the experimental sensitivity down to $10^{-17}$ level using $^{27}_{13}$Al. 
 \begin{table}[t]
\centering
\begin{tabular}{|c|c|c|c|c|c|}
 \hline
 $N$ & $\Gamma^N_{\rm capture}$ (MeV) & $S^p$ & $S^n$ & $V^p$ & $V^n$ \\
 \hline
 $^{197}_{71}$Au & $8.7\times 10^{-15}$ & 0.05023 & 0.0610 & 0.08059 & 0.108 \\
 $^{27}_{13}$Al & $4.6\times 10^{-16}$ & 0.0153 & 0.0163 & 0.0159 & 0.0169 \\ 
 \hline
\end{tabular}
\caption{Muon capture rates~\cite{Suzuki:1987jf} and overlap integral values~\cite{Kitano:2002mt} relevant for $\mu\to e$ conversion in $^{197}_{79}$Au and $^{27}_{13}$Al.}
\label{tab:conv}
\end{table}

\subsubsection{\texorpdfstring{$\ell_\alpha \to \ell_\beta \gamma$}{l\_alpha to l\_beta gamma}}

As mentioned in Section~\ref{sec:massgeneration}, we have leptoquarks with two different generalized fermion numbers, i.e. $F_{S_1} = -2\, , \quad F_{\widetilde{R}_2}=0$. The interaction of these leptoquarks with the charged lepton can be written as \cite{Dorsner:2019itg}
\beqa
\mathcal{L}^{|F|=2} & \supset &  \bar{u}_i^c \left(Y_1^L \right)_{ij} \, P_L \, e_{j} \, S_{1}^{1/3} +  \bar{u}_i^c \left(Y_1^R \right)_{ij} \, P_R \, e_{j} \, S_{1}^{1/3} + \hc, \\
\mathcal{L}^{|F|=0} & \supset & \bar{d}_{i} \left( Y_2 \right)_{ij} P_L e_{j} \, \widetilde{R}_2^{2/3}  \, .
\eeqa
The $\ell_\alpha\to \ell_\beta \gamma$ partial decay rate is given by 
\beqa
\Gamma \left( l_{\alpha} \rightarrow l_{\beta} \gamma\right) & =& \frac{\alpha_{\rm em} \, m_{\alpha}^3}{4}  \left( 1- \frac{m_{\beta}^2}{m_{\alpha}^2} \right)^3\left( \left| \sigma_L^{\alpha\beta} \right|^2 + \left|\sigma_R^{\alpha\beta}\right|^2 \right) \, .
\eeqa
Here $\alpha_{\rm em}$ is the fine structure constant, and the $\sigma_L$ and $\sigma_R$ can be expressed in the leptoquark mass basis as   
\beqa
\sigma_{L}^{X_1} & = & \frac{i\, N_c \cos^2 \theta_{\rm LQ}}{16\pi^2 m_{X_1}^2} \sum_{q=u,c,t} \left[\left( {Y_1^L }\right)_{q\alpha} \left( Y_1^R\right)^\ast_{q\beta} \, m_q \,\mathcal{G}(x_q)   + m_{\alpha } \,\left(Y_1^R \right)^\ast_{q\beta} \left(Y_1^R \right)_{q\alpha}  \,\mathcal{F} (x_q)\right] ,\\
\sigma_{L}^{X_2} & = & \frac{i\, N_c \sin^2 \theta_{\rm LQ}}{16\pi^2 m_{X_2}^2} \sum_{q=u,c,t} \left[\left( {Y_1^L }\right)_{q\alpha} \left(Y_1^R \right)^\ast_{q\beta} \, m_q \,\mathcal{G}(x_q)   + m_{\alpha } \,\left(Y_1^R \right)^\ast_{q\beta} \left(Y_1^R \right)_{q\alpha}  \,\mathcal{F} (x_q)\right] , \\
\sigma_{R}^{X_1} & = & \frac{i\, N_c \,\cos^2 \theta_{\rm LQ}}{16\pi^2 m_{X_1}^2} \sum_{q=u,c,t} \left[\left( {Y_1^L }^* \right)_{q\beta} \left( Y_1^R \right)_{q\alpha} \, m_q \,\mathcal{G}(x_q) + m_{\alpha } \,\left({ Y_1^L } \right)^\ast_{q\beta} \left( {Y_1^L }\right)_{q\alpha}  \,\mathcal{F} (x_q)\right],  \\ 
\sigma_{R}^{X_2} & = & \frac{i\, N_c \,\sin^2 \theta_{\rm LQ}}{16\pi^2 m_{X_2}^2} \sum_{q=u,c,t} \left[\left( { Y_1^L }^* \right)_{q\beta} \left( Y_1^R \right)_{q\alpha} \, m_q \,\mathcal{G}(x_q) + m_{\alpha } \,\left( {Y_1^L }^* \right)_{q\beta} \left( {Y_1^L }\right)_{q\alpha}  \,\mathcal{F} (x_q) \right]. \label{eq:LFVdecaysigmafinal}
\eeqa
Here $x_q=m_q^2/m_{\rm LQ}^2$, $N_c=3$ is the number of colors, and the loop functions $\mathcal{F}(x)$ and $\mathcal{G}(x)$ are given in Appendix~\ref{app:LFV}. For all cLFV processes $\ell_\alpha\to \ell_\beta\gamma$, the mass of $\ell_\beta$ can be neglected with respect to the mass of $\ell_\alpha$. 

Then the total $\sigma$ (for a given $\alpha\beta$ combination) can be written as 
\beqa
\sigma_L & = & \sigma_{L}^{X_1} + \sigma_{L}^{X_2} , \qquad \sigma_R  =  \sigma_{R}^{X_1} + \sigma_{R}^{X_2} \,.
\eeqa

The cLFV decay branching ratios (BRs) can be written as
\beqa
{\rm BR} \left( \ell_{\alpha} \rightarrow \ell_{\beta} \gamma\right) & =& \frac{\alpha_{\rm em} \, m_{\alpha}^3}{4 \, \Gamma_{\alpha}}  \left( \left| \sigma_L^{\alpha\beta} \right|^2 + \left|\sigma_R^{\alpha\beta}\right|^2 \right) \, .
\eeqa
Here $\Gamma_\alpha$ is the total decay width for lepton flavor $\alpha$, which is   
$2.996 \times 10^{-19}~ \rm GeV $ for $\mu$ and $2.267 \times 10^{-12}~ \rm GeV $ for $\tau$.

These theoretical predictions for the LFV BRs are to be compared with the current experimental upper limits:
\begin{center}
\begin{tabular}{ll}
${\rm BR} \left( \mu \rightarrow \, e \gamma \right)  < 3.1 \times 10^{-13}$ & \quad MEG-II~\cite{MEGII:2023ltw}, \\
${\rm BR} \left(\tau \rightarrow \, \mu\, \gamma\right) <4.2 \times 10^{-8}$ & \quad Belle~\cite{Belle:2021ysv}, \\
${\rm BR} \left(\tau \rightarrow \, e\, \gamma\right)<3.3 \times 10^{-8}$ & \quad BaBar~\cite{BaBar:2009hkt} .
\end{tabular}
\end{center}
Belle-II is expected to improve the tau LFV limits by a factor of few~\cite{Belle-II:2022cgf}. 
\subsubsection{\texorpdfstring{Lepton $g-2$}{Lepton g-2}}
\label{sec:gm2}
The same one-loop diagrams that contribute to $\ell_\alpha\to \ell_\beta \gamma$ also give rise to the anomalous magnetic moment of $\ell_\alpha$ for $\alpha=\beta$. The leptoquark contribution can thus be written as 
\begin{equation}
	\Delta a_{\alpha}= i \, m_{\alpha} \left( \sigma_L^\alpha + \sigma_R^\alpha \right),
\end{equation}
where 
\beqa
\sigma_L^\alpha &=& \frac{i\, N_c m_{\alpha}}{16\pi^2 }\left(\frac{ \cos^2 \theta_{\rm LQ}}{m_{X_1}^2} + \frac{ \sin^2 \theta_{\rm LQ}}{m_{X_2}^2}  \right) \sum_{q=u,c,t} \left[ \Re\left[\left( Y_1^L \right)_{q\alpha} \left( Y_1^R \right)^\ast_{q\alpha} \right] \, m_q \,\mathcal{G}(x_q) \right. \nn \\
& & \qquad \qquad\qquad\qquad\qquad\qquad\qquad\qquad\left. + m_{\alpha } \,\left( \left| \left(Y_1^R \right)_{q\alpha} \right|^2 + \left| \left( Y_1^L \right)_{q\alpha}\right|^2 \right) \,\mathcal{F} (x_q) \right] , \label{eq:sigmaL1} \\
\sigma_R^\alpha &=&  \frac{i\, N_c m_{\alpha}}{16\pi^2 }\left(\frac{ \cos^2 \theta_{\rm LQ}}{m_{X_1}^2} + \frac{ \sin^2 \theta_{\rm LQ}}{m_{X_2}^2}  \right) \sum_{q=u,c,t} \left[\Re\left[\left( {Y_1^L }\right)^\ast_{q\alpha} \left( Y_1^R \right)_{q\alpha} \right] \, m_q \,\mathcal{G}(x_q) \right. \nn\\
& & \qquad \qquad\qquad\qquad\qquad\qquad\qquad\qquad\left. + m_{\alpha } \,\left( \left| \left(Y_1^R \right)_{q\alpha} \right|^2 + \left| \left( Y_1^L \right)_{q\alpha}\right|^2 \right) \,\mathcal{F} (x_q)\right] \, \label{eq:sigmaR2}. 
\eeqa

For muon $(g-2)$, the old BNL experiment reported a $3.7\sigma$ discrepancy with respect to the SM prediction~\cite{Muong-2:2006rrc}. The BNL result was recently confirmed by the Fermilab Muon $g-2$ collaboration~\cite{Muong-2:2023cdq}, which increased the discrepancy to $5\sigma$ level, if one uses the 2020 world average of the SM prediction~\cite{Aoyama:2020ynm}. However, the BMW lattice result~\cite{Borsanyi:2020mff} disagrees with the world average~\cite{Aoyama:2020ynm}. Other lattice calculations now seem to agree with the BMW result at least in the `intermediate distance regime'~\cite{gm2theory}, but a more thorough analysis is ongoing. In this murky situation, we choose to use the BMW result, which gives a $1.5\sigma$ deviation from the experimental result: $\Delta a_{\mu} = \left(1.07\pm 0.70 \right) \times 10^{-9}$~\cite{Wittig:2023pcl}.  

The situation for electron $(g-2)$ is no better. Although the experimental value of $a_e$ has been measured very precisely~\cite{Fan:2022eto}, the SM prediction~\cite{Aoyama:2019ryr} relies on the measurement of the fine-structure constant, and currently there is $>5\sigma$ discrepancy between the results derived using two different measurements based on Rb~\cite{Morel:2020dww} and Cs~\cite{Parker:2018vye} atoms. Here we will use both Rb and Cs results in our analysis: $\Delta a_e ({\rm Rb}) =(4.4 \pm 3.0) \times 10^{-13}$~\cite{Morel:2020dww} and 
$\Delta a_e ({\rm Cs}) =(-8.8 \pm 3.6) \times 10^{-13}$~\cite{Parker:2018vye}.
\subsubsection{\texorpdfstring{$\ell_\alpha \to \ell_\beta \ell_\gamma \ell_\delta$}{l\_alpha to l\_beta l\_gamma l\_delta}} \label{sec:MDM}
The four-lepton LFV amplitude is mediated by penguin and box diagrams involving leptoquarks and quarks. The amplitude typically scales as $y^2/m_{\rm LQ}^2$ for the penguin diagrams and as $y^4/m_{\rm LQ}^2$ for the box diagrams, where $y$ is the relevant leptoquark coupling. As the Yukawa couplings involved in this processes are small due to the constraints from cLFVs, the bounds coming from the box diagrams turn out to be subdominant to those from the penguins. However, typically the photon penguin contribution to $\ell_\alpha\to \ell_\beta\ell_\gamma\ell_\delta $ is dominant over the $Z$ penguins. Now, as the leptoquark couples to both left and the right chiral fermions, the dipole contributions are dominant for $\ell_\alpha\to \ell_\beta\ell_\gamma\ell_\delta $. The branching ratio of $\ell_\alpha\to 3\ell_\beta $ decay is found as ~\cite{Benbrik:2010cf,Khasianevich:2023duu}
\beqa
\frac{ BR\left(\ell_\alpha\to 3\,\ell_\beta  \right) }{BR\left(l_{\alpha} \rightarrow \, l_{\beta} \, \gamma\right)} &=& \frac{\alpha_{\rm em}}{3\pi} \left[ \ln \left(\frac{m_{\alpha}}{m_{\beta}} \right) - \frac{11}{4}\right] \, .
\eeqa
This is to be compared with the current limis:
\begin{center}
\begin{tabular}{ll}
${\rm BR} \left( \mu \rightarrow \, 3\,e  \right)  < 1.0 \times 10^{-12}$ & \quad SINDRUM~\cite{SINDRUM:1987nra}, \\
${\rm BR} \left(\tau \rightarrow \, 3\,\mu\right) < 2.1 \times 10^{-8}$ & \quad Belle~\cite{Hayasaka:2010np}, \\
${\rm BR} \left(\tau \rightarrow \, 3\,e\, \right)< 2.7 \times 10^{-8}$ & \quad Belle~\cite{Hayasaka:2010np} .
\end{tabular}
\end{center}
We find that the corresponding limits in the leptoquarks are weaker than those from $\mu\to e$ conversion in nuclei. Therefore, we do not further discuss the results for $\ell_{\alpha}\to 3\,\ell_{\beta}$ type processes here.      
\subsection{Neutrino electric and magnetic dipole moments
}
In this leptoquark model, neutrinos can interact with the photon at 1-loop (by just attaching a photon line to Fig.~\ref{fig:numass} ) which generates magnetic dipole moment (MDM) and electric dipole moment (EDM) for the neutrinos~\cite{Chua:1998yk, Brdar:2020quo, Sanchez-Velez:2022nwm, Bolanos-Carrera:2023ppu}. The most general electromagnetic vertex function can be written as~\cite{Fujikawa:2003ww,Bolanos-Carrera:2023ppu}
\beqa
\Gamma^{\mu} &=& i\,e\Big[ \gamma^{\mu} \left\{ F_1^V(q^2) + F_1^A(q^2) \gamma_5 \right\} + \frac{i}{2\, m_{\nu}} \sigma^{\mu\nu} \, q_{\nu} \, \left\{ F_2^V(q^2) + F_2^A(q^2) \, \gamma_5 \right\}  \nn \\
&& \qquad  + q^{\mu} \, \left\{  F_3^V (q^2) + F_3^A(q^2) \, \gamma_5 \, \right\}\Big] \, , 
\eeqa
where $q$ is the momentum of the outgoing photon, $F_1^V, F_1^A, F_2^V, F_2^A$ are the electric charge, anapole, magnetic dipole and electric dipole form factors, respectively. The MDM and EDM of the neutrino are then given as,
\begin{align}
\mu^{ij}_{\nu} &= e\,\frac{F_2^V(0)}{2\,m_{\nu}} \, , \qquad \qquad 
d^{ij}_{\nu} = -i\,e\, \frac{F_2^A(0)}{2\, m_{\nu}} \, .
\end{align}
If the neutrinos are majorana, the dipole moments are anti-symmetric which means only the transition elements i.e off-diagonal elements are non-vanishing. 

Stringent limits on the neutrino MDM come from the brightness measurement of the tip of the red giant branch in $\omega$-Centauri globular cluster using Gaia DR2 data.  Ref.~\cite{Capozzi:2020cbu} quotes this limit as $\mu_{\nu} < 1.2 \times 10^{-12} \, \mu_{B}$, where $\mu_{B} = \frac{e}{2\, m_e}$ is the Bohr magneton. As far as the neutrino EDM is concerned, no experimental limit exists yet, but several theoretical and indirect bounds have been obtained. The most stringent limit on electron and muon neutrino EDM is quoted as $d_{\nu_e,\nu_{\mu}} < 10^{-21}e$-cm~\cite{delAguila:1990jg} whereas, for the tau neutrino,  $d_{\nu_{\tau}} < 10^{-17}e$-cm~\cite{Akama:2001fp}.

In our leptoquark model, there are two 1-loop diagrams, similar to Fig.~\ref{fig:numass}, that contribute to MDM and EDM, depending on whether the photon radiates off either from the scalar leptoquark with charge 1/3 or from the down-type quark. 
If the photon is emitted from the leptoquark, the amplitude can be written as
\beqa
{\cal{M}}^{ij}_a &=& i\,e \int \frac{d^4\,k}{\left(2\pi\right)^4} \, \frac{\left[ \bar{u} (q_2)\, P_L \, \left( \slashed{k} + m_{d_m}\right)\, P_L\, u(q_1) \,\left( q_1^\mu + q_2^\mu \right)\right] }{\left( k^2 - m_{d_m}^2 \right) \,\left( (q_1-k)^2 - m_{X_i}^2 \right)\,\left( (q_2-k)^2 - m_{X_i}^2 \right)}\, \nonumber \\
&& \times \sum_{\alpha,\beta,m} \left(\widetilde{Y_1^L }\right)_{m\alpha} \, \left(Y_2 \right)_{m\beta} \, U_{j\alpha} \, U_{i\beta} \, ,
\label{eq:50}
\eeqa
where $i,j$ denote the initial and final neutrino state, $k$ is the loop momentum, $q_1$,$q_2$ are the momenta of external neutrinos, $m_{d_m}$ denotes the mass of down-type quark with flavor $m$, 
and  $\left(\alpha,\,\beta\right)$ run over the flavors of neutrinos, i.e. $\left\{e,\mu,\tau\right\}$.  
Similarly, the amplitude of the diagram where the photon is emitted from the down-quark leg can be written as
\beqa
{\cal{M}}^{ij}_b &=& i e \int \frac{d^4\,k}{\left(2\pi\right)^4}  \frac{\left[ \bar{u} (q_2)\, P_L \, \left( \slashed{q_2}-\slashed{k} + m_{d_m} \right) \gamma^{\mu
} \,\left( \slashed{q_1}-\slashed{k} + m_{d_m} \right) \, P_L\, u(q_1)\right] }{\left( k^2 - m_{X_i}^2 \right) \,\left( (q_1-k)^2 - m_{d_m}^2 \right)\,\left( (q_2-k)^2 -  m_{d_m}^2 \right)} \nonumber \\
&& \times \sum_{\alpha,\beta,m} \left(\widetilde{Y_1^L }\right)_{m\alpha}  \left(Y_2 \right)_{m\beta} \, U_{j\alpha} \, U_{i\beta} \, .
\label{eq:51}
\eeqa
Note that the electromagnetic charge of $1/3$ is canceled by the color factor of 3 in both Eqs.~\eqref{eq:50} and \eqref{eq:51}.

From the total amplitude, ${\cal{M}} = {\cal M}_a + {\cal M}_b$ , we identify the dominant contribution to the MDM as  
\beqa
\mu_{\nu}^{ij} & =& \frac{m_b\, m_e}{32\, \pi^2}\, \sin\left( 2\theta_{\rm LQ}\right)\, L(x_b)\,\left(\frac{1}{m_{X_1}^2}-\frac{1}{m_{X_2}^2}\right)\, {\rm Im}\left[ \sum_{\alpha,\beta}  \left(\widetilde{Y_1^L }\right)_{3\alpha}  \left(Y_2 \right)_{3\beta} \, \left(  \, U_{i\beta}\,U^\ast_{j\alpha}  - U^\ast_{i\alpha}\,U_{j\beta} \right)\right] \, ,\nn \\
\eeqa
where, $x_b = m_b^2/m_{\rm LQ}^2$.  Note that as MDM depends on the mass of the down-type quark, the dominant contribution comes from the bottom quark. The loop function $L(x)$ is defined as 
\beqa
L(x) &=& \frac{\log x}{\left(1-x \right)^2} + \frac{1}{\left(1-x \right)} \, .
\eeqa
Similarly, the dominant contribution to the EDM can be identified as
\beqa
d_{\nu}^{ij} & =& - \frac{m_b\,\sin\left( 2\theta_{\rm LQ}\right)}{32\, \pi^2}\, L(x_b)\left(\frac{1}{m_{X_1}^2}-\frac{1}{m_{X_2}^2}\right) {\rm Re}\,\left[ \sum_{\alpha,\beta}  \left(\widetilde{Y_1^L }\right)_{3\alpha}  \left(Y_2 \right)_{3\beta}  \left(   U_{i\beta}U^\ast_{j\alpha}  - U^\ast_{i\alpha}U_{j\beta} \right)\right] . 
\eeqa
Assuming ${\cal O}(1)$ Yukawa coupling, we can express both MDM and EDM in terms of neutrino masses, as follows: 
\beqa
\mu^{ij}_{\nu} & \approx & 5 \times 10^{-20}\, \mu_{B}\, \left(\frac{m_\nu}{0.01 \,{\rm eV}} \right) \left( \frac{{\rm TeV}}{m_{\rm LQ}} \right)^2\, \, , \quad \left|d^{ij}_{\nu} \right|  \approx 2 \times 10^{-30} e\textrm{-cm}\,  \left(\frac{m_\nu}{0.01\, {\rm eV}} \right) \left( \frac{{\rm TeV}}{m_{\rm LQ}} \right)^2. 
\eeqa
This implies that for TeV-scale LQs, both MDM and EDM of neutrinos are far below the current experimental limits. 

We should also comment here that there are no leptoquark contributions to the electron, proton, or neutron EDM in our model, if we assume the leptoquark Yukawa couplings to be real~\cite{Arnold:2013cva, Dekens:2018bci}. 
\subsection{Rare Meson Decays}
\label{sec:meson}
Here we discuss the leptoquark contributions to  the (semi)leptonic rare decays of $B$ and $K$ mesons. We will consider the neutral-current $B$ decays involving $b\to s\ell^+\ell^-$ and charged-current $B$ decays involving $b\to c\ell \bar{\nu}$, in particular the LFUV observables parameterized in terms of the ratios of BRs, 
\begin{align}
    R_{D^{(*)}} &=\frac{{\rm BR}(\bar{B}\to D^{(*)}\tau \bar{\nu})}{{\rm BR}(\bar{B}\to D^{(*)}\ell \bar{\nu})} \quad {\rm with}~ \ell=e,\mu, \\
    R_{K} &=\frac{{\rm BR}(B^+\to K^+\mu^+\mu^-)}{{\rm BR}(B^+\to K^+e^+e^-)}, \\
    R_{K^*} &=\frac{{\rm BR}(B^0\to K^{*0}\mu^+\mu^-)}{{\rm BR}(B^0\to K^{*0}e^+e^-)}, 
\end{align}
which are theoretically clean observables with strongly suppressed hadronic and CKM-angle uncertainties. We also consider other rare decays such as $B\to K\nu\bar{\nu}$, $K^+\to \pi^+ \nu\bar{\nu}$, and the rare kaon decays $K_L\to \mu^\pm e^\mp$ and $K^+\to \pi^+\mu^\pm e^\mp$. 
\subsubsection{\texorpdfstring{$R_K \text{ and } R_{K^{*}}$}{RK and RK*}}
The relevant effective Hamiltonian for the process can be written as  
\begin{eqnarray}
{\cal{H}_{\rm eff}} &=& -\frac{4\, G_F \,}{\sqrt{2}}\, \lambda_t^{bs} \, \left[\sum_{X=9,10} \left( C_{X}^{ll}\, \mathcal{O}_{X}^{ll} + C_{X'}^{ll}\, \mathcal{O}_{X'}^{ll} \right)  \right]\, ,
\end{eqnarray}
where $\lambda^{bs}_q = V_{qb} \, V_{qs}^{\ast}$ , $C_{X^{(')}}$ is the Wilson coefficients corresponding to the operator $\mathcal{O}_{X^{(')}}$, defined as  
\begin{eqnarray}
\mathcal{O}_{9} &=& \frac{\alpha_{\rm em}}{4\pi} \left(\bar{s} \,\gamma^{\mu} P_L \, b \right) \, \left(\bar{l} \,\gamma_{\mu} \, l \right) \, ,\nonumber \\
\mathcal{O}_{10} &=& \frac{\alpha_{\rm em}}{4\pi} \left(\bar{s} \,\gamma^{\mu} P_L \, b \right) \, \left(\bar{l} \,\gamma_{\mu} \, \gamma_5 \, l \right) \, , \nonumber \\
\mathcal{O}_{9'} &=& \frac{\alpha_{\rm em}}{4\pi} \left(\bar{s} \,\gamma^{\mu} P_R \, b \right) \, \left(\bar{l} \,\gamma_{\mu} \, l \right) \, , \nonumber \\
\mathcal{O}_{10'} &=& \frac{\alpha_{\rm em}}{4\pi} \left(\bar{s} \,\gamma^{\mu} P_R \, b \right) \, \left(\bar{l} \,\gamma_{\mu} \, \gamma_5 \, l \right) \, .
\end{eqnarray}
At tree level, $\mathcal{O}_{X'}$ operators can be generated by the mediation of $\tilde{R}_2^{2/3}$ and $\mathcal{O}_{X}$ operators are generated at 1-loop from the EW penguin and box diagrams where $S_1$ leptoquark is circulating inside the loop~\cite{Freitas:2022gqs}. When $y_1^L$ Yukawa couplings are $\mathcal{O}(1)$, $y_2 $ needs to be very small to generate a small neutrino mass for TeV scale LQ mass. In this case, the EW penguin and box contributions are greater than the tree-level contribution. 
At the tree-level, the Wilson coefficients are 
\begin{eqnarray}
\delta C'_9 & =&  -\delta C'_{10} = \frac{\pi \, v^2}{\alpha_{\rm em} \, V_{tb}\, V^*_{ts}} \,  \frac{\left( Y_2 \right)_{s\ell} \left( Y_2 ^*\right)_{b\ell}}{2 \,m^2_{\widetilde{R}_2}} .
\end{eqnarray}
For the penguin and box diagrams, the NP Wilson coefficients are given by 
\begin{eqnarray}
\delta C^{\rm penguin}_{9 \,ll} &=& \frac{v^2}{16  \, \lambda_t^{bs}} \sum_{i=1,3} \,  \left( \widetilde{ Y_1^L }\right)_{2i} \, \left( \widetilde{Y_1^L}^*\right)_{3i}\, \left( \frac{\cos^2\theta}{m_{X_1}^2} + \frac{\sin^2\theta}{m_{X_2}^2} \right) \, , \nn \\
\delta C^{\rm box}_{9 \,ll} &=& \frac{v^2}{64 \,\pi\, \alpha_{\rm em} \,\lambda_t^{bs}} \sum_{i=1,3} \,\sum_{j=u,c,t} \,  \left( \widetilde{ Y_1^L }\right)_{2i} \, \left( \widetilde{Y_1^L}^*\right)_{3i}\, \left(\left|(Y_1^L)_{jl}\right|^2 + \left|(Y_1^R)_{jl}\right|^2  \right) \,\left( \frac{\cos^2\theta}{m_{X_1}^2} + \frac{\sin^2\theta}{m_{X_2}^2} \right) \, , \nn \\
\delta C^{\rm box}_{10 \,ll} &=& \frac{v^2}{64 \,\pi\, \alpha_{\rm em} \,\lambda_t^{bs}} \sum_{i=1,3} \,\sum_{j=u,c,t} \,  \left( \widetilde{ Y_1^L }\right)_{2i} \, \left( \widetilde{Y_1^L}^*\right)_{3i}\, \left( \left|(Y_1^L)_{jl}\right|^2 -\left|(Y_1^R)_{jl}\right|^2  \right) \,\left( \frac{\cos^2\theta}{m_{X_1}^2} + \frac{\sin^2\theta}{m_{X_2}^2} \right) .\nn \\ 
\end{eqnarray}
The LHCb experiment recently presented new measurements of the $R_{K^{(*)}}$ ratios which turned out to be compatible with the SM~\cite{LHCb:2022qnv, LHCb:2022vje}.  From a recent global analysis using all available $b\to s\ell^+\ell^-$ data~\cite{Hurth:2023jwr}, it was found that the primed operators ${\cal O}'_9$ and ${\cal O}'_{10}$ (with right chiral quark currents) are loosely constrained with the best-fit values 
\beqa
\delta C_9 &=&  -1.18 \pm 0.19 \, \qquad \delta C_{10} = 0.23 \pm 0.20 \, \nn \\
\delta C_{9'} &=&  0.06 \pm 0.31 \, \qquad \delta C_{10'} = -0.05 \pm 0.19 \,
\eeqa
Therefore, $R_{K^{(*)}}$ does not impose a stringent constraint on our leptoquark model parameter space.

\subsubsection{\texorpdfstring{$R_D \text{ and } R_{D^{*}}$}{RD and RD*}}
Both $S_1$ and $\widetilde{R}_2$ leptoquarks contribute to $R_{D^{(*)}}$. The effective Hamiltonian can be written as~\cite{Becirevic:2024pni} 
\beqa
\mathcal{H}_{\rm eff}& =& \frac{4\, G_F}{\sqrt{2}} V_{cb} \left[ \left( \delta_{\ell\tau}+ C_{V_L}\right)\, {\cal O}_{V_L} + C_{S_L} \, {\cal O}_{S_L} + C_{T_L} \, {\cal O}_{T_L }\right],
\eeqa 
where the operators are defined as 
\beqa
{\cal O}_{V_L} &=& \left( \bar{c} \gamma^{\mu} P_L b\right) \left( \bar{\ell}\gamma_{\mu} P_L \nu \right) , \\
{\cal O}_{S_L} &=& \left(\bar{c} P_L \, b \right) \left( \bar{\ell} \, P_L \, \nu \right) , \\
{\cal O}_{T_L} &=&  \left(\bar{c} \,  \sigma^{\mu\nu}\,P_L \, b \right) \left( \bar{\ell} \,\sigma_{\mu\nu}\, P_L \, \nu \right) ,
\eeqa
and the $\delta_{\ell\tau}$ term denotes the SM-dominant contribution for $b \rightarrow c\tau^-\bar{\nu}$. The Wilson coefficients can be found as 
\beqa
C_{V_L} &=& \frac{v^2}{V_{cb}}\, \frac{\left( Y_1^L\right)^\ast_{c\ell} \left(\widetilde{Y_1^L }\right)_{b\ell  }}{4} \, \left(\frac{\cos^2 \theta_{\rm LQ}}{m_{X_1}^2} + \frac{\sin^2 \theta_{\rm LQ}}{m_{X_2}^2} \right) , \\
C_{S_L} &=& \frac{v^2}{V_{cb}}\, \frac{\left(Y_1^R\right)^\ast_{c\ell} \left( \widetilde{Y_1^L }\right)_{b\ell  }}{4} \, \left(\frac{\cos^2 \theta_{\rm LQ}}{m_{X_1}^2} + \frac{\sin^2 \theta_{\rm LQ}}{m_{X_2}^2} \right) , \\
C_{T_L} &=& -\frac{C_{S_L}}{4} .\label{eq:rdrds}
\eeqa

The Wilson coefficients, mentioned above are extracted at $\mu = \mu_{\rm ew}$. To compare with the experimental observables, we need to run them down to $\mu = m_b =4.18 $ GeV using the renormalization group equations, i.e.
\beqa 
C_i(\mu=m_b) \, =\, \Omega_N(\mu,\mu_{\mathrm{ew}})\; C_i(\mu_{\mathrm{ew}})\, ,
\eeqa
where at the lowest order (leading logarithm), the evolution operator is given by~\cite{Chetyrkin:1997dh,Gracey:2000am,Dorsner:2013tla,Hiller:2016kry} 
\beqa
\Omega_N(\mu,\mu_{\mathrm{ew}})\, =\, 
\left(\frac{\alpha_s^{(5)}(m_b)}{\alpha^{(5)}_s(m_t)}\right)^{-\gamma^J_1/\beta_1^{(5)}} \,
\eeqa
with QCD running $\beta$-function, $\beta_1^{(n_f)}= (2 n_f-33)/6$, where $n_f$ is the relevant number of quark flavors at the hadronic scale. The anomalous dimensions for the currents are
\beqa 
\gamma_1^V = 0\, ,\qquad\qquad
\gamma_1^S = 2\, ,\qquad\qquad
\gamma_1^T = -2/3\, .
\eeqa
Note that the vector currents are not affected, while the scalar and tensor currents renormalize multiplicatively.

With the given effective operators, we can express the ratio of $R_{D} \,(R_{D^*})$ to the SM prediction~\cite{Iguro:2024hyk}:
\beqa
r_D = \frac{R_{D}}{R_{D}^{\rm SM}} &\approx & \left| 1+ C_{V_L}\right|^2 + 1.01 \,\left| C_{S_L} \right|^2 + 0.84\,\left| C_{T_L} \right|^2+ 1.49\, \, {\rm Re} \left[ (1+C_{V_L})\,C_{S_L}^*\right] \nn \\
& & \qquad + 1.08 \, \, \rm Re \left[ (1+C_{V_L})\,C_{T_L}^*\right] , \\
r_{D^*}= \frac{R_{D^*}}{R_{D^*}^{\rm SM }} 
&\approx & \left| 1+C_{V_L}\right|^2 + 0.04  \left| C_{S_L}\right|^2 +16.07 \left| C_{T_L}\right|^2 \nn \\
& & \qquad - 0.11 \, \, \rm Re \left[ (1+C_{V_L})\, C_{S_L}^* \right] - 5.17 \, \, {\rm Re} \left[ (1+ C_{V_L}) \, C_{T_L}^* \right].
\eeqa
Using a recent global fit to the experimental results, Ref.~\cite{Iguro:2024hyk} found a $4.3\sigma$ deviation from the SM prediction: 
\beqa
r_D &=& 1.186 \pm 0.072,  \quad \quad r_{D^*} = 1.149 \pm 0.04 .
\label{eq:rD}
\eeqa
Treating this $R_{D^{(*)}}$ anomaly at the face value, we will analyze the parameter space which can successfully fit this anomaly. Given the volatile situation with flavor anomalies, we will also present our results for a hypothetical case where this anomaly disappears with more data in the future, i.e.~both $R_D$ and $R_{D^*}$ are consistent with the SM predictions.

\subsubsection{\texorpdfstring{$B^+\to K^+ \, \nu\, \bar{\nu}$}{B to K nu nu bar}}
Leptoquarks can also induce rare semileptonic $B$ decays like $B^+ \rightarrow K^+\, \nu\, \bar{\nu}$ and  $B^0\rightarrow {K^{*}}^{0}\, \nu\, \bar{\nu}$, governed by $(\bar{q}\,q\, \bar{\nu}\nu)$ effective interactions. 
The SM predictions for these decays are ${\rm BR}\left(B^+ \rightarrow \, K^+\,\nu\,\bar{\nu} \right)\,=\, (4.65\pm 0.62)\times 10^{-6}$ and  ${\rm BR}\left(B^0 \rightarrow \, {K^{*}}^{0}\,\nu\,\bar{\nu} \right)\,=\, (10.13\pm 0.92)\times 10^{-6}$~\cite{Buras:2022wpw}. Belle-II recently reported the first observation of $B^+ \rightarrow K^+\, \nu\, \bar{\nu}$ decay~\cite{Belle-II:2023esi}. 
Using the weighted average of BABAR, Belle and Belle-II data, Ref.~\cite{Buras:2024ewl} quoted an experimental value for ${\rm BR}\left(B^+ \rightarrow \, K^+\,\nu\,\bar{\nu} \right)_{\rm exp}=(1.3 \pm 0.4)\times\,10^{-5}$. 
As for $B^0 \rightarrow \, {K^{*}}^{0}\,\nu\,\bar{\nu} $, the current experimental upper limit is ${\rm BR}\left(B^0 \rightarrow \, {K^{*}}^{0}\,\nu\,\bar{\nu} \right) < 2.7 \times \, 10^{-5}$ at 90\% CL by the Belle collaboration~\cite{Belle:2017oht}. The constraints on new physics (NP) contributions are expressed in terms of the ratios $R^{\nu\nu}_{K^{(*)}}$, defined as 
\beqa
R^{\nu\nu}_{K^{(*)}}= \frac{{\rm BR}^{\rm SM+NP}\left(B^{+(0)} \rightarrow K^{+(*)}\nu \,\bar{\nu} \right)}{{\rm BR}^{\rm SM}\left(B^{+(0)} \rightarrow K^{+(*)}\nu \,\bar{\nu} \right)} . \label{eq:rk}
\eeqa
From the recent Belle-II measurement~\cite{Belle-II:2023esi}, we get $R_{K}^{\nu\nu} = (2.8 \pm 0.94 )$ and from the Belle upper limit~\cite{Belle:2017oht}, we get $R_{K^{*}}^{\nu\nu} <2.7$. 

The relevant effective Hamiltonian governing these decays can be written as 
\beqa
\mathcal{H}_{\rm eff} &=& \frac{4\, G_F}{\sqrt{2}}V_{tb}\,V^{*}_{ts} \left( C^{ij}_L \, \mathcal{O}^{ij}_L + C^{ij}_R  \mathcal{O}^{ij}_R\right) ,
\eeqa
where the operators are defined as 
\beqa
\mathcal{O}^{ij}_L &=& \frac{\alpha_{\rm em}}{4\pi}\left(\bar{d}\gamma^{\mu}P_L s \right) \left(\bar{\nu}_i\gamma_{\mu}(1-\gamma_5) \nu_j \right) , \\
\mathcal{O}^{ij}_R &=& - \frac{\alpha_{\rm em}}{4\pi}\left(\bar{d}\gamma^{\mu}P_R s \right) \left(\bar{\nu}_i\gamma_{\mu}(1-\gamma_5) \nu_j \right) .
\eeqa
The corresponding Wilson coefficients are given as 
\beqa
C^{ij}_L &=& C_{\rm SM} \delta^{ij} + \delta C_L^{ij} \nn \\
&=&  C_{\rm SM} \delta^{ij} + \frac{\pi \,v^2}{2\,\alpha_{\rm em}\, V_{tb}\, V^{*}_{ts}} \, \left( \frac{\cos^2\theta}{m_{X_1}^2} + \frac{\sin^2\theta}{m_{X_2}^2} \right)\,\left(\widetilde{ Y_1^L }\right)_{2j}\,\left(\widetilde{ Y_1^L }^\ast\right)_{3i} , \\
C^{ij}_R &=& \delta C_R^{ij}= - \frac{\pi \,v^2}{2\,\alpha_{\rm em}\, V_{tb}\, V^{*}_{ts}} \, \left( \frac{\sin^2\theta}{m_{X_1}^2} + \frac{\cos^2\theta}{m_{X_2}^2} \right)\,\left(  Y_2 \right)_{2j}\,\left( Y_2^* \right)_{3i}.
\eeqa
As the experiments cannot tag the neutrino flavor, we need to sum over all possible flavor combinations while calculating the ratio of branching fractions in Eq. (\ref{eq:rk}). The ratios can be expressed as~\cite{Browder:2021hbl, He:2023bnk}
\beqa
R_{K}^{\nu\nu}
&=& \frac{1}{3 \left|C_{\rm SM}\right|^2} \sum_{i,j} \left|C_{L}^{ij} + C_{R}^{ij} \right|^2 , \\
R_{K^*}^{\nu\nu} 
&=& \frac{1}{3 \left|C_{\rm SM}\right|^2} \sum_{i,j} \left|C_{L}^{ij} + C_{R}^{ij} \right|^2 - \frac{2}{3 \left|C_{\rm SM}\right|^2}\left(1+\eta\right) \sum_{i,j} \, {\rm Re} \,\left[ C_{L}^{ij} \, C_{R}^{*ij} \right] \label{eq:rknunu}
\eeqa
The numerical value of $\eta$ was found to be $0.33 \pm 0.09$ using the Light Cone Sum Rule (LCSR) method~\cite{Gubernari:2018wyi}.

\subsubsection{\texorpdfstring{$K^+ \to \pi^+ \, \nu\, \bar{\nu}$}{K to pi nu nubar}}
Apart from $B$ meson decays, leptoquark models are also constrained from the rare $K$ meson decays. One such example is $K^+ \rightarrow \, \pi^+ \, \nu \, \bar{\nu}$  which gets tree-level contribution from the scalar leptoquarks \cite{Bobeth:2017ecx,Fajfer:2018bfj,Mandal:2019gff,Buras:2024ewl}. The experimentally determined value is ${\rm BR} \left( K^{+} \rightarrow \pi^{+} \nu \,\bar{\nu} \right)_{\rm exp}= \left(10.6^{+4.1}_{-3.5} \times 10^{-11}\right)$ by NA62~\cite{NA62:2021zjw} which is consistent with the SM prediction ${\rm BR} \left( K^{+} \rightarrow \pi^{+} \nu \,\bar{\nu} \right)_{\rm SM}= (7.73 \pm 0.61)\times 10^{-11}$~\cite{Brod:2021hsj}. Thus, there is a very small room for the NP contribution for $K^+ \rightarrow \, \pi^+ \, \nu\,\bar{\nu}$ because the SM value is already well inside the experimental range.
The BR of $K^{+}\rightarrow \pi^{+}\nu\,\bar{\nu}$ can be written as ~\cite{Bobeth:2016llm,Bobeth:2017ecx}
\beqa
{\rm BR} \left( K^+ \rightarrow \pi^+ \nu\, \bar{\nu}\right) &=& \kappa_{+} \frac{\left( 1 + \Delta_{\rm em}\right)}{3\, \lambda^{10}} \sum_{i,j=e,\mu,\tau} \left[ {\rm Im}^2 \left( \lambda_t^{sd} \, X_t^{ij}\right) + {\rm Re}^2 \left(\lambda_c^{sd} \, X_c^{ii} + \lambda_t^{sd} \, X_t^{ij}\right) \right] ,
\eeqa
where $i,j$ denote the flavors of the emitted neutrinos,  $\lambda =0.225$ is the Wolfenstein parameter of the CKM matrix, $\Delta_{\rm em} = -0.003$, $\kappa_+ = 0.52 \times 10^{-10} \left( \frac{\lambda}{0.225}\right)$ and $\lambda^{sd}_q = V_{qs} \, V_{qd}^{\ast}$, where $q$ is the quark flavor. The short-distance contribution $X$ is defined as $X_t^{ij} = X_{\rm SM} (x_t) \delta_{ij} + X_{\rm LQ}^{ij}$, where $x_t = m_t^2/m_{\rm LQ}^2$. In the SM, the dominant contribution to the branching ratio comes from the top quark and the corresponding $X_{\rm SM} (x_t) = 1.481 \pm 0.009$ and $X_c \, \approx \, 10^{-3} $ is the charm quark contribution~\cite{Bobeth:2016llm}. NP contributions in our leptoquark model can be written as 
\beqa
X_{\rm LQ}^{ij} = -\frac{\sin^2\theta_{w} \, \pi v^2\, }{\alpha_{\rm em}} \, \frac{\left( C_{L}^{ij} + C_{R}^{ij} \right) }{\lambda_t^{sd}} ,
\eeqa
where $\theta_w$ is the weak mixing angle, and 
\beqa
C_{L}^{ij} &=& \frac{1}{2} \,\left(\widetilde{Y_1^L }\right)_{2i} \left(\widetilde{Y_1^L }\right)_{1j}\, \left( \frac{\cos^2\theta_{\rm LQ}}{m_{X_1}^2} + \frac{\sin^2\theta_{\rm LQ}}{m_{X_2}^2} \right) , \\
C_{R}^{ij} &=& \frac{1}{2} \left(Y_2 \right)_{2i} \left(Y_2 \right)_{1j}\, \left( \frac{\sin^2\theta_{\rm LQ}}{m_{X_1}^2} + \frac{\cos^2\theta_{\rm LQ}}{m_{X_2}^2} \right) .
\eeqa

We can also consider the analogous decay of the neutral kaon, $K_L\to \pi^0\,\nu\,\bar\nu$. However, this has not been observed yet, and there only exists an upper bound on  ${\rm BR}(K_L\to \pi^0\,\nu\,\bar\nu)_{\rm exp}<4.9\times 10^{-9}$ from KOTO~\cite{KOTO:2020prk}. This is more than two orders of magnitude larger than the SM prediction: ${\rm BR}(K_L\to \pi^0\,\nu\,\bar\nu)_{\rm SM}=(2.94\pm 0.15)\times 10^{-11}$~\cite{Buras:2021nns}. Therefore, this process cannot put any meaningful constraint on the NP scenario at the moment.

\subsubsection{\texorpdfstring{$K_L\to \mu^\pm e^\mp$ \rm{and} $K^+\to \pi^+ \mu^\pm e^\mp$}{KL to mu e and K+ to pi mu e}}

The effective Hamiltonian for these purely leptonic decays of the pseudoscalar meson $P\rightarrow\, \ell_{i}\, \ell_{j}$ and $P\rightarrow \, P'\, \ell_{i} \, \ell_{j}$  can be written as 
\beqa
\mathcal{H}_{\rm LEFT}= \sqrt{2}\, G_F \sum_{I \in \left\{V,S,T \right\} } C_{I_{XY}}\, \mathcal{O}_{I_{XY}} ,
\eeqa
where $X,Y\in \{ L,R\}$. The relevant operator generated at tree level $K_L \rightarrow \, \ell^-_{i} \, \ell^+_j $ and $K^+ \rightarrow \, \pi^{+} \, \ell^-_i \, \ell^+_j$ is 
$
\mathcal{O}_{V_{LR}}^{ij21} = \left(\bar{\ell}_i \gamma^{\mu} \, P_L \, \ell_j\right) \,\left(\bar{s} \gamma_{\mu} \, P_R \, d\right) 
$ and the corresponding Wilson coefficient is denoted as 
\begin{align}
    C_{V_{LR}}^{ij} = \frac{1 }{2 \, G_F \, m^2_{\widetilde{R}_2^{2/3}}} \left( \left( Y_2 \right)_{2j} \left( Y_2 \right)_{1i}^{\ast} +  \left( Y_2 \right)_{1j} \left( Y_2 \right)_{2i}^{\ast} \right).
\end{align}
Here $1,2$ correspond to the first (down) and second (strange) generation quarks. Besides the tree level diagram, one loop box diagrams mediated by $S_1$ leptoquark also contribute to these processes. The penguin diagrams are not possible in this case because of the absence of flavor-changing neutral current (FCNC) for $Z_{\mu}$ and $A_{\mu}$.  
The Wilson coefficients generated from the box diagrams are
\beqa
C_{V_{LL}}^{ij} &=& \frac{v^2}{64\,\sqrt{2}\, \pi^2} \sum_{l=1,3}\sum_{m=u,c,t} \,\left( \widetilde{ Y_1^L }\right)_{2l} \, \left( \widetilde{Y_1^L}^*\right)_{3l}\, \left(Y_1^L\right)_{mi}\,\left( Y_1^L\right)_{mj}^\ast \,\left( \frac{\cos^2\theta}{m_{X_1}^2} + \frac{\sin^2\theta}{m_{X_2}^2} \right) \, , \nn \\
C_{V_{RL}}^{ij} &=& \frac{v^2}{64\,\sqrt{2}\, \pi^2} \sum_{l=1,3}\sum_{m=u,c,t} \,\left( \widetilde{ Y_1^L }\right)_{2l} \, \left( \widetilde{Y_1^L}^*\right)_{3l}\, \left(Y_1^R \right)_{mi}\,\left( Y_1^R \right)_{mj}^\ast \,\left( \frac{\cos^2\theta}{m_{X_1}^2} + \frac{\sin^2\theta}{m_{X_2}^2} \right) \, .
\eeqa
The branching ratio of $K_L \rightarrow \ell_i^-  \ell_j^+$ can be computed as \cite{Plakias:2023esq}
\beqa
{\rm BR} \left(K_L \rightarrow \ell_i^-  \ell_j^+ \right) &=& \tau_K \frac{f_K^2 \, m_K\, m^2_{\ell}}{128 \, \pi \, v^4} \left(1- \frac{m_{\ell}^2}{m_K^2} \right)^2 \left\{\left|C_{VA}^{ij} \right|^2 +\left|C_{AA}^{ij} \right|^2 \right\} , \label{eq:Kll}
\eeqa
where $m_K$ is the kaon mass, $\tau_{K}$ is its lifetime, and $f_K$ is the kaon decay constant.
The Wilson coefficients in Eq. (\ref{eq:Kll}) are defined as
\beqa
C_{VA}^{ij} &=& C_{V_{LR}}^{ij} -  C_{V_{LL}}^{ij} -  C_{V_{RL}}^{ij}\, ,  \nn \\
C_{AA}^{ij} &=&  C_{V_{LL}}^{ij} - C_{V_{LR}}^{ij} -   C_{V_{RL}}^{ij} \, .
\eeqa
Similarly, the branching ratio for $K^+ \rightarrow \, \pi^+ \, \ell_i \, \ell_j$ can be written as~\cite{Plakias:2023esq}
\beqa
{\rm BR}\left(K^+ \rightarrow \, \pi^+ \, \ell_i \, \ell_j \right) &=& a_{VV} \left|C^{ij}_{VV}\right|^2 + a_{AV}\, \left| C^{ij}_{AV} \right|^2 \, , 
\eeqa
where $a_{VV} \simeq a_{AV} \simeq 0.157 $ are kaon form factors in the limit where the light lepton mass is neglected, and the Wilson coefficients are given as
\beqa
C^{ij}_{VV} &=& C^{ij}_{V_{LR}} + C^{ij}_{V_{LL}} +C^{ij}_{V_{RL}} \, , \qquad C^{ij}_{AV} = C^{ij}_{V_{RL}} - C^{ij}_{V_{LL}} -C_{V_{LR}}^{ij} .
\eeqa
These BRs are subjected to stringent experimental limits: ${\rm BR} \left(K_L \rightarrow \mu^\pm e^\mp \right)  < 4.7 \times 10^{-12}$~\cite{BNL:1998apv} and  ${\rm BR}\left(K^+ \rightarrow \, \pi^+ \, \mu^+ \, e^- \right) < 1.3 \times 10^{-11}$~\cite{Sher:2005sp} at 90\% CL.

\section{Numerical Analysis}
\label{sec:results}
In this section, we use all the above mentioned constraints and perform a multidimensional scan over all the leptoquark parameters to carve out the allowed parameter space. 
\subsection{Parametrization of the Yukawa Coupling Matrix} \label{casas-ibarra}
In the canonical seesaw mechanism, the Yukawa couplings can be expressed in terms of the PMNS mixing matrix elements and light neutrino masses using the Casas-Ibarra (CI) parametrization~\cite{Casas:2001sr}. However, in the leptoquark scenario with multiple Yukawa couplings, the standard CI approach does not work. However, one can use a similar approach to parameterize one Yukawa coupling in terms of the other Yukawa couplings, PMNS matrix elements and neutrino masses~\cite{Dolan:2018qpy}, as follows: 
\begin{align}
     Y_2 = \frac{1}{\mathbf{C}_1} \,  \mathcal{M}_{d}^{-1}  \, \left( \widetilde{ Y_1^L }^T\right)^{-1} \,  U^{*} \,\sqrt{\mathcal{M}_{\nu}^{\rm diag}}\, \mathbb{R} \, \sqrt{\mathcal{M}_{\nu}^{\rm diag}} \, U^{\dagger} \, ,  \label{eq:casas-Ibarra} 
\end{align}
where $\mathbf{C}_{1} = \frac{3\sin 2\theta_{\rm{LQ}}}{32 \pi^2}\ln\left( \frac{m_{X_1}^2}{m_{X_2}^2}\right)$ is the prefactor in the neutrino mass matrix [cf.~Eq.~\eqref{eq:mrel}], and $\mathbb{R}$ is an arbitrary $3\times 3$ complex matrix with $\mathbb{R} +\mathbb{R}^T =\mathbb{I}$. 
Thus, using Eq.~(\ref{eq:casas-Ibarra}), $Y_2 $ can be determined in terms of $Y_1^L $, $\mathbb{R}$, neutrino masses and PMNS matrix elements. In general, the complex $\mathbb{R}$ matrix can be parameterized in several ways; one such choice is~\cite{Dolan:2018qpy} 
\begin{align*}
\mathbb{R} &= \begin{pmatrix}
\frac{1}{2} & r_1 & r_2 \\ -r_1 & \frac 12 & r_3 \\ -r_2 & -r_3 & \frac 12
\end{pmatrix} .
\end{align*}
In general, $r_1$, $r_2$ and $r_3 $ can be any complex number. For simplicity, we assume 
\begin{align}
    r_1=r_2=r_3=r \, ,
    \label{eq:r}
\end{align}
which is an additional unknown parameter in the model. 
For our numerical analysis, we vary $r$ within the range $(-10^{2} : 10^{2})$. For concreteness, we also choose a benchmark point for the leptoquark masses, which satisfies the current LHC constraints~\cite{Parashar:2022wrd, Allwicher:2022mcg}:
\beqa
m_1 = m_2 = m_{\rm LQ} = 2.0 \, {\rm TeV}, \quad \alpha_1 = \alpha_2 =\alpha'_2=0.2, \quad \kappa= m_{\rm LQ}/10 \, ,
\eeqa
which yields  
\beqa
&&\theta_{\rm LQ} = 0.7420 \, \, {\rm rad}, \quad 
m_{X_1} = 1.99 \, {\rm TeV} , \quad m_{X_2} = 2.011\, {\rm TeV},  \quad m_{\widetilde{R}_2^{2/3}} = 2.00 \, {\rm TeV} \, .
\label{eq:benchmark}
\eeqa
Later, we will relax this condition to examine the dependence of the observables on the leptoquark mass. As we will see later, for suitable benchark values of Yukawa couplings, large values of $r$ will be constrained from observables like $\mu \rightarrow$ e conversion and $B \rightarrow \, K^+ \, \nu \, \bar{\nu}$, and for the benchmark value of $m_{\rm LQ}$ chosen here, $r$ cannot be larger than ${\cal O}(1)$.

In order to determine the Yukawa matrix $Y_2 $, we take the following textures for $Y_1^L$ and $Y_1^R $:
\beqa
Y_1^L &=&\begin{pmatrix}
\yl{11} & \yl{12} & \yl{13}\\
0 & 0 & \yl{23}\\ 
\yl{31}& \yl{32}& \yl{33}
\end{pmatrix} , \quad \quad 
Y_1^R=\begin{pmatrix} \yr{11} & \yr{12} & \yr{13}\\
0 & 0 & \yr{23}\\ 
\yr{31} & \yr{32} & \yr{33}
\end{pmatrix} .
\eeqa
Here, we explicitly assume $\yl{21},\yl{22}, \yr{21}, \yr{22}$ as zero to prevent NP contributions to $B\to D\ell \nu$ (with $\ell=e,\mu$), which does not give a good fit to the $b\to c\ell\nu$ observables, as noted in Ref.~\cite{Fedele:2022iib}. With the above textures, we vary the elements of the matrices $Y_1^L$ and $Y_1^R$ randomly between their perturbative limits $\left( -\sqrt{4\pi} \,:\, \sqrt{4\pi} \right)$. The oscillation parameters required in Eq.~(\ref{eq:casas-Ibarra}) are given as inputs in their currently allowed $3\, \sigma$ range~\cite{Esteban:2020cvm, nufit}, assuming NO for concreteness.  The Majorana phases are included in $U$ and are varied in the range $(0: \pi)$ and the lightest neutrino mass is varied over $(10^{-5} : 1 )$ eV. The $Y_2 $ matrix determined with these input values is subjected to perturbativity bounds and as well as constraints from rare meson decays, cLFV processes and lepton $g-2$, as discussed in Section~\ref{sec:flavor}, and  tabulated in Tables~\ref{tab:Bdecays},  \ref{tab:LFVs} and \ref{tab:g-2}, respectively. In Tables~\ref{tab:LFVs} and \ref{tab:g-2}, we give the current experimental limits on the observables and the corresponding upper bounds on the relevant Yukawa couplings for the benchmark value of the leptoquark mass chosen in Eq.~\eqref{eq:benchmark}, assuming these to contribute maximally to that particular observable.
\begin{table}[t!]
\centering
\scriptsize
\begin{tabular}{||c|c|c||}
\hline
Process  &   Observable   & Yukawa couplings involved \\ \hline
\multirow{2}{*}{$ B \, \rightarrow K\, l^+ \, l^-$} &   $\delta C_9 =  -1.18 \pm 0.19 $ ~\cite{Hurth:2023jwr}  & $\left|\left( \widetilde{ Y_1^L }\right)_{2i} \, \left( \widetilde{Y_1^L}^*\right)_{3i} \right| \times \left(\frac{2 \rm TeV}{m_{\rm LQ}} \right)^{2} < 10 $ \\
&   $\delta C_{10} = 0.23 \pm 0.20  $~\cite{Hurth:2023jwr}  &  $ \left| \left( \widetilde{ Y_1^L }\right)_{2i} \, \left( \widetilde{Y_1^L}^*\right)_{3i}\, \left( \left|(Y_1^L)_{jl}\right|^2 \pm \left|(Y_1^R)_{jl}\right|^2  \right) \right| \times \left(\frac{2 \rm TeV}{m_{\rm LQ}} \right)^{2} < 1 $  \\
\multirow{2}{*} {$B^* \, \rightarrow K^*\,l^+ \, l^-$} &  $\delta C_9' =0.06\pm 0.31 $ ~\cite{Hurth:2023jwr}  & \multirow{2}{*}{$\left| \left(Y_2 \right)_{2i} \left(Y_2 \right)_{3i} \right|  \times \left(\frac{2 \rm TeV}{m_{\rm LQ}} \right)^2 \, \lesssim 10^{-4} $}    \\ 
 &  $\delta C_{10}' = -0.05\pm 0.19 $~\cite{Hurth:2023jwr}  &  \\ \hline
$B \, \rightarrow D\, \ell \, \bar{\nu_l}$ &  $r_D = [1.114-1.258]$~\cite{Iguro:2024hyk} &  $ 0.594 \lesssim \Yl{23} (\widetilde{Y_1^L})_{33} \times \left(\frac{2 \rm TeV}{m_{\rm LQ}} \right)^2 \,  \lesssim 0.969$  \\ 
$B^* \, \rightarrow D^*\, \ell \, \bar{\nu}$ &  $r_{D^*}=[1.109-1.189]$~\cite{Iguro:2024hyk} &  $ 0.618 \lesssim \Yr{23} (\widetilde{Y_1^L})_{33} \times \left(\frac{2 \rm TeV}{m_{\rm LQ}} \right)^2 \,  \lesssim 1.013 $  \\ \hline
\multirow{2}{*}{$B^+\rightarrow K^{+} \,\nu\, \bar\nu$} & \multirow{2}{*}{$R_{K}^{\nu\nu}=2.8 \pm 0.92$~\cite{Belle-II:2023esi}}  &  $ 0.208\,\lesssim \left | \Ylt{2i} \Ylt{3i} \right| \times \left(\frac{2 \rm TeV}{m_{\rm LQ}} \right)^2 \, \lesssim 0.051$  \\ 
& &  $ -0.07 \lesssim  \Yt{2i} \Yt{3i} \times \left(\frac{2 \rm TeV}{m_{\rm LQ}} \right)^2 \, \lesssim \, 0.02 $ \\
\multirow{2}{*}{$B\rightarrow K^{*} \,\nu \, \bar\nu$} & \multirow{2}{*}{$R_{K^*}^{\nu\nu}< 2.7 $~\cite{Belle:2017oht} }& $ 0.107 \lesssim\left | \Ylt{2i} \Ylt{3j} \right| \times \left(\frac{2 \rm TeV}{m_{\rm LQ}} \right)^2 \,  \lesssim 0.129$ \\
& &  $ 0.107 \lesssim \left | \Yt{2i} \Yt{3j} \right|\times \left(\frac{2 \rm TeV}{m_{\rm LQ}} \right)^2  \,  \lesssim 0.129$ \\ \hline
\multirow{2}{*}{$K^{+} \rightarrow \pi ^{+} \,\nu \,\bar\nu$} &  \multirow{2}{*}{${\rm BR} \left( K^{+} \rightarrow \pi ^{+} \nu \nu\right) = \left(10.6^{+4.1}_{-3.5} \times 10^{-11}\right)$~\cite{NA62:2021zjw}}  &  $ 1.0 \times 10^{-3}\lesssim (\widetilde{Y_1^L})_{2i} (\widetilde{Y_1^L})_{1j} \times \left(\frac{2 \rm TeV}{m_{\rm LQ}} \right)^2 \, \lesssim 1.6 \times 10^{-3}$  \\
& &  $ 1.0 \times 10^{-4}\lesssim \left( Y_2 \right)_{2i} \left( Y_2 \right)_{1j} \times \left(\frac{2 \rm TeV}{m_{\rm LQ}} \right)^2 \,  \lesssim 1.6 \times 10^{-3}$ \\ \hline
\multirow{3}{*}{$K_L \rightarrow \, \mu^{\pm} \, e^{\mp} $} &  \multirow{2}{*}{${\rm BR} (K_L \rightarrow \, \mu^{\pm} \, e^{\mp})< 4.7 \times 10^{-12}$~\cite{BNL:1998apv}}&   $ \left| \left( \widetilde{ Y_1^L }\right)_{2l} \, \left( \widetilde{Y_1^L}\right)_{3l}\, \left(Y_1^L\right)_{m1}\,\left( Y_1^L\right)_{m2}  \right| \times \left(\frac{2 \rm TeV}{m_{\rm LQ}} \right)^2 \,  < 0.0354$ \\
& & $\left| \left( Y_2 \right)_{11} \left( Y_2 \right)_{22} \right|\, \times \left(\frac{2 \rm TeV}{m_{\rm LQ}} \right)^2 $ \, $ < 7.93 \times 10^{-5}$ \\
& &   $\left| \left( Y_2 \right)_{12} \Yt{21} \right| \times \left(\frac{2 \rm TeV}{m_{\rm LQ}} \right)^2 \, < 7.93 \times 10^{-5}$ \\ \hline
\multirow{2}{*}{
$K^+ \rightarrow \, \pi^+ \, \mu^{\pm} \, e^{\mp}$} & \multirow{2}{*}{${\rm BR} (K^+ \rightarrow \, \pi^+ \, \mu^{\pm} \, e^{\mp})< 1.3 \times 10^{-11}$~\cite{Sher:2005sp} } & $ \left| \left( \widetilde{ Y_1^L }\right)_{2l} \, \left( \widetilde{Y_1^L}\right)_{3l}\, \left(Y_1^L\right)_{m1}\,\left( Y_1^L\right)_{m2}  \right| \times \left(\frac{2 \rm TeV}{m_{\rm LQ}} \right)^2 \, < 0.538 $\\ 
& & $  \left| \left( Y_2 \right)_{11} \left( Y_2 \right)_{22} \right| \times \left(\frac{2 \rm TeV}{m_{\rm LQ}} \right)^2 \,  < 6.02 \times 10^{-4}$ \\ 
\hline
\end{tabular}
\caption{List of rare meson decay constraints on the Yukawa couplings.}
\label{tab:Bdecays}
\end{table}


\begin{table}[t!]
\centering 
\footnotesize
\begin{tabular}{||c|c|c||}
\hline
Process & Observable & Limits on Yukawa \\ \hline
\multirow{3}{*}{$\mu N\to eN$} & \multirow{3}{*}{$\mathcal {R}|_{\mu \to e}^{\rm Au}<7\times 10^{-13}$~\cite{SINDRUMII:2006dvw}} & $ \Big[ \left| \Yl{12} \Yl{11}\right|  \, ,\,\left|\Yr{12} \Yr{11}\right| \Big] \,\times \left(\frac{2 \rm TeV }{m_{\rm LQ} } \right)^2 < 4.06 \times 10^{-5} $  \\ 
& & $ \Big[ \left| \Yl{12} \Yr{11}\right|  \, , \, \left|\Yr{12} \Yl{11}\right| \Big] \,\times \left(\frac{2 \rm TeV }{m_{\rm LQ} } \right)^2  < 4.21 \times 10^{-5} $ \\
& & $\left| \Yt{12} \Yt{11}\right|  \,\times \left(\frac{2 \rm TeV }{m_{\rm LQ} } \right)^2  <  3.68 \times 10^{-5} $ \\
\hline
\multirow{4}{*}{$\mu \rightarrow \, e \gamma $} & \multirow{4}{*}{$ {\rm BR} \left( \mu \rightarrow \, e \gamma \right)  < 3.1 \times 10^{-13} $ ~\cite{MEGII:2023ltw}}  &  $\Big[ \left|\Yl{11} \Yl{12} \right|,\left|\Yr{11} \Yr{12}\right|\Big] \,\times \left(\frac{2 \rm TeV }{m_{\rm LQ} } \right)^2  < 2.74 \times 10^{-3}  $\\ 
& & $\Big[\left|\Yl{31}\Yl{32}\right|, \left|\Yr{31}\Yr{32}\right| \Big] \,\times \left(\frac{2 \rm TeV }{m_{\rm LQ} } \right)^2 < 2.49 \times 10^{-3}  $\\ 
& & $ \Big[\left|\Yr{11}\Yl{12} \right|, \left|\Yl{11}\Yr{12} \right|\Big] \,\times \left(\frac{2 \rm TeV }{m_{\rm LQ} } \right)^2  < 5.92 \times 10^{-4}  $\\
& & $\Big[\left|\Yl{31}\Yr{32}\right|, \left| \Yl{32}\Yr{31}\right| \Big] \,\times \left(\frac{2 \rm TeV }{m_{\rm LQ} } \right)^2 < 5.86 \times 10^{-8}  $\\
\hline
\multirow{4}{*}{$\tau \rightarrow \, \mu \,  \gamma $} & \multirow{4}{*}{$ {\rm BR} \left(\tau \rightarrow \, \mu\, \gamma\right) <4.2 \times 10^{-8} $~\cite{Belle:2021ysv}} & $\Big[ \left|\Yl{12} \Yr{13} \right|,\left|\Yl{13} \Yr{12}\right|\Big]\times \left(\frac{2 \rm TeV }{m_{\rm LQ} } \right)^2 < 8.7 $ \\ 
 &  &$\Big[ \left|\Yl{12} \Yl{13} \right|,\left|\Yr{12} \Yr{13}\right|\Big]\times \left(\frac{2 \rm TeV }{m_{\rm LQ} } \right)^2 < 2.2  $ \\
  &  & $\Big[ \left|\Yl{32} \Yl{33} \right|,\left|\Yr{32} \Yr{33}\right|\Big]\times \left(\frac{2 \rm TeV }{m_{\rm LQ} } \right)^2 < 2.39 $  \\
  &  & $\Big[ \left|\Yl{32} \Yr{33} \right|,\left|\Yl{33} \Yr{32}\right|\Big] \times \left(\frac{2 \rm TeV }{m_{\rm LQ} } \right)^2 < 8.60 \times 10^{-4}$ \\ 
\hline
\multirow{3}{*}{$\tau \rightarrow \, e\,  \gamma $} & \multirow{2}{*}{$ {\rm BR} \left(\tau \rightarrow \, e\, \gamma\right)<3.3 \times 10^{-8} $ ~\cite{BaBar:2009hkt} } & $ \Big[ \left|\Yl{31}\Yr{33}\right|, \left|\Yr{31}\Yl{33} \right| \Big] \times \left(\frac{2 \rm TeV }{m_{\rm LQ} } \right)^2 < 7.61 \times 10^{-4} $ \\
& & $\Big[ \left|\Yl{11}  \Yl{13} \right| \Big] \times \left(\frac{2 \rm TeV }{m_{\rm LQ} } \right)^2< 0.5   $  \\
& & $\Big[  \left|\Yr{11} \Yr{13} \right| \Big]  \times \left(\frac{2 \rm TeV }{m_{\rm LQ} } \right)^2< 2.11 $ \\
 \hline
\multirow{2}{*}{$0\nu\beta\beta$} & \multirow{2}{*}{$m_{\beta\beta} < 0.036 \, \rm eV $~\cite{KamLAND-Zen:2024eml}} & $-0.0005 < \, \Yl{11} \, \Yt{11}\times \left(\frac{2 \rm TeV}{m_{\rm LQ}} \right)^2 \, \,< 0.0003$ \\
 & & $-0.0006 < \, \Yr{11} \, \Yt{11} \times \left(\frac{2 \rm TeV}{m_{\rm LQ}} \right)^2 \,< 0.0003$ \\\hline
\end{tabular}
\caption{Constraints on Yukawa couplings put by the corresponding cLFV processes and $0\nu\beta\beta$ decay.}
\label{tab:LFVs}
\end{table}
\begin{table}[t!]
\centering 
\footnotesize
\begin{tabular}{||c|c|c||}
\hline
Lepton $g-2$ & Experimental value & Yukawa couplings needed \\ \hline
$\left(g-2\right)_{\mu}$ & $\Delta a_{\mu} = \left(1.07\pm 0.70 \right) \times 10^{-9}$ ~\cite{Wittig:2023pcl} & $ 0.99 \times 10^{-3} \lesssim \Yl{32}\Yr{32}  \,  \times \left(\frac{2 \rm TeV}{m_{\rm LQ}} \right)^2  \lesssim 4.71 \times 10^{-3} $ \\
 \hline
\multirow{4}{*}{ $\left(g-2\right)_e$} &  \multirow{2}{*}{$\Delta a_e ({\rm Rb}) =(4.4 \pm 3.0) \times 10^{-13}~$\cite{Morel:2020dww}}  & $0.78 \lesssim  \Yl{11}\Yr{11} \,  \times \left(\frac{2 \rm TeV}{m_{\rm LQ}} \right)^2 \lesssim  4.13 $ \\ 
 &  & $ 0.77 \times 10^{-4} \lesssim  \Yl{31}\Yr{31}  \,  \times \left(\frac{2 \rm TeV}{m_{\rm LQ}} \right)^2  \lesssim  4.08 \times 10^{-4 } $ \\ \cline{2-3} 
  & \multirow{2}{*}{$\Delta a_e ({\rm Cs}) =(-8.8 \pm 3.6) \times 10^{-13}$~\cite{Parker:2018vye}}   & $ -6.92 \lesssim  \Yl{11}\Yr{11}  \,  \times \left(\frac{2 \rm TeV}{m_{\rm LQ}} \right)^2 \lesssim -2.90 $ \\ 
 &  & $- 6.84 \times 10^{-4} \lesssim  \Yl{31}\Yr{31} \,  \times \left(\frac{2 \rm TeV}{m_{\rm LQ}} \right)^2  \lesssim - 2.87 \times 10^{-4} $ \\ \hline 
\end{tabular}
\caption{Yukawa couplings needed to satisfy $1\sigma$ values of lepton $g-2$.}
\label{tab:g-2}
\end{table}

\begin{figure}
     \centering
\includegraphics[width=1.0\textwidth]{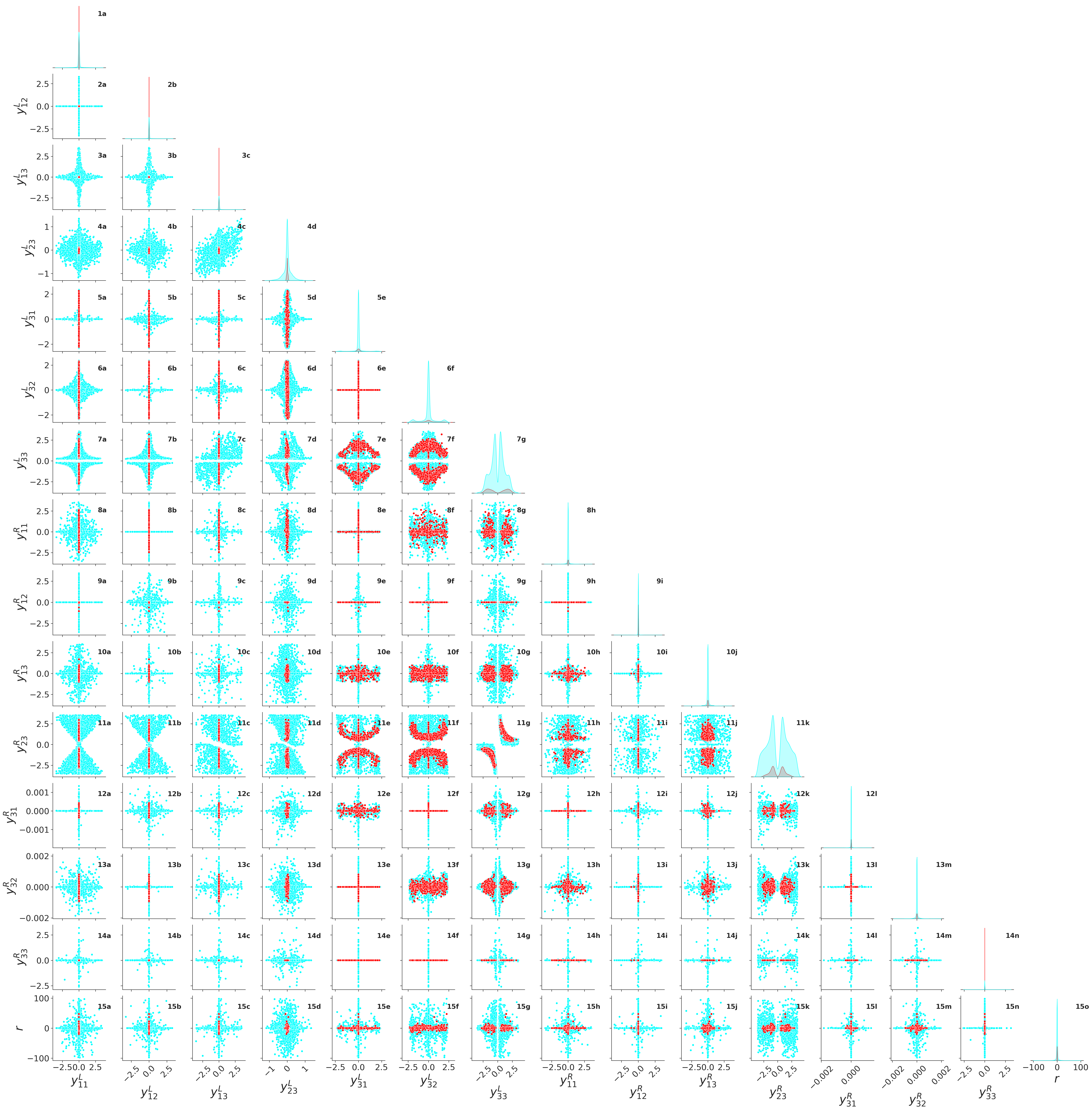}
    \caption{Pairwise correlation plot between various Yukawa matrix $Y_1^L, Y_1^R$ elements, as well as the parameter $r$ of the $\mathbb{R}$ matrix, in the presence of $R_D, R_D^\ast$ anomaly (Case I). The cyan points satisfy the constraints from cLFV decays, LFUV ratios, $ B \, \rightarrow \, K \, \nu \, \bar\nu $ constraints and generate the correct order of magnitude for neutrino masses and oscillation parameters. Red points survive after putting the constraint from $\mu\to e$ conversion in Au nucleus.}
    \label{fig:cp-w-rd}
\end{figure}

\begin{figure}[t!]
    \centering
    \includegraphics[width=1.0\textwidth]{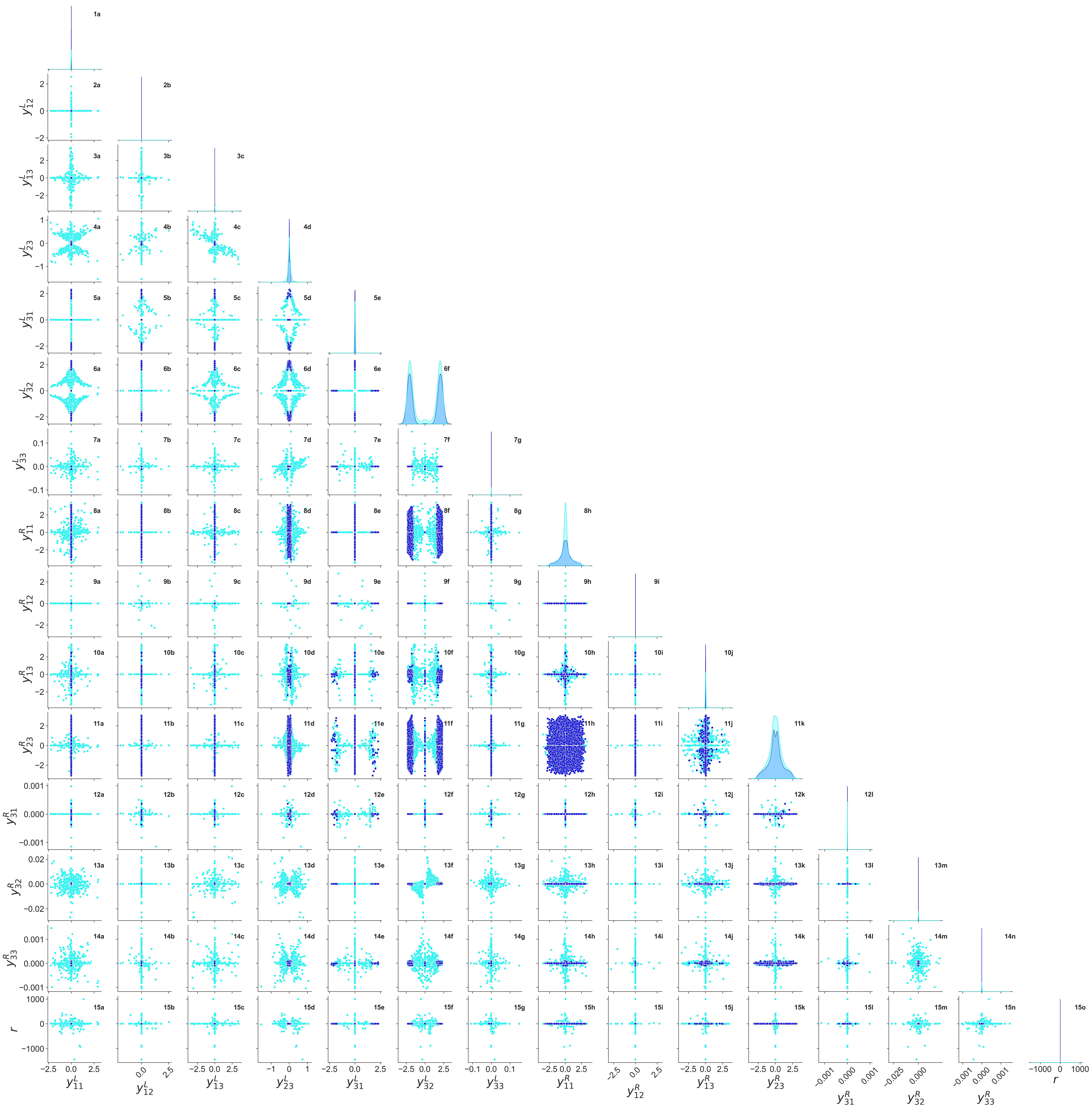}
    \caption{Pairwise correlation plot between various Yukawa matrix $Y_1^L, Y_1^R$ elements as well as  the parameter $r$ of the $\mathbb{R}$ matrix, in the hypothetical scenario where $R_D, R_D^\ast$ anomaly is gone (Case II). The cyan points satisfy the constraints from cLFV decays, LFUV ratios in Case II, $ B \, \rightarrow \, K \, \nu \, \bar \nu $ constraints and generate the correct order of magnitude for neutrino masses and oscillation parameters. Blue points survive after putting the constraint from $\mu\to e$ conversion in Au nucleus.}
    \label{fig:cp-wo-rd}
\end{figure}

\subsection{Results} 
Now, to capture the interplay between the different elements of Yukawa matrices, we have presented Figs.~\ref{fig:cp-w-rd} and \ref{fig:cp-wo-rd} where we study the pairwise correlations between different elements of the Yukawa matrices $\YL$ and $\YR$, as well as the parameter $r$ appearing in the $\mathbb{R}$ matrix [cf.~Eq.~\eqref{eq:r}]. In both figures, the rightmost panel in each row has the same parameter on the $x$ and $y$ axes; hence, these are just the distributions over which a particular parameter is varied. In our analysis, we have considered two different cases for the $R_{D^{(*)}}$ observables:  
\begin{enumerate}
    \item  \textbf{Case I}: $R_{D},R_{D^*}$ are anomalous, i.e the experimental values of  $R_{D},R_{D^*}$ are not consistent with the SM predictions, which seems to be the current situation [cf.~Eq.~\eqref{eq:rD}]. Fig.~\ref{fig:cp-w-rd} refers to this case. 
    \item \textbf{Case II}: This is a hypothetical scenario where the experimental values of $R_{D},R_{D^*}$ are consistent with the SM predictions within the $1\sigma$ error bars. In this case, we consider $\Delta r_{D^{(*)}} = \left| r_{D^{(*)}}- r_{D^{(*)}}^{\rm SM} \right| < 10^{-3}$. Fig.~\ref{fig:cp-wo-rd} refers to this case.
\end{enumerate}

 In both Figs.~\ref{fig:cp-w-rd} and \ref{fig:cp-wo-rd}, the cyan points produce the correct neutrino mass as well as oscillation parameters assuming NO, satisfy the current constraints from cLFV decays ($\mu\rightarrow \, e\, \gamma \,\, , \, \tau \rightarrow \, \mu \, \gamma $ and $\tau \rightarrow \, e \, \gamma$), the rare $B$  and $K$ meson decays listed in Subsection~\ref{sec:meson} and also can reproduce the observed value of electron $(g-2)$ (either Cs or Rb-based value). 
 The red points are obtained after putting the constraint coming from $\mu\to e$ conversion in Gold (Au) nucleus. From here, we can infer that this process gives the most stringent constraint on the Yukawa couplings and much of the otherwise allowed parameter space gets disfavored. 
The strongest anti-correlation is seen in panel ($11g$) which indicates that the dominant contribution to $R_{D}$ and $R_{D^\ast}$ comes from $C_{S_L}$ [cf. Eq.~(\ref{eq:rdrds})] which involves $\yl{33}$ and $\yr{23}$. Panel (11g) also implies that $\yl{33} \,\yr{23} > 0$, otherwise, the leptoquark contribution will destructively interfere with the SM contribution and Eq.~(\ref{eq:rD}) will be smaller than one.

Fig.~\ref{fig:cp-wo-rd} shows the correlation between the same Yukawa matrix elements and $r$, like in Fig.~\ref{fig:cp-w-rd}, but with the assumption that $R_{D}-R_{D^\ast}$ are consistent with the SM predictions (Case II). As a result, certain correlations or anti-correlations among the pair of respective Yukawa couplings, observed in the earlier figure, vanish. In panel (11g), now the points near to origin (0,0) are allowed as the $R_{D}-R_{D^\ast} $ anomaly is absent in this case. However, the $\mu \to e$ conversion still puts the most stringent constraint in this case also and disfavors a large part of the parameter space. The final surviving points are shown in blue. 

\begin{figure}[t!]
    \centering
    \includegraphics[width=0.49\textwidth]{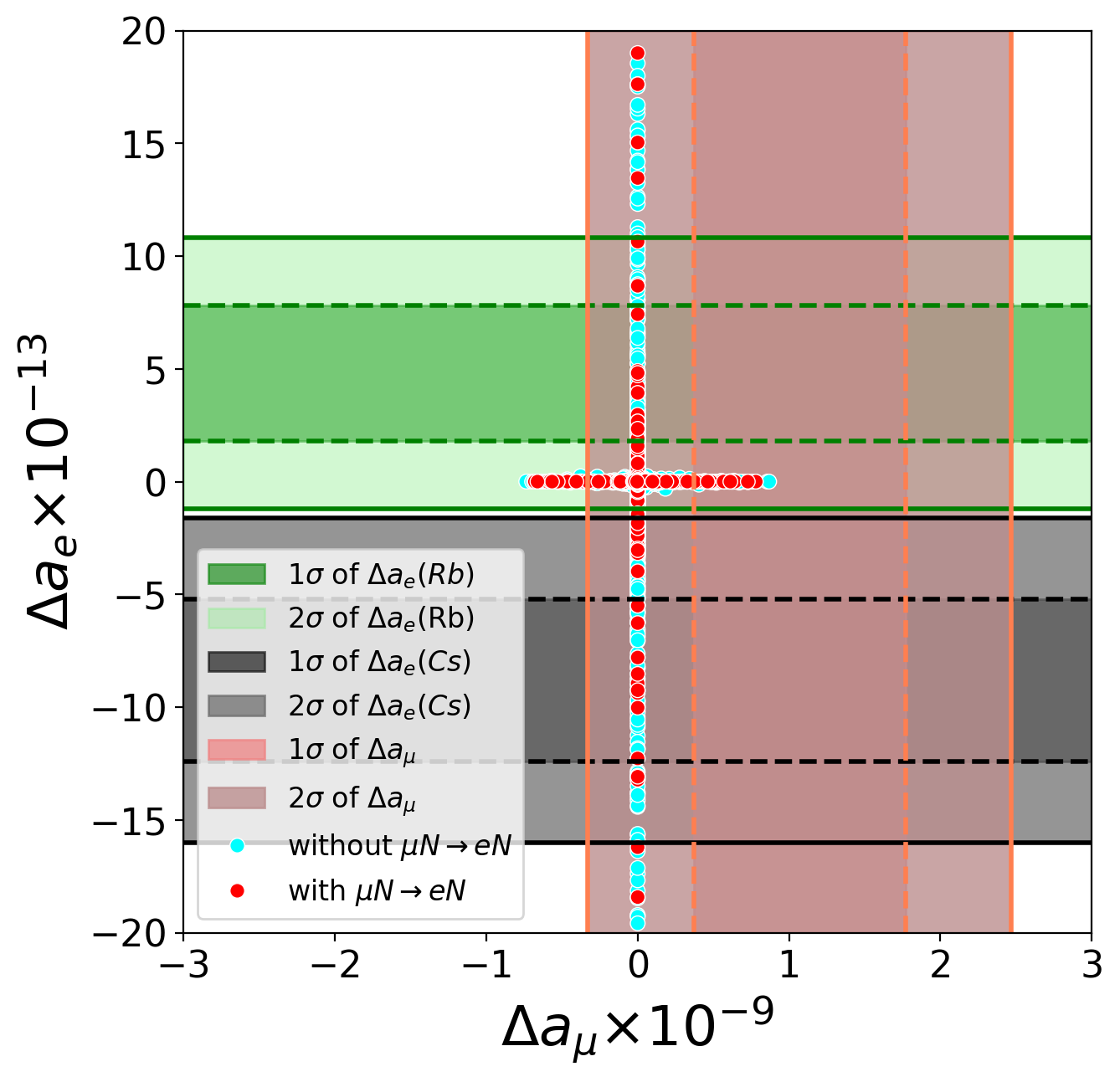}
    \includegraphics[width=0.49\textwidth]{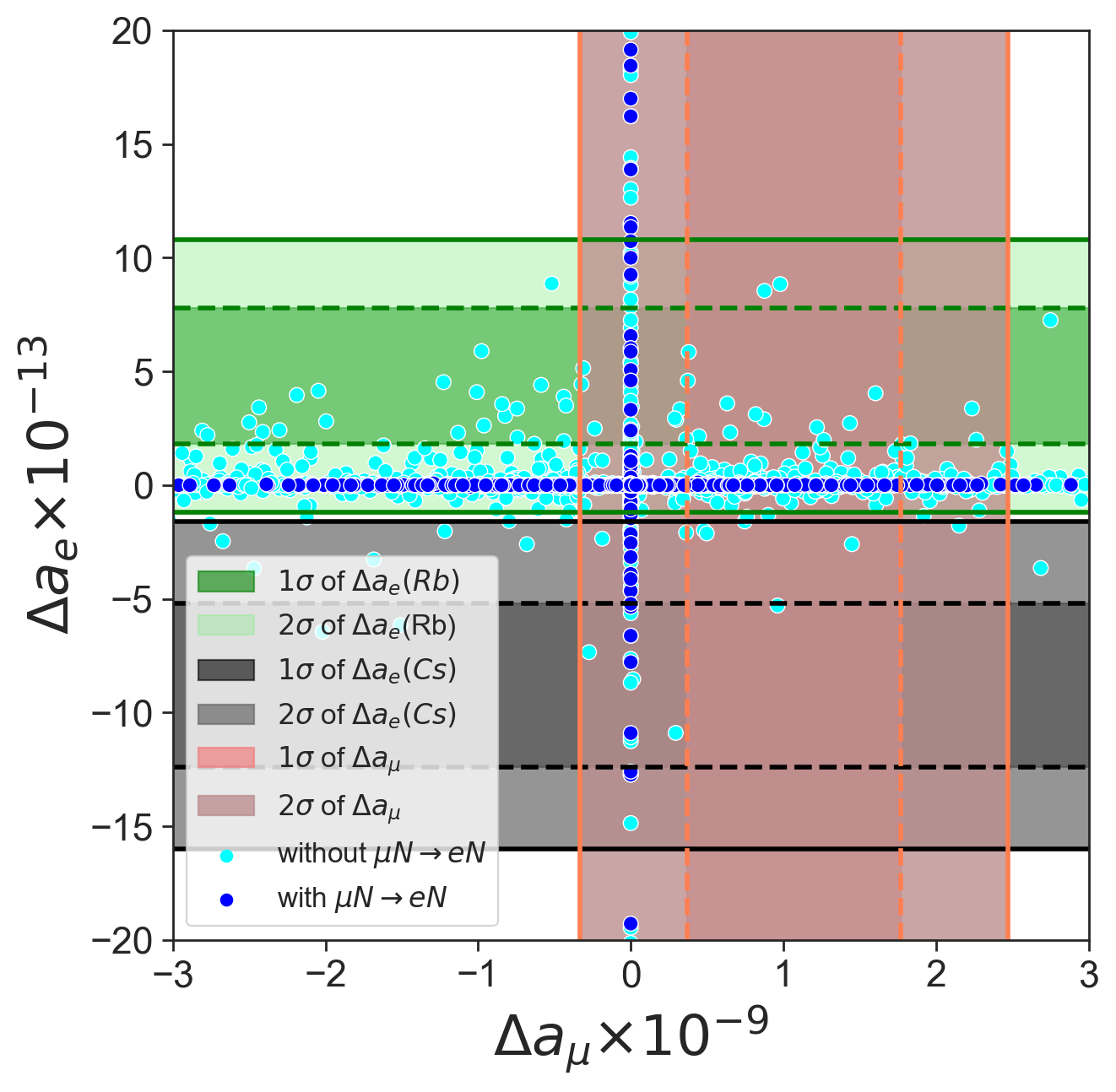}
    \caption{{\it Left Panel:} Plot of $\Delta a_{\mu}$ \textit{vs} $\Delta a_{e}$ in our leptoquark scenario satisfying current experimental limits on $R_D$-$R_{D^*}$ (Case I). {\it Right Panel:} Assuming $R_D$-$R_{D^*}$ is no longer anomalous (Case II). The cyan points satisfy all constraints except $\mu\to e$ conversion, and red (blue) points also satisfy $\mu\to e$ conversion bound for Case I (Case II). The vertical yellow shaded bands show the $1\sigma$ and $2\sigma$ allowed ranges of $\Delta a_\mu$, whereas the horizontal coral (grey) shaded bands show the $1\sigma$ and $2\sigma$ allowed ranges of $\Delta a_e$ for Rb (Cs).}
    \label{fig:rd_g2}
\end{figure}

In Fig.~\ref{fig:rd_g2}, we demonstrate  the tension between the electron and muon $(g-2)$ anomalies for both Case I and II. The left panel is for Case I (with $R_{D^{(*)}}$ being anomalous) and the right panel is for Case II (where $R_{D^{(*)}}$ is consistent with the SM). In both figures, the cyan points satisfy the cLFV decay constraints (excluding the $\mu\to e$ conversion) and rare meson decays, whereas the red (blue) points in Case I (II) also include the $\mu\to e$ conversion bound. The dark (light) coral shaded region correspond to the $1\sigma$ ($2\sigma$) region of $\Delta a_{\mu}$ taking the BMW result for the SM prediction. 
 It is seen from the left  panel that with  $R_{D},R_{D^*}$ not consistent with the SM prediction, the allowed  points of muon $(g-2)$  can not be arbitrarily large. This occurs because, to satisfy $R_{D}-R_{D^{\ast}}$, $\yl{33}$ must be at least $\sim \mathcal{O}(0.1)$. This constraint forces $\yl{32}$ and $\yr{32}$ to remain small to comply with the $\tau \rightarrow \mu \, \gamma$ limit, resulting in very small values of $\Delta a_{\mu}$.
If the constraints coming from $R_{D},R_{D^*}$ are relaxed as in the right panel, then the muon $(g-2)$ can be large as seen from the right panel.
This is because the absence of the $R_{D}-R_{D^\ast}$ anomaly suggests small values of $\yl{33}$. Thus, even with the sizable values of $\yl{23}$ and $\yr{23}$, constraints from  $\tau \rightarrow \mu \gamma$ can be avoided.  
In Fig.~\ref{fig:rd_g2}, the dark (light) green shaded region correspond to the  $1\sigma$ ($2\sigma$) range of $\Delta a_e$ for Rb. The black (grey) shaded region represent the same for the Cs. It is seen from both panels that the cyan points can reach the $1 \sigma$ range of $\Delta a_e$ value. However, when $\mu\to e$ conversion constraint is applied, there are points which can still satisfy $\Delta a_e$ for Rb or Cs.
It is seen from both the panel that one can not simultaneously satisfy for $\Delta a_{\mu}$ and $\Delta a_e$. This is due to the constraints coming from cLFV decays especially from $\mu \rightarrow \, e \gamma $ and $B^{+} \rightarrow \, K^{+} \, \nu \,\bar{\nu}$.  
To explain the $\Delta a_{\mu}$ anomaly $\yl{32}$ and $\yr{32}$ need to satisfy the relation mentioned in Table~\ref{tab:g-2}. Therefore, the constraints from $\mu \rightarrow \, e \, \gamma$ require both $\yl{31}$ and $\yr{31}$ very small,~${\cal O}(<10^{-6})$ and similarly, the constraint from $B \rightarrow \,K^{+} \, \nu\, \bar{\nu}$ requires $\yl{11}$ to smaller than $\mathcal{O} (0.1)$. Hence, $\Delta a_e$ is small in this case. On the other hand, for lowest values of $\yl{32}$ and $\yr{32}$ i.e when $\Delta a_{\mu}$ is small, $\yl{31}$ and $\yr{31}$ can be large and one can have large $\Delta a_e$ values. Therefore, $\Delta a_{\mu}$ is small, $\Delta a_e$ is large, and vice versa, indicating a contrast between the two lepton g-2 values.

\begin{figure}[t!]
    \centering
    \includegraphics[width=0.49\textwidth]{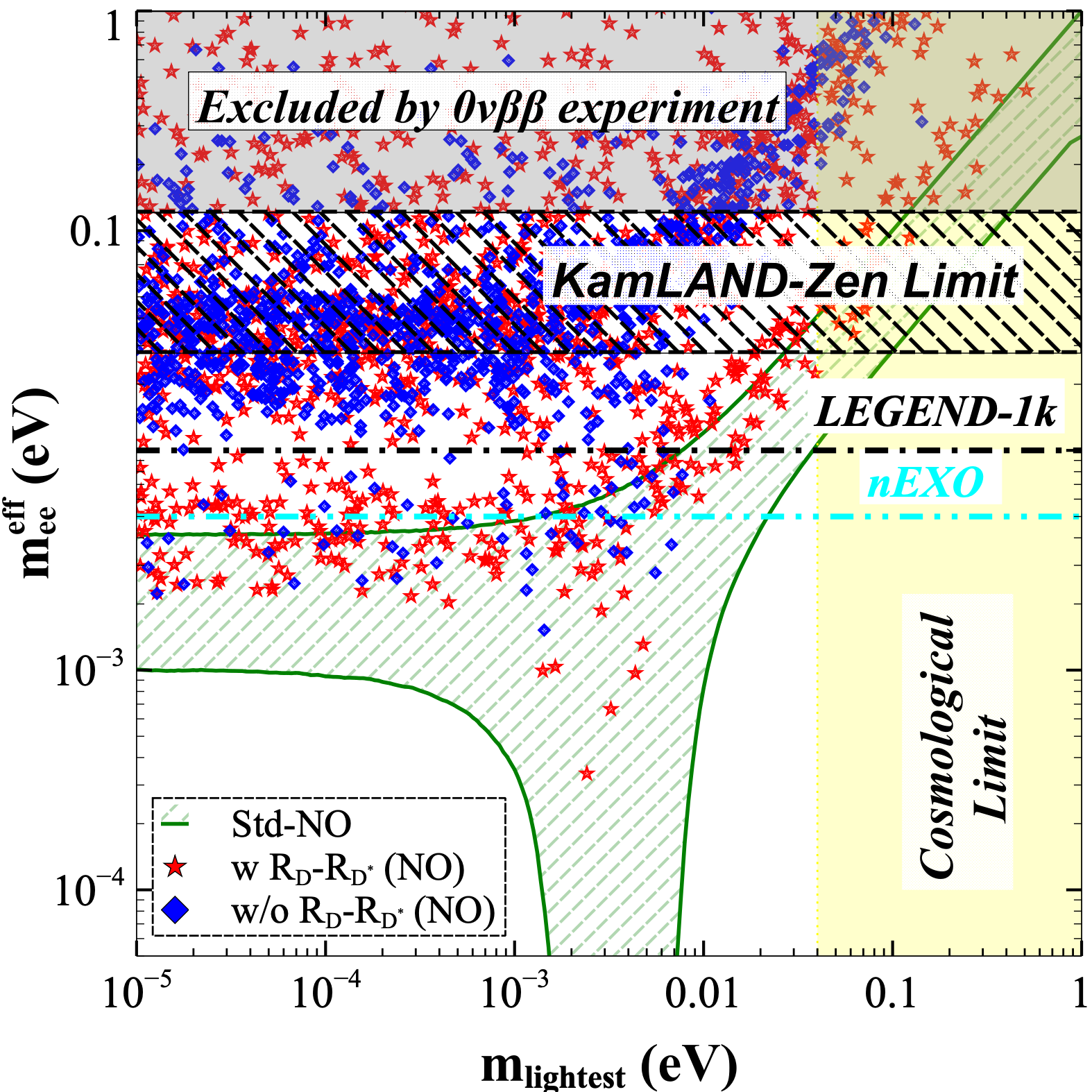}
    \includegraphics[width=0.49\textwidth]{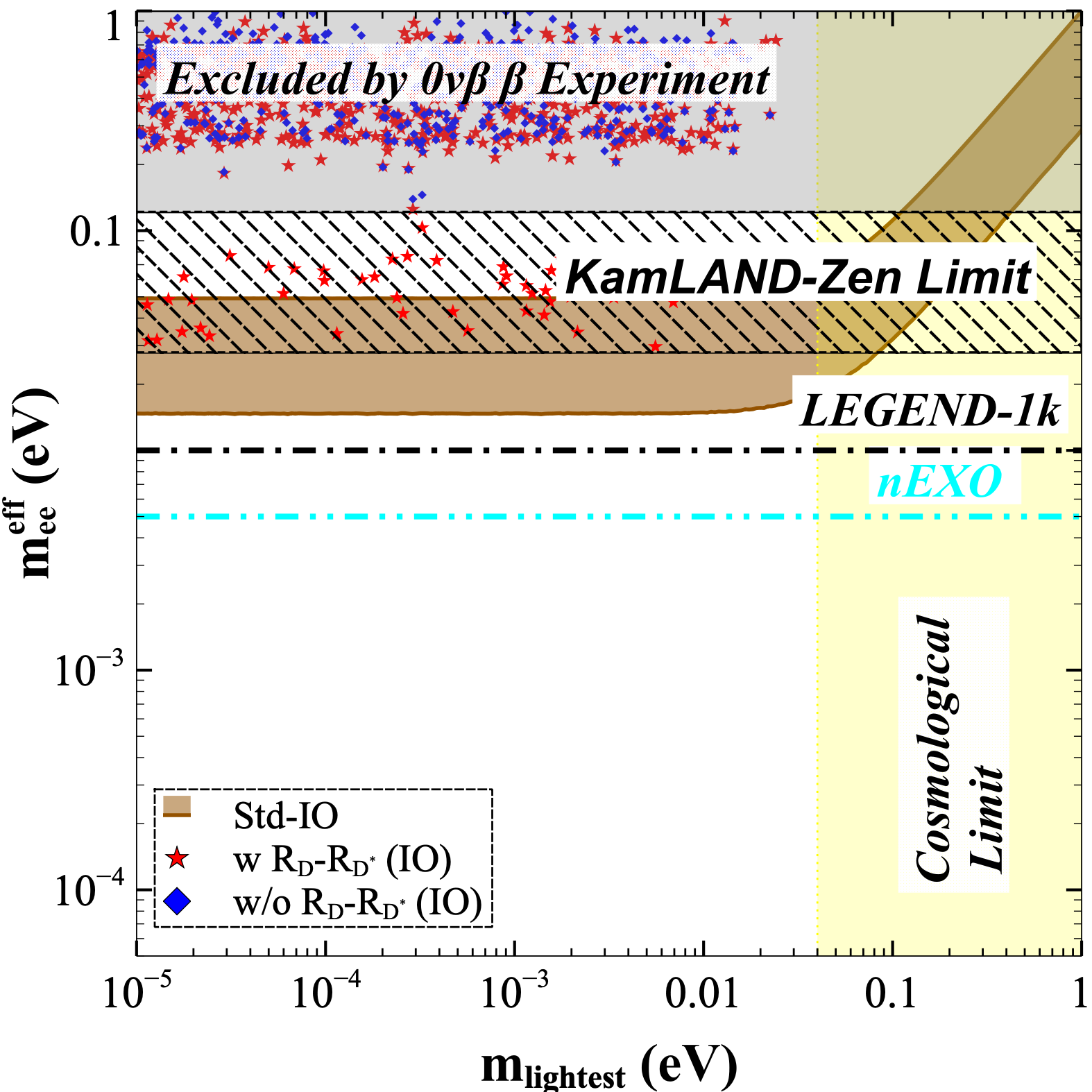}
    \caption{Plot of $m_{ee}^{\rm eff}$ as a function of the lightest neutrino mass where active neutrino masses obey NO (left panel) and IO (right panel). The standard NO and IO regions are denoted by lime green hatched and brown shaded regions, respectively. The blue diamonds (red stars) denote the $m_{ee}^{\rm eff}$ contributions in Case I and II, respectively. The magenta hatched horizontal region denotes the current KamLAND-Zen upper bound on $m_{ee}$ for different NMEs while the black dashed-dot (cyan dashed-dot-dot) line denotes the future reach of LEGEND-1000 (nEXO) on $m_{ee}$. The vertical light yellow shaded region represents the excluded range from the cosmological limit on the sum of light neutrino masses.}
    \label{fig:0nbb}
\end{figure}
\begin{figure}[h!]
    \centering
    \includegraphics[width=0.49\linewidth]{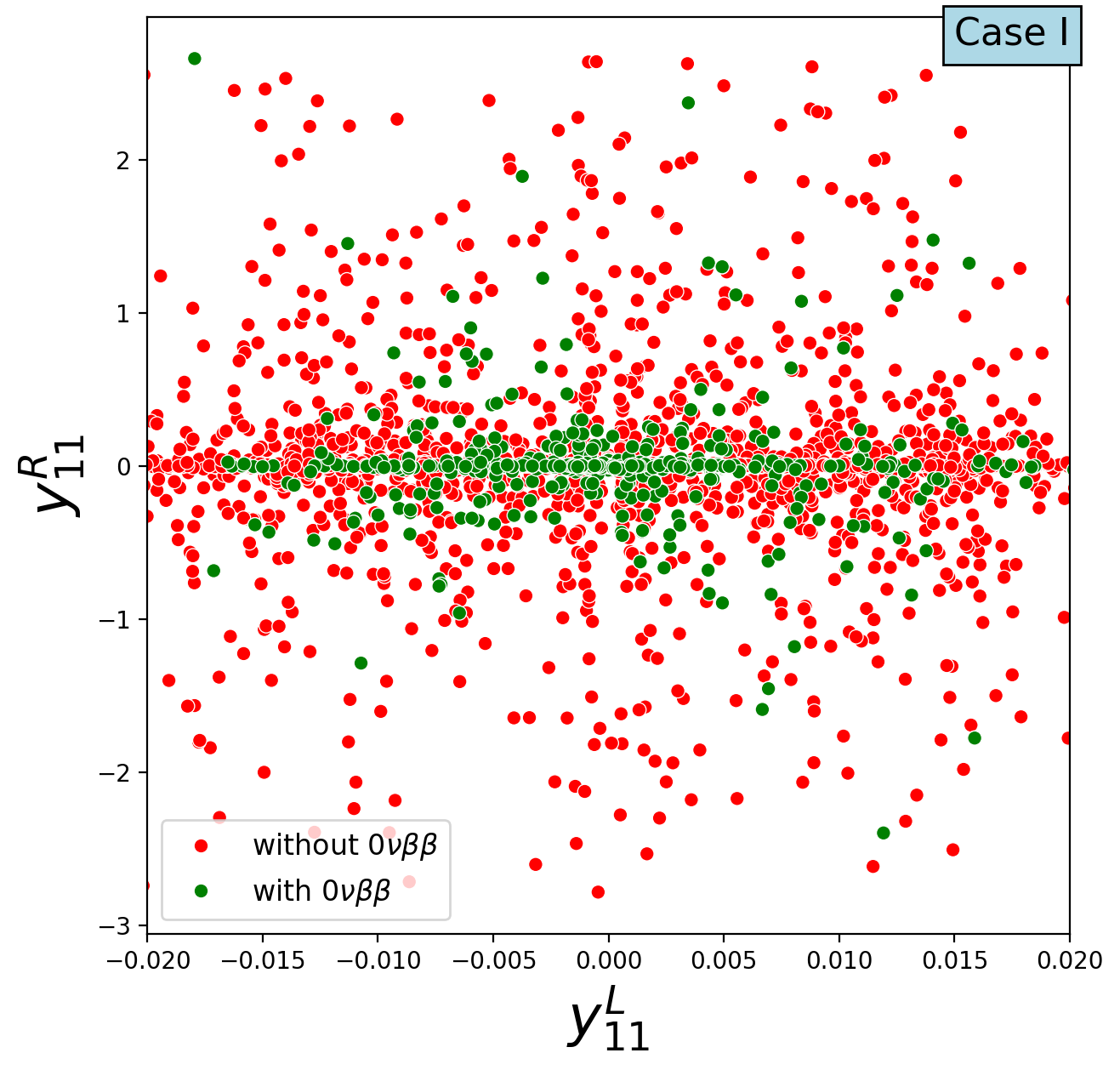}
    \includegraphics[width=0.49\textwidth]{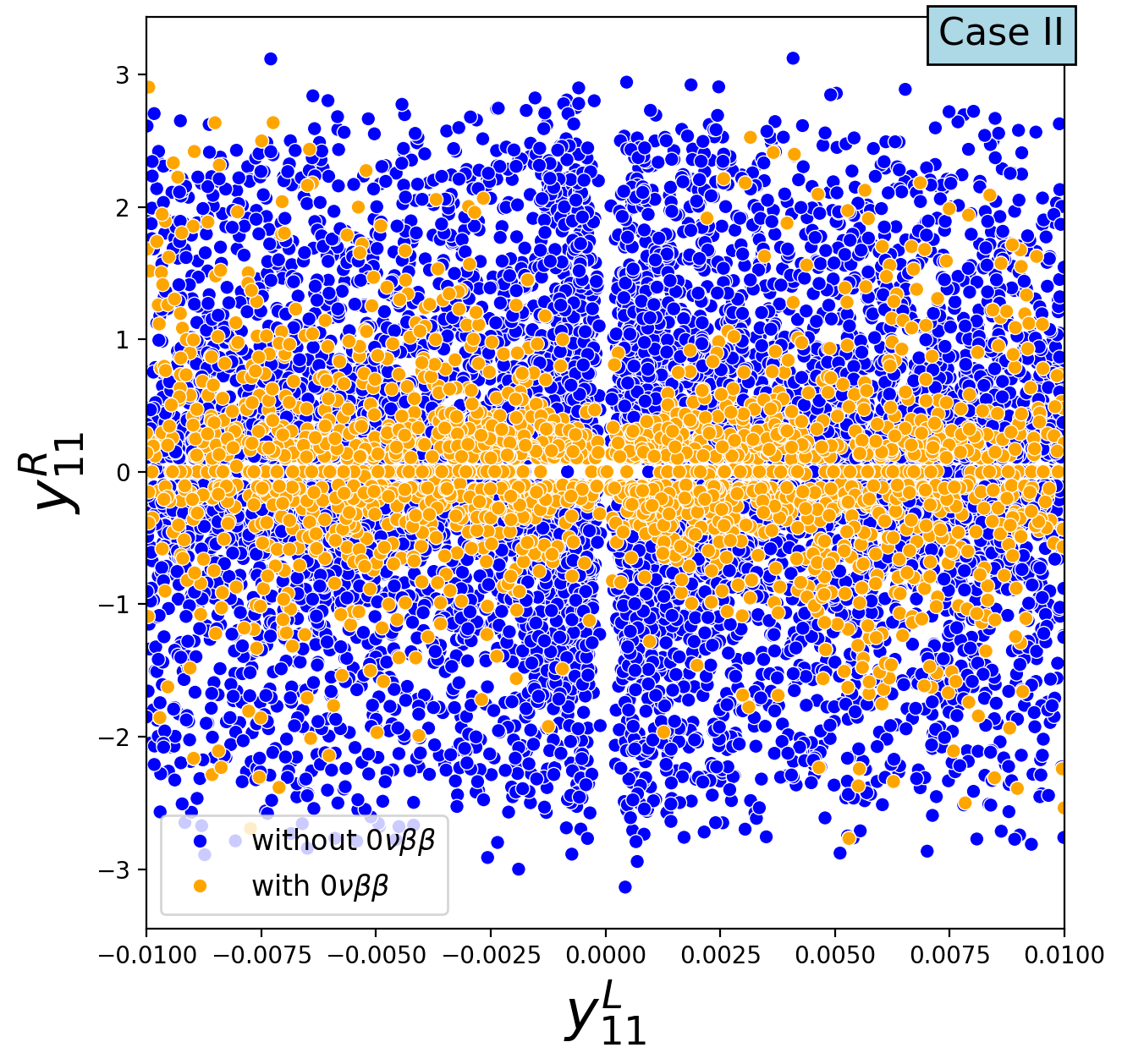}
    \caption{left panel: Plot of $\yl{11}-\yr{11}$ with and without the $0\nu\beta\beta$ decay (KamLAND-Zen) constraint on the parameter space for Case I. The red points correspond to the parameter space, indicated in Fig.~\ref{fig:cp-w-rd} and the green points are the allowed parameter space by the $0\nu\beta\beta$ decay.  right panel: Plot of $\yl{11}-\yr{11}$ with and without the $0\nu\beta\beta$ decay (KamLAND-Zen) constraint on the parameter space for Case II. The blue points correspond to the parameter space, indicated in Fig.~\ref{fig:cp-wo-rd} and the orange points denote the parameter space allowed by the $0\nu\beta\beta$ decay.}
    \label{fig:yl11-yr11}
\end{figure}

\subsection{Impact of $0\nu\beta\beta$ constraint on the leptoquark parameter space}
Now that we have identified the allowed leptoquark parameter space from the flavor observables and neutrino mass constraints, we proceed to study the $0\nu\beta\beta$ predictions. In Fig.~\ref{fig:0nbb}, we have plotted $m_{ee}^{\rm eff}$ [cf.~Eq.~\eqref{eq:thalf1}] as a function of the lightest neutrino mass for values of Yukawa couplings that are allowed from the combined analysis as presented in Figs.~\ref{fig:cp-w-rd} and \ref{fig:cp-wo-rd}. The oscillation parameters are varied in their allowed $3\sigma$ range~\cite{nufit}, and the Majorana phases are varied in the range ($0:\pi$). In the left (right) panel the red stars [blue diamonds] denote the values of $m_{ee}^{\rm eff}$ after including the leptoquark contributions for NO (IO) for Case I [II]. The standard contribution to $m_{ee}^{\rm eff}$ for NO (IO) is displayed in green (brown) for comparison purpose. Also, the current bound on $m_{ee}$ from KamLAND-Zen experiment~\cite{KamLAND-Zen:2024eml} and the future projections from LEGEND-1000~\cite{LEGEND:2021bnm} and nEXO~\cite{nEXO:2021ujk} are shown using magenta hatched filled band (with the band representing NME uncertainties), black dashed-dot and cyan dashed-dot-dot lines respectively. The cosmologically disfavored region is denoted by the yellow vertical band from Planck data~\cite{Planck:2018vyg}. From the left panel, we see that the cancellation region found in the standard 3-flavor picture is no longer there when leptoquark contributions are added.  
For some values of parameters, $m_{ee}^{\rm eff}$ can exceed the maximum value of  the standard scenario. Most of these regions can be explored in the future ton-scale experiments like LEGEND-1000 and nEXO, as can be seen from the figure.  In the scanned region,  $m_{ee}^{\rm eff}$ for NO including the leptoquark contribution goes into the standard IO region. In order to distinguish the standard IO region without leptoquarks from NO region with leptoquarks, just $0\nu\beta\beta$ is not enough, and we need additional flavor and/or collider observables. This is why the study of $0\nu\beta\beta$ correlations with other observables is so important. 

It is found that for certain parameter space, one can get large values of $m_{ee}^{\rm eff}$ which are already ruled out by $0\nu\beta\beta$ experiments. This suggests that $0\nu\beta\beta$ can further constrain the leptoquark parameter space, particularly $\yl{11} $ and $\yr{11}$ ( shown in Fig.~\ref{fig:yl11-yr11}), depending on the whether diagram (c) or diagram (d) gives dominant contribution. 
Similarly for IO (right panel), adding the leptoquark contribution  will enhance the value of $m_{ee}^{\rm eff}$ and there are a few points which give very high contributions and are disfavored by the current KamLAND-Zen upper bound on $m_{ee}$. 
When $R_D - R_{D^\ast}$ values are consistent with the SM, the values of $\yl{33}$ are small. Since $ \Y2 $ depends on the inverse of $ \YL $, $\Y2 $ becomes large, resulting in larger values of $m_{ee}^{\rm eff}$. This is also evident from Fig.~\ref{fig:0nbb}, which shows that the values of $m_{ee}^{\rm eff}$ are slightly higher in the scenario without $R_D - R_{D^\ast}$ anomaly present than in the scenario with $R_D - R_{D^\ast}$ anomaly. Additionally, the inverted mass ordering for neutrinos is relatively disfavored (compared to the normal ordering) by $0\nu\beta\beta$ decay constraints in the vanishing $R_D - R_{D^\ast}$ scenario.

\subsection{Effect of Varying Leptoquark Mass}
In the previous section, we have fixed the leptoquark mass at $2.0$ TeV for the parameter scanning. To see the effect of varying leptoquark mass, in this section, we fix the $Y_1^L$ and $Y_1^R$ matrix elements to a suitable representative benchmark value and vary the leptoquark mass to see what is its maximum allowed value that can explain the $R_{D^{(*)}}$ and $\Delta a_\ell$ anomalies. It is important to mention here that the following analysis is strictly valid only for the chosen benchmark values of the Yukawa couplings: 
\beqa
Y_1^L &=& \begin{pmatrix}
    -0.02 & 0 & -0.001 \\
    0 & 0 & 0.001 \\
    0 & 1.5 & 0.5
\end{pmatrix} , 
    \quad \quad Y_1^R = \begin{pmatrix}
    1.0 & 0 & 0.01 \\
    0 & 0 & 1.8 \\
    0 & 0.001  & 10^{-4}
\end{pmatrix} . \label{eq:bp}
\eeqa 

\begin{figure}[t!]
    \centering
    \includegraphics[scale=0.5]{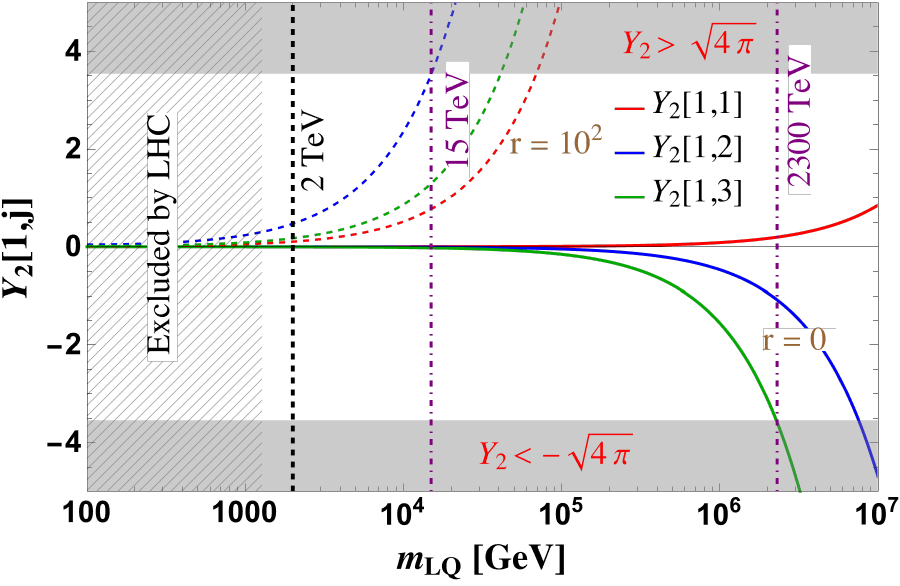}
    \includegraphics[scale=0.5]{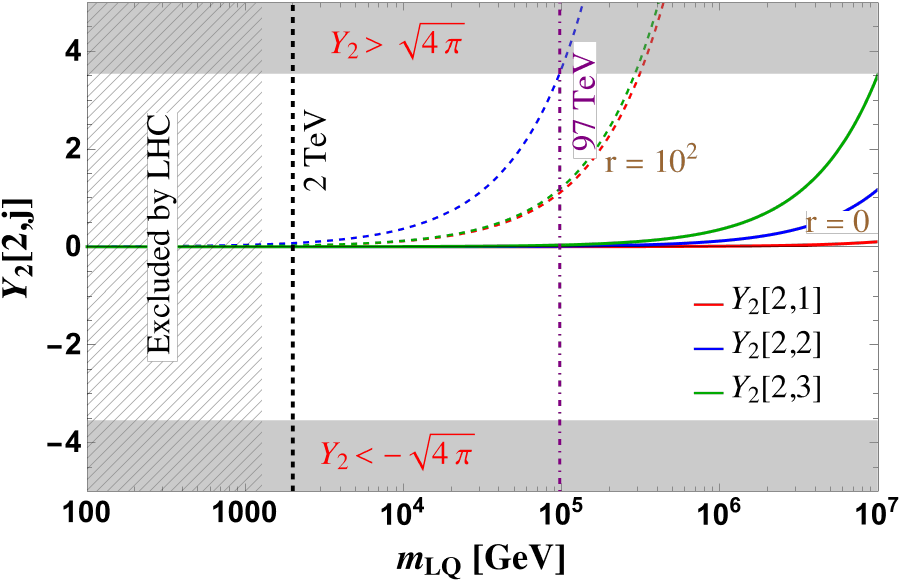}
    \includegraphics[scale=0.5]{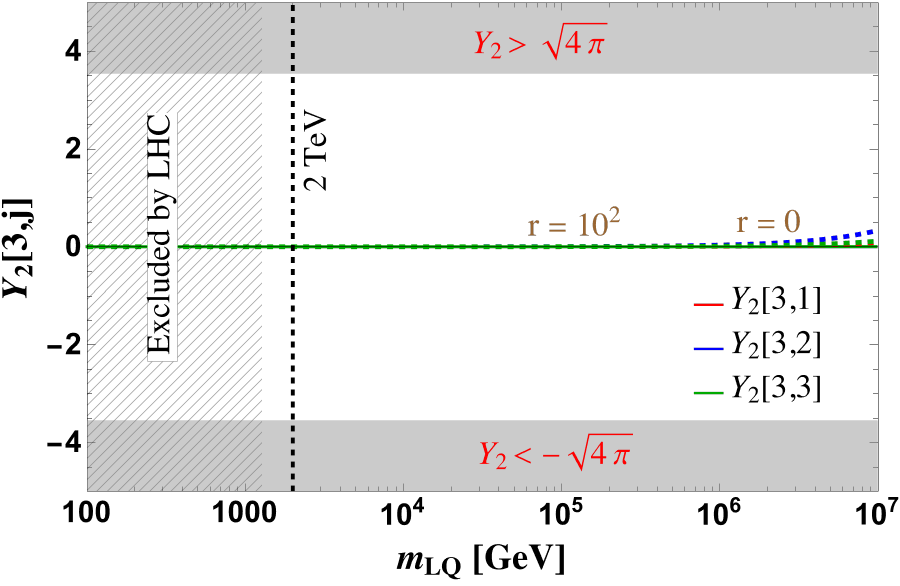}
    \caption{Variation of $Y_2$ elements with leptoquark mass when all the elements of $Y_1^L$ and $Y_1^R$ are fixed as given by Eq.~(\ref{eq:bp}). The solid (dashed) curves are for $r=0~(10^2)$. The horizontal shaded regions represent the perturbativity limit on the Yukawa couplings. The vertical hatched region is excluded by LHC searches for leptoquarks.}
    \label{fig:yukawa}
\end{figure}
Once $Y_1^L$ and $Y_1^R$ are fixed to the above values, the $Y_2 $ elements are determined from Eq.~(\ref{eq:casas-Ibarra}), giving the input parameters as discussed in the earlier Section. In determining the value of $Y_2 $, we have fixed the mass of the lightest neutrino as $\left( m_{\rm lightest}\right) \, 0\, \rm eV$. The $U_{\rm PMNS}$ parameters are fixed to their best fit values mentioned in Ref.~\cite{nufit} and the Majorana phases are fixed as $\alpha = 0 \,$ and $\beta = 0$.
In Fig.~(\ref{fig:yukawa}), we have shown the variation of the elements of $Y_2 $ as a function of the leptoquark mass for two different values of $r$. The hatched region on the left corresponds to the collider bound on $m_{\rm LQ}>1280$ GeV from direct LHC searches~\cite{ParticleDataGroup:2022pth}. The vertical lines correspond to the different leptoquark masses as depicted in the figure. It is seen that the magnitude of the $Y_2 $ matrix elements increases with increasing leptoquark mass. This is because as $m_{\rm LQ}$ grows, $\mathbf{C}_1 \approx \kappa \, v/m_{\rm LQ}^2$  [cf. Eq.~(\ref{eq:casas-Ibarra})]; therefore, the value of $Y_2 $ increase as the LQ mass increases. This can also be understood from Eq.~\ref{eq:BP_point} that $Y_2 $ must increase as $m_{\rm LQ}$ increases to generate neutrino masses of the $\mathcal{O} (10^{-2})$ eV. As $Y_2 $ is proportional to $\mathbb{R}$ [cf. Eq.~(\ref{eq:casas-Ibarra})], it also increases with $r$ for a given value of the leptoquark mass. From this figure, we find that the elements of $ Y_2 $ matrix exceed the perturbativity limit for $m_{\rm LQ} \, > 15 $ TeV for $r=100$. So, large values of $r$ are not allowed by the perturbativity conditions of the elements of $Y_2 $ matrix. 

\begin{figure}[t!]
    \centering
    \includegraphics[scale=0.5]{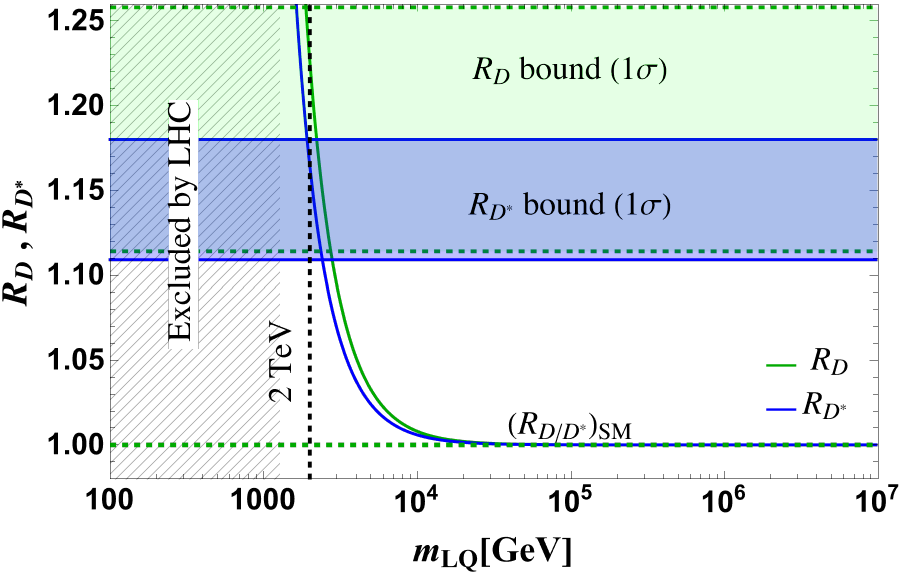}
    \includegraphics[scale=0.48]{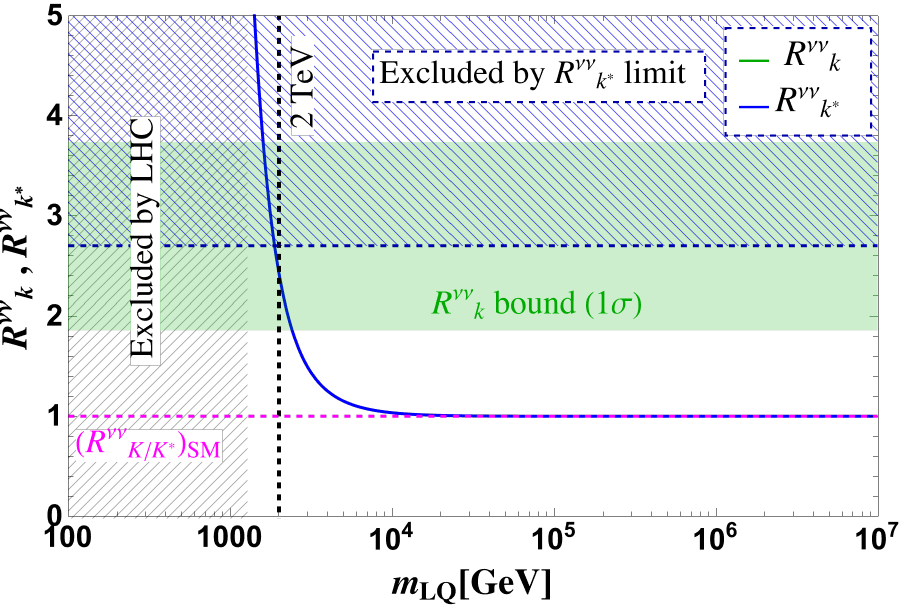}\\[2pt]
    \includegraphics[scale=0.5]{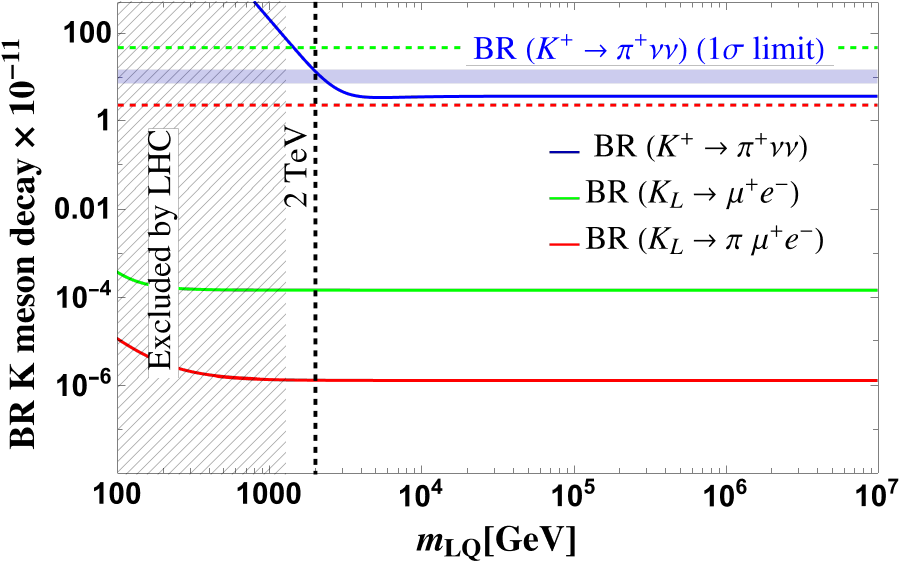}
    \includegraphics[scale=0.5]{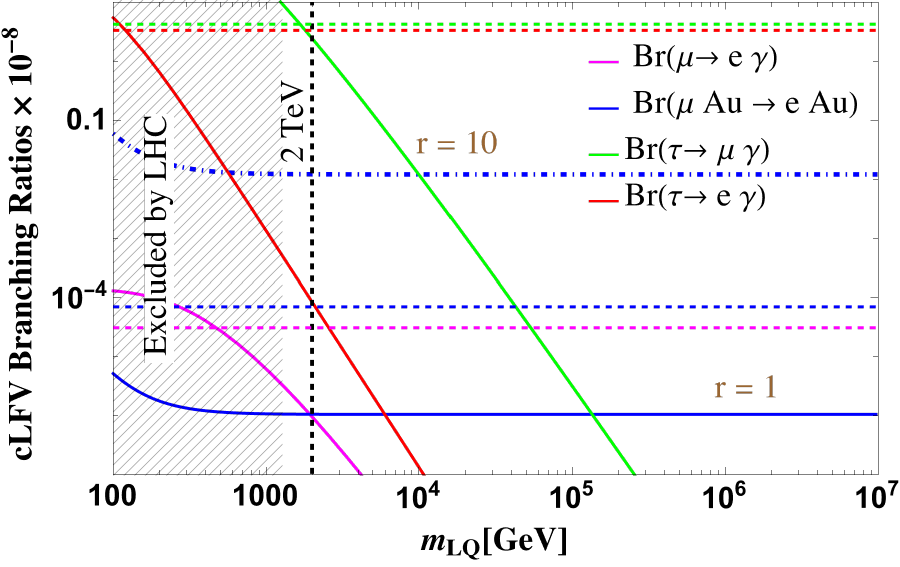}
    \includegraphics[scale=0.5]{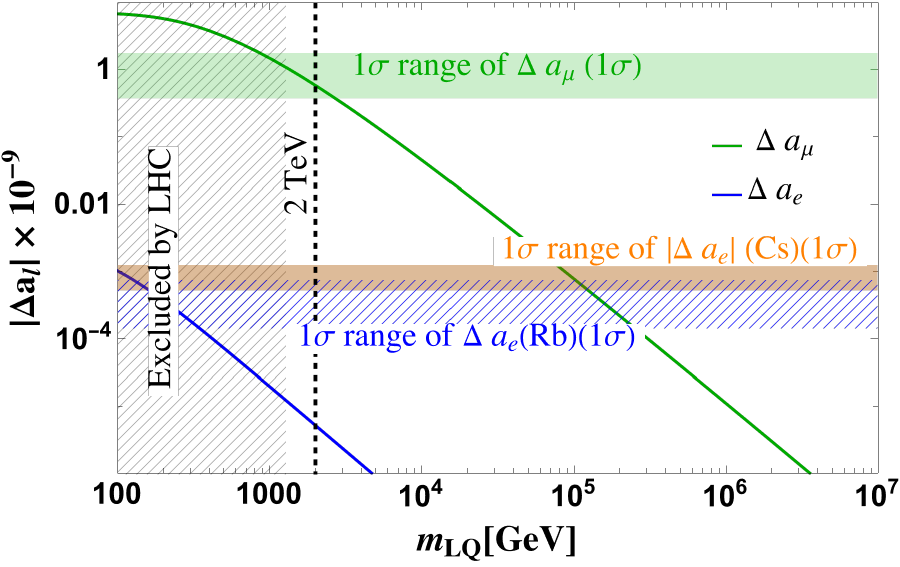}
    \includegraphics[scale=0.5]{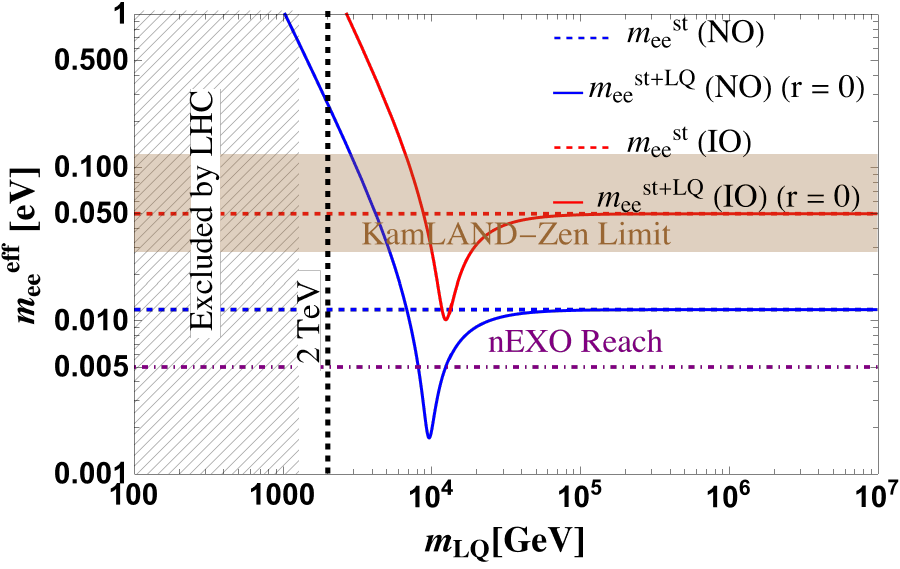}
    \caption{{\it Upper left:} Variation of $R_{D}$ (green) and $R_{D^\ast}$ (blue) as a function of leptoquark mass. The green (light blue) shaded region corresponds to the current experimental values on $R_{D}$ ($R_{D^\ast}$) at $1\sigma$. The green dashed line corresponds to the SM prediction. {\it Upper right:} $R_{K}^{\nu\nu}$ (green) and  $R_{K^\ast}^{\nu\nu}$ (blue) as a function of leptoquark mass for three values of $r$. The green-shaded region corresponds to the current experimental value on $R_{K}^{\nu\nu}$ at $1\sigma$, and the light blue-shaded region shows the experimental limit on ($R_{K^\ast}^{\nu\nu}$). {\it Middle left:} BR of K meson decays ($K^{+}\rightarrow \pi^+ \, \nu  \, \bar \nu$ (blue, $1\sigma$), $K_L \rightarrow \mu^{+}\, e^-$ (green), $K_L \rightarrow \pi \mu^+\,e^-$ (red)) as a function of leptoquark mass. Blue band denotes the experimental value of $K^{+}\rightarrow \pi^+ \, \nu \,\bar \nu $. Green and red lines correspond to the current experimental limits on the corresponding BRs.  {\it Middle right:}  BRs of cLFV decays as a function of the leptoquark mass. Magenta, blue, green and brown curves correspond to BRs of $\mu \rightarrow \, e \, \gamma$, $\tau \rightarrow\, \mu \, \gamma$, $\tau \rightarrow \, e\, \gamma$ and $\mu\to e$ conversion ratio respectively, and the horizontal lines denote their experimental limits. {\it Lower left:} Variation $\Delta a_{\mu}$ and $\left|\Delta a_e \right|$ as a function of leptoquark mass. Green bands correspond to the $1\sigma$ value of $\Delta a_{\mu}$ whereas, blue (brown) band shows the same for $\Delta a_e \, \left(\Delta a_e  \right) $ for Rb (Cs) .  {\it Lower right:} Variation of $m_{ee}^{\rm eff}$ as a function of $m_{\rm LQ}$ for $r = 1$ (allowed by $\mu \rightarrow \, e $ conversion). The solid blue (red) dashed line denotes the standard contribution for NO. The horizontal brown shaded band is the current KamLAND-Zen limit at 90\% CL including NME uncertainties, and the purple dashed-dot line is the future nEXO projection.}
    \label{fig:rdrklfvg2}
\end{figure}

Fig.~\ref{fig:rdrklfvg2} shows the variation of $R_D$, $R_{D^\ast}$, $R_{K}^{\nu\nu}$, $R_{K^\ast}^{\nu\nu}$, cLFV decay BRs, $\mu\to e$ conversion rate, rare $K$ meson decay BRs and the $m_{ee}^{\rm eff}$ as a function of the leptoquark mass. From the left most panel, it is inferred that only a small region of $m_{\rm LQ}$ around 2 TeV scale is allowed for the current values of $R_D$ and $R_{D^\ast}$. As $m_{\rm LQ}$ increases, the value $R_D$-$R_{D^\ast}$ decreases and gradually converge to the SM value, and hence, the anomaly can not be explained for higher values of leptoquark mass. In the case of $R_{K}^{\nu\nu}$ and $R_{K^\ast}^{\nu\nu}$, the nature is similar to $R_{D}-R_{D^\ast}$. As leptoquark mass increases, $R_{K^{(*)}}^{\nu\nu}$ decreases and gradually goes out of the green allowed region for $R_{K}^{\nu\nu}$. Here, both $R_{K}^{\nu\nu}$ and $R_{K^{(*)}}^{\nu\nu}$ have the same value because throughout the leptoquark mass range the value of $Y_2 $ is small and $R_{K^{(*)}}^{\nu\nu}$ is dominated by only $C_L$ [cf. Eq.~(\ref{eq:rknunu})].  

The  BRs of cLFV decay are  inversely proportional to the leptoquark mass and gradually decrease as $m_{\rm LQ}$ increases as can be seen in the lower  left panel of Fig. \ref{fig:rdrklfvg2}.
However, the $\mu\to e$ conversion rate shows a different trend as it  is seen  to be independent of leptoquark mass for higher values of the leptoquark mass.
This can be understood from Eq.~(\ref{eq:cVLR}) where we see that
for large leptoquark mass and fixed $Y_1^L $ values, all the Wilson coefficients in Eq.~(\ref{eq:mueconversion}) except the parameter $ C_{V_{LR}}^{d} \sim y_2^2/m_{\rm LQ}^2$ is independent of $m_{\rm LQ}$ [(c.f Eq.~(\ref{eq:BP_point})] making the conversion ratio independent of $m_{\rm LQ}$

From  the lower right panel of Fig.~\ref{fig:rdrklfvg2}, it is observed  that  $m_{ee}^{\rm eff}$ for NO~(IO) decreases as $m_{\rm LQ}$ increases, and  for higher values of leptoquark mass, approaches the standard contribution [blue~(red) dashed line]. For certain range of $m_{\rm LQ}$, we find that $m_{ee}^{\rm eff} < m_{ee}^{\rm std}$ which means that in this range the leptoquark contribution is similar to the standard contribution and destructively interferes with the standard contribution for the chosen BP values. 
Similar behavior is also seen for the IO case. We find that for $m_{\rm LQ} = 2$ TeV, the chosen BP values satisfy all the low-energy constraints but $m_{ee}^{\rm eff} $ is higher than the KamLAND-Zen limit and hence the given benchmark point is discarded for both orderings of neutrino masses. Therefore, $0\nu\beta\beta$ decay  can be used in conjunction with the other flavor observables mentioned in Section~\ref{sec:flavor} to further constrain the leptoquark parameter space. Future ton-scale $0\nu\beta\beta$ experiments like nEXO and LEGEND-1000 will be important in probing the unexplored leptoquark parameter space for both NO and IO.
  
\section{Conclusions}  
\label{sec:con}
Scalar leptoquarks provide an attractive framework for neutrino mass generation radiatively. At the same time, the new  leptoquark interactions give rise to new contributions to various lepton number and flavor violating processes. We have considered the Standard Model augmented with two scalar leptoquarks. While a single leptoquark cannot generate correct neutrino mass, the combination of singlet-doublet leptoquarks ($S_1-\tilde{R}_2$) can generate the neutrino mass radiatively. Such models have a rich phenomenology since leptoquarks can couple to both quarks and leptons. Our main aim was to study the implications of this combination in context of $0\nu\beta\beta$ decay. To find the allowed values of the Yukawa couplings, we imposed the 
constraints from neutrino mass and mixing, as well as demanded compliance  with bounds coming from charged lepton flavor violation, lepton flavor universality violation and low-energy rare meson decays. We performed a comprehensive  global parameter scan satisfying all available experimental data arising from the above-mentioned constraints, as well as existing anomalies in $R_{D^{(*)}}$ and $\Delta a_{\ell}$ to constrain the allowed parameter space for various Yukawa matrix elements with a TeV-scale leptoquark mass. Our combined analysis reveals some interesting interplay and tensions coming from different constraints. For instance, in this two leptoquark scenario, the muon and electron $g-2$ values are in tension with each other. We found that the most stringent constraints on the allowed Yukawa couplings are obtained from $\mu \rightarrow e$ conversion in nuclei. We obtained the prediction for $m_{ee}^{\rm eff}$, arising from the combined contributions of standard and leptoquark contributions, for the Yukawa coupling values that pass all the above mentioned constraints. We find that the contribution from leptoquark mediated diagrams to $m_{ee}^{\rm eff}$ can be significant and even greater than the standard contribution for normal ordering of the light neutrinos and the cancellation region is no longer present around $m_{\rm{lightest}} \sim {\rm meV}$. The total value of $m_{ee}^{\rm eff}$ can lie in the desert region between the standard NO and IO regions and hence can be probed by future experiments like LEGEND-1000 and nEXO. On the other hand, for IO, we find that most of the points otherwise allowed by low-energy flavor constraints are disfavored by the current $0\nu\beta\beta$ limit from KamLAND-Zen. 

We have also varied the leptoquark mass fixing some benchmark values of Yukawa couplings  and  discussed the constraints on this from  different observables, as well as from the perturbativity bounds. For the chosen benchmark values, we find very stringent constraint on leptoquark mass, $m_{\rm LQ} > 6.6 $ TeV for IO, from the $0\nu\beta\beta$ decay. 

\section*{Acknowledgments}
S.G and D.P. thank Frank  Deppisch, Namit Mahajan, and  Saurabh Shukla for discussions. This work of BD was partly supported by the U.S. Department of Energy under grant No. DE-SC 0017987. BD also acknowledges the Center for Theoretical Underground Physics and Related Areas (CETUP* 2024) and the Institute for Underground Science at SURF for hospitality and for providing a stimulating environment, where this work was finalized. SG acknowledges the J.C. Bose Fellowship (JCB/2020/000011) from the Anusandhan National Research Foundation, Government of India. She also acknowledges Northwestern University (NU), where a part of  this work was done, for hospitality and Fullbright-Nehru Academic and Professional Excellence fellowship for funding the visit to NU. SG also acknowledges the hospitality at Technical University of Munich during the final stage of the work  and a grant from   the Excellence Cluster ORIGINS of the Deutsche Forschungsgemeinschaft (DFG,
German Research Foundation). CM acknowledges the support of the Royal Society, UK, through the Newton International Fellowship (grant number NIF$\backslash$R1$\backslash$221737). He also extends his gratitude to the Physical Research Laboratory (PRL), India, for their hospitality where this work was initiated, and for the support provided by the J.C. Bose Fellowship during his visit to PRL. The computations were performed on the Param Vikram-1000 High Performance Computing Cluster of the Physical Research Laboratory (PRL).

{\bf Note Added:} While we were finalizing this work, Ref.~\cite{Fajfer:2024uut} appeared,  which also studies the correlation between $0\nu\beta\beta$ and flavor observables in scalar leptoquark models. However, they have only focused on leptoquark masses beyond 100 TeV, whereas our main focus is on TeV-scale leptoquarks, where additional flavor and collider constraints come into play. 

\appendix
\section{Fierz Transformation List} \label{app:fierz}

Calculations of low-energy weak interactions with fermions often involve superposition of quartic products of Dirac spinors, where the order of the spinors varies among the terms. A common technique to standardize their ordering is known as the Fierz transformation. 
The usual Fierz relation can be written as~\cite{Nieves:2003in}
\beqa
e_{\Gamma} (1234) & = & \sum_{\Gamma'} F_{\Gamma\, \Gamma'} e_{\Gamma'} (1432) .
\eeqa
Here $\Gamma, \Gamma' \in \, \{S, V, T, A, P\}$ , $F_{\Gamma\, \Gamma'} $ are numerical coefficients and 
$
e_{\Gamma} (1234) \equiv   \left( \bar{w}_1 \Gamma w_2 \right) \left( \bar{w}_3 \Gamma w_4 \right) $, where $w_i$ are the spinors. 
In case of leptoquark interactions, we use the following Fierz relation
\beqa
e_{S}(1234) = \frac 14 \left[ e_{S}(1432) + e_V (1432) +\frac 12 e_T (1432) - e_A (1432) + e_P (1432) \right] .
\eeqa
Then the Fierz transformation list is as follows: 
\begin{align}
    \left( \bar{d}^c P_L \nu \right) \left( \bar{e} P_R u^c \right) & =  - \frac 12 \left( \bar{u} \gamma^{\mu}P_L d \right) \left( \bar{e} \gamma_{\mu} P_L \nu_L \right), \\
    \left( \bar{d}^c P_L \nu \right) \left( \bar{e} P_L u^c \right) & =  \frac 12 \left( \bar{d}^c P_L u^c \right) \,  \left( \bar{e} P_L \nu \right) + \frac 18 \left( \bar{d}^c \sigma^{\mu\nu}P_L u^c  \right) \,  \left( \bar{e}\sigma_{\mu\nu} P_L \nu \right) \nn 
\\ 
& =\frac 12 \left( \bar{u} P_L d \right) \,  \left( \bar{e} P_L \nu \right) - \frac 18 \left( \bar{u} \sigma^{\mu\nu}P_L d  \right) \,  \left( \bar{e}\sigma_{\mu\nu} P_L \nu \right) ,\\
\left( \bar{d}^cP_R \nu^c \right) \left( \bar{e}P_R u^c \right) &=  \frac 12	 \left( \bar{d}^c P_R u^c \right) \left( \bar{e} P_R \nu^c \right) + \frac 18 \left( \bar{d}^c \sigma^{\mu\nu}P_R u^c\right) \left( \bar{e} \sigma_{\mu\nu} P_R \nu^c\right) \nn \\
& = \frac 12	 \left( \bar{u} P_R d \right) \left( \bar{e} P_R \nu^c \right) - \frac 18 \left( \bar{u} \sigma^{\mu\nu}P_R d\right) \left( \bar{e} \sigma_{\mu\nu} P_R \nu^c\right), \\
\left( \bar{d}^c P_R \nu^c \right) \left( \bar{e} P_L u^c\right) & =  \frac 12 \left(\bar{d}^c \gamma^{\mu} P_L  u^c \right) \left( \bar{e} \gamma_{\mu} P_R \nu^c \right) \, = \, -\frac 12 \left( \bar{u} \gamma^{\mu} P_R d\right) \left( \bar{e} \gamma_{\mu} P_R \nu^c \right) .
\end{align}

\section{Sub-amplitudes, NMEs and PSFs for $0\nu\beta\beta$ decay}\label{app:sub-amplitudes}
In this model, operators that contribute to the $0\nu\beta\beta$ decay are ${\cal O}_{S+P}^{S+P},\, {\cal O}_{T+T_5}^{T+T_5}$ and ${\cal O}_{V+A}^{V+A}$ and their corresponding Wilson coefficients are $\epsilon_{S+P}^{S+P},\, \epsilon_{T+T_5}^{T+T_5}$ and $\epsilon_{V+A}^{V+A}$ respectively which is given in Eqs.~(\ref{eq:S+P}),~(\ref{eq:TR}) and (\ref{eq:v+A}). With these Wilson coefficients, we can write Eq.~(\ref{eq:half-life}) as 
\beqa
\left( T_{1/2}^{0\nu}\right)^{-1} &=& g_A^4 \Big[ G_{01} \, \left| {\cal A}_{\nu}\right|^2 + 4\, G_{02}\,\left| {\cal A}_{E}\right|^2 + 2\,G_{04} \left( \left|{\cal A}_{m_e}\right|^2 + \text{Re }\left[ {\cal A}_{m_e}^\ast \, {\cal A}_{\nu}\right] \right) \Big. \nn \\
& & \Big. - 2\, G_{03}\, \text{Re}\left[ {\cal A}_{\nu} \, {\cal A}_{E}^\ast + 2\, {\cal A}_{m_e}\, {\cal A}_{E}^\ast\right]  \Big],
\eeqa
where the sub-amplitudes ${\cal A}_i$ defined as \cite{Cirigliano:2018yza}
\beqa
\cal A_{\nu} &=& \frac{m^{st}_{ee}}{m_e}\, {\cal M}_{\nu}^{(3)} + \frac{m_N}{m_e} {\cal M}_{\nu}^{(6)} ,\\
{\cal A}_E &=& {\cal M}_{E,R}^{(6)} ,\\
{\cal A}_{m_e} &=& {\cal M}_{m_e,R}^{(6)} \, .
\eeqa
Here, $m_N$ is the nucleon mass  ($\sim 1 $ GeV), $m_e$ is the electron mass, ${\cal M}_{\nu}^{(3)} $ denotes the contribution induced by the light Majorana neutrinos and other ${\cal M}_{i}^{(6)} $ encapsulate the contributions from other dimension-6 LNV operators present in this model; c.f.~Eq.~\eqref{eq:S+P} and \eqref{eq:v+A}). 
These ${\cal M}_i's$ can be expressed in terms of Wilson coefficients and NMEs, 
\beqa
{\cal M}_{\nu}^{(3)} & =& -\frac{M_F}{g_A^2}+M_{GT} + M_T + \frac{2m_{\pi}^2 g_{\nu}^{NN}}{g_A^2}\,M_{F,sd} \,, \\
 {\cal M}_{\nu}^{(6)} &=& -\frac{B}{m_N}\, \epsilon_{S+P}^{S+P} \,M_{PS} + 
 \epsilon_{T+T_5}^{T+T_5} \,M_{T6} \, , \\
\mathcal{M}_{m_e,R}^{(6)} &=&  \epsilon_{V+A}^{V+A}\, M_{m_e,R}\,,\\
\mathcal{M}_{E,R}^{(6)} &=& \epsilon_{V+A}^{V+A}\,\, M_{E,R}\,,
\eeqa
where $g_{\nu}^{NN} \sim  {\cal O} (F_\pi^{-2}) $ and $B = 2.7 $ GeV  at $\mu=2$ GeV in the $\overline{\rm MS}$ scheme \cite{Cirigliano:2018yza} . The NMEs can be calculated via \cite{Cirigliano:2018yza} 
\beqa
M_{GT} &=&  M_{GT}^{AA} + M_{GT}^{AP} + M_{GT}^{PP} +  M_{GT}^{MM}\, ,\\
    M_{T}  &= & M_{T}^{AP} + M_{T}^{PP} + M_{T}^{MM} \, ,\\
    M_{PS}  &= & \frac{1}{2} M_{GT}^{AP} + M_{GT}^{PP} + \frac{1}{2} M_{T}^{AP} + M_{T}^{PP}\, , \\
    M_{T6}  &=&  2\frac{g_T^\prime-g_T^{NN}}{g_A^2} \frac{m_\pi^2}{m_N^2} M_{F,sd} - 8 \frac{g_T}{g_M} (M_{GT}^{MM}+ M_T^{MM}) \nonumber \\
    & & + g_T^{\pi N} \frac{m_\pi^2}{4m_N^2}(M_{GT,sd}^{AP}+ M_{T,sd}^{AP}) +  g_T^{\pi \pi} \frac{m_\pi^2}{4m_N^2}(M_{GT,sd}^{PP}+ M_{T,sd}^{PP}) , \\ 
      M_{m_e,R} & = &\frac{1}{6} \left( \frac{g_V^2}{g_A^2} M_{F} + \frac{1}{3} \, (M_{GT}^{AA}- 4 M_{T}^{AA}) +3 \, (M_{GT}^{AP}+ M_{GT}^{PP} + M_T^{AP}+M_T^{PP})\right . \nonumber \\
     & & \left . - 12 \, \frac{g_{VR}^{m_e}}{g_A^2} M_{F,sd}\right) \, , \\
      M_{E,R} & =& -\frac{1}{3} \left( \frac{g_V^2}{g_A^2} M_F - \frac{1}{3} (2 M_{GT}^{AA}+M_T^{AA}) + 6 \frac{g_{VR}^E}{g^2_A} M_{F,sd}\right) \, .
\eeqa
The values of the LECs ($g_i$'s) are taken from Table 1 in Ref.~\cite{Cirigliano:2018yza}. 
The same values are also employed in the $\tt \nu DoBe$ package \cite{Scholer:2023bnn} which we use to calculate the half-lives. The values of the NMEs are given in Table~\ref{tab:NMEs} and the phase space factors are given  in Table~\ref{tab:PSFs}.

\begin{table}[!t]
\renewcommand{\arraystretch}{1.3}
\centering
\begin{tabular}{|c|c|c|c|c|c|c|c|}
\hline
$M_F$ & $M_{GT}^{AA}$ & $M_{GT}^{AP}$ & $M_{GT}^{PP}$ & $M_{GT}^{MM}$ & $M_{T}^{AA}$ & $M_{T}^{AP}$ & $M_{T}^{PP}$ \\ \hline
-0.52 & 3.203 & -0.45 & 0.09 & 0.10 & 0.00 & 0.12 & -0.03  \\ \hline\hline
$M_{T}^{MM}$ & $M_{F,sd}$ & $M_{GT,sd}^{AA}$ & $M_{GT,sd}^{PP}$ & $M_{T,sd}^{AP}$ & $M_{T,sd}^{AP}$ & $M_{T,sd}^{PP}$ & \\ \hline
0.02 & -0.76 & 2.40 & -0.71 & 0.17 & -0.38 & 0.12 &   \\ \hline
\end{tabular}
\caption{\label{tab:NMEs} Nuclear matrix elements of $^{136}{\rm Xe}$ obtained using IBM2 \cite{Scholer:2023bnn}.}
\end{table}

\begin{table}[!t]
\renewcommand{\arraystretch}{1.3}
\centering
\begin{tabular}{|c|c|c|c|c|c|}
\hline
$G_{01}$ & $G_{02}$ & $G_{03}$ & $G_{04}$ & $G_{06}$ & $G_{09}$ \\ \hline\hline
2.09 & 5.15 & 1.40 & 1.88 & 2.86 & 4.59 \\ \hline
\end{tabular}
\caption{\label{tab:PSFs} Phase space factors of $^{136}{\rm Xe}$ given in units of $10^{-14} {\, \rm yr}^{-1}$~\cite{Horoi:2017gmj,Scholer:2023bnn}. }
\end{table}

\section{Loop Functions for cLFV and g-2}

\label{app:LFV}

The loop functions $ \mathcal{F}(x)$ and $ \mathcal{G}(x)$ appearing in Eqs.~\eqref{eq:sigmaL1}-\eqref{eq:sigmaR2} can be defined as 
\beqa
\mathcal{F}(x) &=& Q_S f_S(x) - f_F(x) \, , \qquad 
\mathcal{G}(x)  =  Q_S \, g_S(x) - g_F(x) , 
\eeqa
where $Q_S$ is the electromagnetic charge for the scalar leptoquark and 
\beqa 
f_S(x) & =& \frac{x+1}{4(1-x)^2} + \frac{x \, \ln x}{2(1-x)^3} ,  \\
f_F(x) &=& \frac{x^2 -5x-2}{12(x-1)^3} + \frac{x\, \ln x}{2(x-1)^4} , \\
g_S(x) &=& \frac{1}{x-1} - \frac{\ln x}{(x-1)^2}  , \\
g_F(x) &=&\frac{x-3}{2(x-1)^2} + \frac{\ln x}{(x-1)^3} .
\label{eq:FGfn}
\eeqa
For $x\ll 1$, these loop functions can be approximated as 
\beqa
f_S(x) &  \simeq & \frac 14 , \quad 
f_F(x)  \simeq  \frac 16, \quad 
g_S(x) \simeq  -\ln (x) , \quad 
g_F(x) \simeq  -\ln (x) \, .\label{eq:FGfn2}
\eeqa
Note that for leptoquark $\widetilde{R}_2^{2/3}$ with $Q_S =2/3$, we have $\mathcal{F} (x) \sim  \frac{2}{3} \times \frac 14 - \frac 16 = 0$.  

\bibliographystyle{utcaps_mod}
\bibliography{reference}

\end{document}